    \documentclass[
        aps,
        prd,
        groupdaddress,
        reprint,
    ]{revtex4-2}
    \usepackage{
        amsmath,
        amssymb,
        newtxtext,
        newtxmath,
        graphicx,
        ae,
        aecompl,
        booktabs,
        wasysym,
        xcolor,
        soul,
    }
    \usepackage[T1]{fontenc}

    \newcommand*{\rd}[2]{\frac{\mathrm{d}#1}{\mathrm{d}#2}}

    \newcommand*{\rdil}[2]{\mathrm{d}#1 / \mathrm{d}#2}
    
    \newcommand*{\at}[1]{\left.#1\right|}
    \newcommand*{\abs}[1]{\left|#1\right|}
    
    \newcommand*{\bv}[1]{\boldsymbol{\mathbf{#1}}}
    \newcommand*{\uv}[1]{\hat{\bv{#1}}}
    \newcommand*{\p}[1]{\left(#1\right)}
    \newcommand*{\s}[1]{\left[#1\right]}
    \newcommand*{\z}[1]{\left\{#1\right\}}

    \colorlet{Corr}{red}

    \newlength{\colummwidth}
\begin{document}

\title{Spin-Orbit Misalignments in Tertiary-Induced Black-Hole Binary Mergers:
Theoretical Analysis}
\author{Yubo Su}
\affiliation{Cornell Center for Astrophysics and Planetary Science, Department
of Astronomy, Cornell University, Ithaca, NY 14853, USA}
\email{yubosu@astro.cornell.edu}
\author{Dong Lai}
\affiliation{Cornell Center for Astrophysics and Planetary Science, Department
of Astronomy, Cornell University, Ithaca, NY 14853, USA}
\affiliation{Tsung-Dao Lee Institute \& School of Physics and Astronomy, Shanghai
Jiao Tong University, 200240 Shanghai, China}
\author{Bin Liu}
\affiliation{Cornell Center for Astrophysics and Planetary Science, Department
of Astronomy, Cornell University, Ithaca, NY 14853, USA}

\date{\today}

\begin{abstract}
    Black-hole (BH) binary mergers driven by gravitational perturbations of
    tertiary companions constitute an important class of dynamical formation
    channels for compact binaries detected by LIGO/VIRGO\@. Recent works have
    examined numerically the combined orbital and spin dynamics of BH binaries
    that undergo large Lidov-Kozai (LK) eccentricity oscillations induced by a
    highly inclined companion and merge via gravitational wave radiation.
    However, the extreme eccentricity variations make such systems difficult to
    characterize analytically. In this paper, we develop an analytical formalism
    for understanding the spin dynamics of binary BHs undergoing LK-induced
    mergers. We show that, under certain conditions, the eccentricity
    oscillations of the binary can be averaged over to determine the long-term
    behavior of the BH spin in a smooth way. In particular, we demonstrate that
    the final spin-orbit misalignment angle $\theta_{\rm sl}$ is often related
    to the binary's primordial spin orientation through an approximate adiabatic
    invariant. Our theory explains the ``$90^\circ$ attractor'' (as found in
    recent numerical studies) for the evolution of $\theta_{\rm sl}$ when the
    initial BH spin is aligned with the orbital axis and the octupole LK effects
    are negligible---such a ``$90^\circ$ attractor'' would lead to a small
    binary effective spin parameter $\chi_{\rm eff}\sim 0$ even for large
    intrinsic BH spins. We calculate the deviation from adiabaticity in closed
    form as a function of the initial conditions. We also place accurate
    constraints on when this adiabatic invariant breaks down due to resonant
    spin-orbit interactions. We consider both stellar-mass and supermassive BH
    tertiary companions, and provide simple prescriptions for determining
    analytically the final spin-orbit misalignment angles of the merging BH
    binaries.
\end{abstract}

\keywords{
black hole physics -- gravitational waves -- binaries: general 
-- stars: kinematics and dynamics 
}

\maketitle

\section{Introduction}\label{s:intro}

As LIGO/VIRGO continues to detect mergers of black hole (BH) binaries
\citep[e.g.][]{Abbott:2016blz, abbott2019binary}, it is increasingly important
to systematically study various formation channels of BH binaries and their
observable signatures. The canonical channel consists of isolated binary
evolution, in which mass transfer and friction in the common envelope phase
cause the binary orbit to decay sufficiently that it subsequently merges via
emission of gravitational waves (GW) within a Hubble time
\citep[e.g.][]{lipunov1997black, lipunov2017first, podsiadlowski2003formation,
belczynski2010effect, belczynski2016first, dominik2012double, dominik2013double,
dominik2015double}. BH binaries formed via isolated binary evolution are
generally expected to have small misalignment between the BH spin axis and the
orbital angular momentum axis \citep{postnov2019black,
belczynski2020evolutionary}. On the other hand, various flavors of dynamical
formation channels of BH binaries have also been studied. These involve either
strong gravitational scatterings in dense clusters
\citep[e.g.][]{zwart1999black, o2006binary, miller2009mergers,
banerjee2010stellar, downing2010compact, ziosi2014dynamics, rodriguez2015binary,
samsing2017assembly, samsing2018black, rodriguez2018post, gondan2018eccentric}
or more gentle ``tertiary-induced mergers'' \citep[e.g.][]{
bin_misc1, bin_misc2, bin_misc3, bin_misc4, bin_misc5, blaes2002kozai,
miller2002four, wen2003eccentricity, antonini2012secular, antonini2017binary,
silsbee2016lidov, bin1, bin2, randall2018induced, hoang2018black}. The dynamical
formation channels generally produce BH binaries with misaligned spins
with respect to the orbital axes.

GW observations of binary inspirals can put constraints on BH masses and spins.
Typically, the spin constraints come in the form of two dimensionless
mass-weighted combinations of the component BH spins: (i) the aligned spin
parameter
\begin{equation}
    \chi_{\rm eff} \equiv \frac{m_1 \chi_1 \cos \theta_{\rm s_1 l}
            + m_2 \chi_2 \cos \theta_{\rm s_2 l}}{m_1 + m_2},
            \label{eq:chi_eff}
\end{equation}
where $m_{1,2}$ are the masses of the BHs, $\theta_{\rm s_i l}$ is the angle
between the $i$-th spin and the binary orbital angular momentum axis, and
$\chi_i \equiv cS_i / (Gm_i^2)$ is the dimensionless Kerr spin parameter; and
(ii) the perpendicular spin parameter \citep{schmidt2015towards}
\begin{equation}
    \chi_{\rm p} \equiv \max\z{
        \chi_1 \sin \theta_{\rm s_1 l}, \frac{q\p{4q + 3}}{4 + 3q} \chi_2 \sin
        \theta_{\rm s_2 l}},
\end{equation}
where $q \equiv m_2 / m_1$ and $m_1 \geq m_2$. The systems
detected in the first and second observing runs (O1 and O2) of
LIGO/VIRGO  have $\chi_{\rm eff} \sim 0$ (but see
\citep{zackay2019highly, venumadhav2020new} for exceptions). In the third
observing run (O3) of LIGO/VIRGO, two events exhibit substantial spin-orbit
misalignment. In GW190412 \citep{GW190412}, the two BH component masses are
$29^{+5.0}_{-5.3}M_\odot$ and $8.4^{+1.7}_{-1.0}M_\odot$. The primary (more
massive) BH is inferred to have $\chi_1 = 0.43^{+0.16}_{-0.26}$, and the
effective spin parameter of the binary is constrained to be $\chi_{\rm eff} =
0.25^{+0.09}_{-0.11}$, indicating a non-negligible spin-orbit misalignment
angle. In GW190521 \citep{190521}, the two component BHs have masses of
$85^{+21}_{-14}M_{\odot}$ and $66^{+17}_{-18}M_{\odot}$ and spins of $\chi_1 =
0.69^{+0.27}_{-0.62}$ and $\chi_2 = 0.72^{+0.24}_{-0.64}$. The binary's aligned
spin is $\chi_{\rm eff} = 0.08^{+0.27}_{-0.36}$ while the perpendicular spin is
$\chi_{\rm p} = 0.68^{+0.25}_{-0.37}$, again suggesting significant spin-orbit
misalignments.

\begin{figure}
    \centering
    \includegraphics[width=\colummwidth]{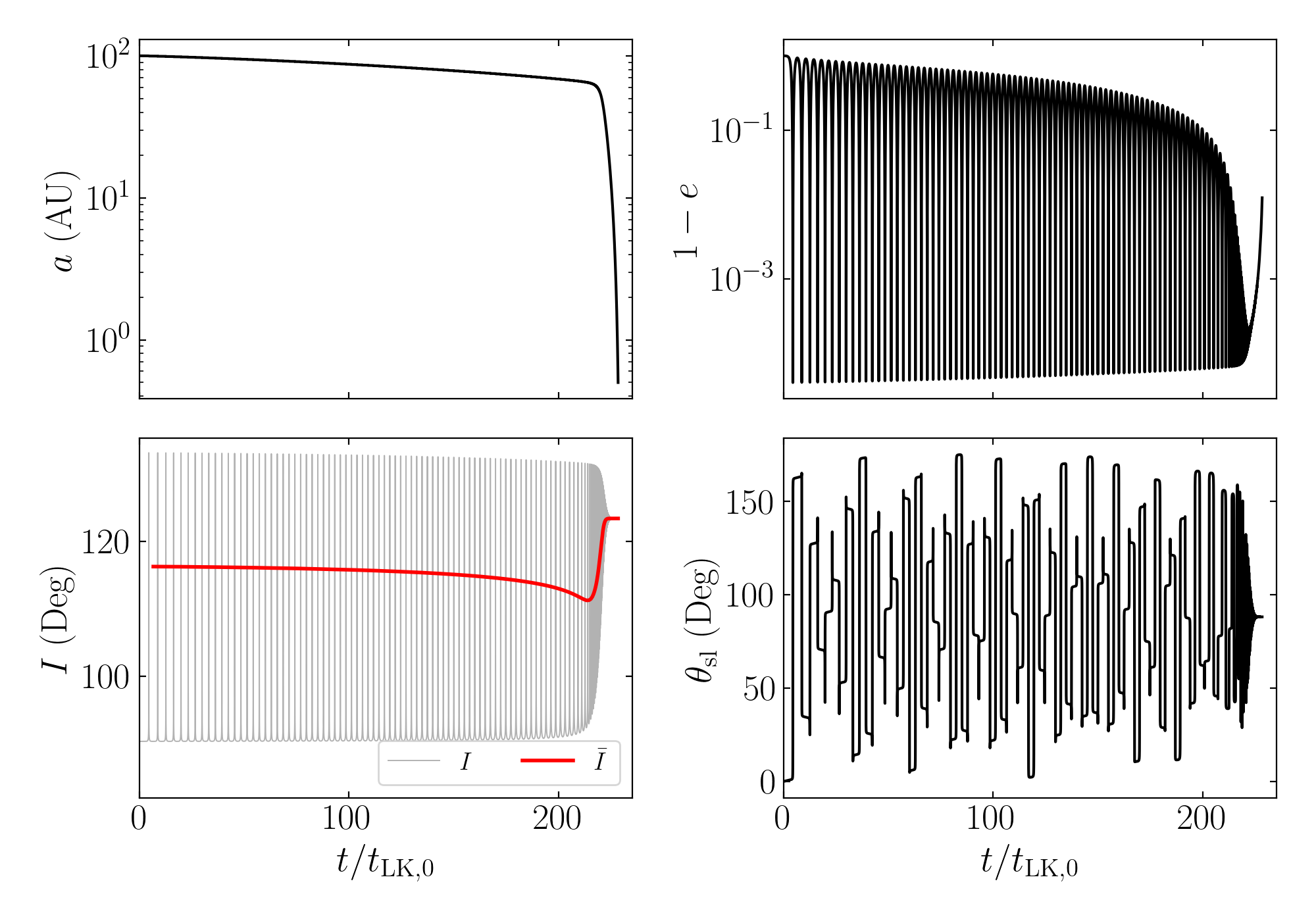}
    \caption{An example of the ``$90^\circ$ spin attractor'' in LK-induced BH
    binary mergers. The four panes show the time evolution of the binary
    semi-major axis $a$, eccentricity $e$, inclination $I$ [the red line denotes
    the averaged $\bar{I}$ given by Eq.~\eqref{eq:barI}], and spin-orbit
    misalignment angle $\theta_{\rm sl}$. The unit of time $t_{\rm LK, 0}$ is
    the LK timescale [Eq.~\eqref{eq:t_lk}] evaluated for the initial conditions.
    The inner binary has $m_1 = 30M_{\odot}$, $m_2 = 20M_{\odot}$, and initial
    $a_0 = 100\;\mathrm{AU}$, $e_0 = 0.001$, $I_0 = 90.35^\circ$ (with respect
    to the outer binary), and $\theta_{\rm sl, 0} = 0$. The tertiary has $a_{\rm
    out} = 2.2\;\mathrm{pc}$, $e_{\rm out} = 0$, and $m_3 = 3 \times 10^7
    M_{\odot}$ (The result depends only on $m_3 /
    \tilde{a}_{\rm out}^3$, provided that $L_{\rm out} \gg L$). It can be seen
    that $\theta_{\rm sl}$ evolves to $\sim 90^\circ$ as $a$ decays to smaller
    values, and we stop the simulation when $a = 0.5\;\mathrm{AU}$ as
    the LK oscillation ``freezes'' and $\theta_{\rm sl}$ has
    converged to a constant value.}\label{fig:4sim_90_350}
\end{figure}

\begin{figure}
    \centering
    \includegraphics[width=\colummwidth]{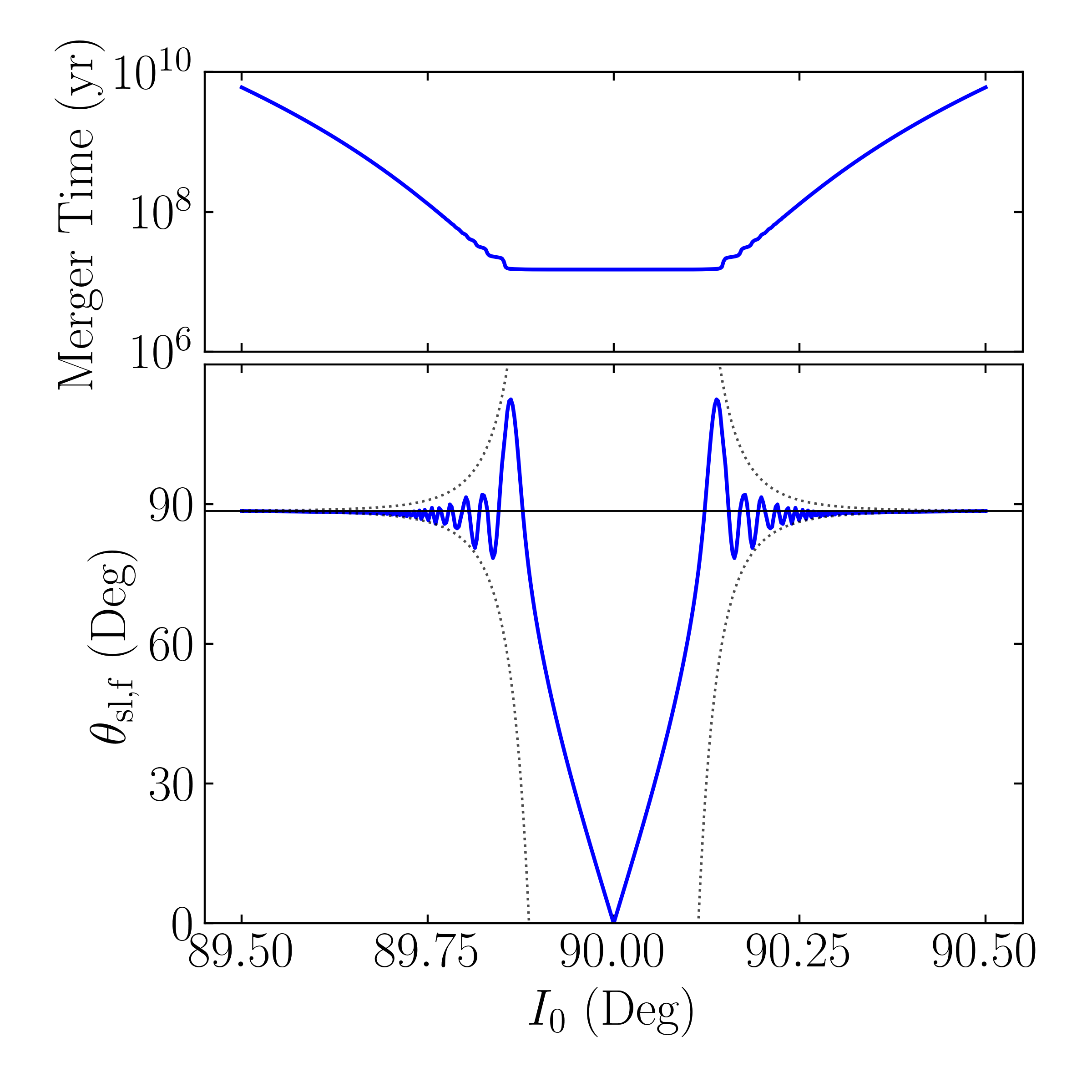}
    \caption{The merger time and the final spin-orbit misalignment angle
    $\theta_{\rm sl, f}$ as a function of the initial inclination $I_0$
    for LK-induced mergers. The other parameters are the same as those in
    Fig.~\ref{fig:4sim_90_350}. For $I_0$ somewhat far away from
    $90^\circ$, the resulting $\theta_{\rm sl, f}$ are all quite near
    $90^\circ$. In the lower panel, the horizontal black solid
    line shows the predicted $\theta_{\rm sl, f}$ if $\bar{\theta}_{\rm e}$ is
    conserved, i.e.\ Eq.~\eqref{eq:4_4_general}, and the black dashed line shows
    Eq.~\eqref{eq:qslf_plot_black}, which provides an estimate for the deviation
    from the $90^\circ$ attractor.}\label{fig:qslscan}
\end{figure}

\citet[][hereafter LL17, LL18]{bin1, bin2}, and \citet{bin3} carried out a
systematic study of binary BH mergers in the presence of a tertiary companion.
LL17 pointed out the important effect of spin-orbit coupling (de-Sitter
precession) in determining the final spin-orbit misalignment angles of BH
binaries in triple systems. They considered binaries with sufficiently compact
orbits (so that mergers are possible even without a tertiary) and showed that
the combination of LK oscillations (induced by a modestly inclined tertiary) and
spin-orbit coupling gives rise to a broad range of final spin-orbit
misalignments in the merging binary BHs. We call these mergers \emph{LK-enhanced
mergers}. On the other hand, LL18 considered the more interesting case of \emph{
LK-induced mergers}, in which an initially wide BH binary (too wide to merge in
isolation) is pushed to extreme eccentricities (close to unity) by a highly
inclined tertiary and merges within a few Gyrs. LL18 examined a wide range of
orbital and spin evolution behaviors and found that LK-induced mergers can
sometimes exhibit a ``$90^\circ$ attractor'' in the spin evolution: when the BH
spin is initially aligned with the inner binary angular momentum axis
($\theta_{\rm sl, 0} = 0$), it evolves towards a perpendicular state
($\theta_{\rm sl, f}\simeq90^\circ$) near merger (when the inner binary becomes
gravitationally decoupled from the tertiary, and $\theta_{\rm sl}$ freezes).
Qualitatively, they found that the attractor exists when the LK-induced orbital
decay is sufficiently ``gentle'' and the octupole LK effects are unimportant.
Figure~\ref{fig:4sim_90_350} gives an example of a system evolving towards this
attractor, where $\theta_{\rm sl}$ converges to $\approx 90^\circ$ at late times
in the bottom right panel. Figure~\ref{fig:qslscan} shows how $\theta_{\rm sl,
f}$ varies when the initial inclination of the tertiary orbit $I_0$ (relative to
the inner orbit) is varied. Note that for rapid mergers (when $I_0$ is close to
$90^\circ$), the attractor does not exist; as $I_0$ deviates more from
$90^\circ$, the merger time increases and $\theta_{\rm sl, f}$ becomes close to
$90^\circ$. This ``$90^\circ$ attractor'' gives rise to a peak around $\chi_{\rm
eff} = 0$ in the final $\chi_{\rm eff}$ distribution in tertiary-induced mergers
[LL18; \citealp{bin3}]. This peak was also found in the population studies of
\citet{antonini2018precessional}.

The physical origin of this ``$90^\circ$ attractor'' and under what conditions
it can be achieved are not well understood. LL18 proposed an explanation based
on analogy with an adiabatic invariant in systems where the inner binary remains
circular throughout the inspiral (LL17; see also \citep{yu2020spin}). However,
the validity of this analogy is hard to justify, as significant eccentricity
excitation is a necessary ingredient in LK-induced mergers. In addition, the
LK-\emph{enhanced} mergers considered in LL17 show no $90^\circ$ attractor even
though the orbital evolution is slow and regular.

In this paper, we present an analytic theory to explain the $90^\circ$ attractor
and to characterize its regime of validity. More generally, we develop a
theoretical framework to help understand the BH spin evolution in LK-induced
mergers. In Sections~\ref{s:setup_orbital} and~\ref{s:setup_spin}, we set up the
relevant equations of motion for the orbital and spin evolution of the system.
To simplify the theoretical analysis, we initially consider the cases where the
tertiary mass is much larger than the binary mass. In
Sections~\ref{s:fast_merger} and~\ref{s:harmonic}, we develop an analytic
understanding of the spin evolution. In Section~\ref{s:stellar}, we generalize
our results to stellar-mass tertiary companions. We discuss and conclude in
Section~\ref{s:discussion}.


\section{LK-Induced Mergers: Orbital Evolution}\label{s:setup_orbital}

In this section we summarize the key features and relevant equations for
LK-induced mergers to be used for our analysis in later sections. Consider a
black hole (BH) binary with masses $m_1$ and $m_2$ having total mass $m_{12}$,
reduced mass $\mu = m_1 m_2 / m_{12}$, semimajor axis $a$ and eccentricity $e$.
This inner binary orbits around a tertiary with mass $m_3$, semimajor axis
$a_{\rm out}$ and eccentricity $e_{\rm out}$ in a hierarchical configuration
($a_{\rm out}\gg a$). Unless explicitly stated, we assume $m_3 \gg m_1, m_2$
(e.g.\ the tertiary can be a supermassive black hole, or
SMBH), although our analysis can be easily generalized to comparable masses
(see Section~\ref{s:stellar}). We denote the orbital angular
momentum of the inner binary by $\bv{L} \equiv L \uv{L}$ and the angular
momentum of the outer binary by $\bv{L}_{\rm out} \equiv L_{\rm out} \uv{L}_{\rm
out}$. Since $L_{\rm out} \gg L$, we take $\bv{L}_{\rm out}$ to be fixed.

The equations of motion governing the orbital elements $a$, $e$, $\ascnode$,
$I$, $\omega$ (where $\ascnode$, $I$, $\omega$ are the longitude of the
ascending node, inclination, and argument of periapsis respectively) of the
inner binary are
\begin{align}
    \rd{a}{t} &= \p{\rd{a}{t}}_{\rm GW}\label{eq:dadt},\\
    \rd{e}{t} &= \frac{15}{8t_{\rm LK}} e\,j(e)\sin 2\omega
        \sin^2 I + \p{\rd{e}{t}}_{\rm GW}\label{eq:dedt},\\
    \rd{\ascnode}{t} &= \frac{3}{4t_{\rm LK}}
        \frac{\cos I\p{5e^2 \cos^2\omega - 4e^2 - 1}}{j(e)}
            \label{eq:dWdt},\\
    \rd{I}{t} &= -\frac{15}{16t_{\rm LK}}\frac{e^2\sin 2\omega \sin
        2I}{j(e)},\label{eq:dIdt}\\
    \rd{\omega}{t} &= \frac{3}{4t_{\rm LK}}
        \frac{2j^2(e) + 5\sin^2\omega (e^2 - \sin^2 I)}{j(e)}
        + \Omega_{\rm GR},\label{eq:dwdt}
\end{align}
where we have defined
\begin{align}
    j(e) &= \sqrt{1 - e^2},\\
    t_{\rm LK}^{-1} &\equiv n\p{\frac{m_3}{m_{12}}}
        \p{\frac{a}{\tilde{a}_{\rm out}}}^3,\label{eq:t_lk}
\end{align}
with $n \equiv \sqrt{G m_{12} / a^3}$ the mean motion of the inner binary, and
$\tilde{a}_{\rm out} = a_{\rm out} \sqrt{1 - e_{\rm out}^2}$. The
GR-induced apsidal precession of the inner binary is given by
\begin{equation}
    \Omega_{\rm GR}(e) = \frac{3Gnm_{12}}{c^2aj^2(e)}.
\end{equation}
The dissipative terms due to gravitational radiation are
\begin{align}
    \p{\rd{a}{t}}_{\rm GW} &= -\frac{a}{t_{\rm GW}(e)},\label{eq:dadt_gw}\\
    \p{\rd{e}{t}}_{\rm GW} &= -\frac{304}{15}\frac{G^3 \mu m_{12}^2}{c^5a^4}
        \frac{1}{j^{5}(e)}\p{1 + \frac{121}{304}e^2}\label{eq:dedt_gw},
\end{align}
where
\begin{equation}
    t_{\rm GW}^{-1}(e) \equiv \frac{64}{5}\frac{G^3 \mu m_{12}^2}{c^5a^4}
            \frac{1}{j^{7}(e)}\p{1 + \frac{73}{24}e^2
                + \frac{37}{96}e^4}.\label{eq:t_gw}
\end{equation}

Equations (\ref{eq:dedt}--\ref{eq:dwdt}) include only the
effects of quadrupole perturbations from $m_3$ on the binary. The octupole
effects depend on the parameter
\begin{equation}
    \epsilon_{\rm oct} = \frac{m_1 - m_2}{m_{12}}
        \frac{ae_{\rm out}}{a_{\rm out}(1 - e_{\rm out}^2)},\label{eq:epsoct}
\end{equation}
\citep[see][]{bin_diego}. Throughout this paper, we consider systems where
$\tau_{\rm oct} \ll 1$ so that the octupole effects are neglected.

In Fig.~\ref{fig:4sim_90_350}, we provide an example of an LK-induced merger for
our fiducial parameters (described in the figure caption). The inner binary
initially decays slowly as its eccentricity undergoes LK
oscillations, with a nearly constant maximum eccentricity close to unity.
Then, as $a$ decreases and the minimum eccentricity within each LK cycle
increases, the orbital decay of the inner binary accelerates.
As apsidal precession further suppresses eccentricity
oscillations, the eccentricity rapidly decays. We terminate the simulation at $a
= 0.5\;\mathrm{AU}$ as the spin-orbit misalignment angle $\theta_{\rm sl}$ has
reached its final value, even though the binary still has non-negligible
eccentricity. We refer to the example depicted in Fig.~\ref{fig:4sim_90_350}
as the fiducial example, and much of our analysis in later sections will be
based on this example unless otherwise noted. Also note that the parameters of
Fig.~\ref{fig:4sim_90_350} give the same $t_{\rm LK}$ as Fig.~4 of LL18.

We next discuss the key analytical properties of the orbital evolution.

\subsection{Analytical Results Without GW Radiation}

First, neglecting the GW radiation terms, the system admits two
conservation laws, the ``Kozai constant'' and energy conservation,
\begin{align}
    &j(e) \cos I = \mathrm{const},\label{eq:const1}\\
    &\frac{3}{8}\s{2e^2 + j^2(e) \cos^2 I - 5e^2 \sin^2 I \sin^2 \omega}
        + \frac{\epsilon_{\rm GR}}{j(e)} =
            \mathrm{const},\label{eq:const2}
\end{align}
(see \citep{anderson2016formation} and LL18 for more general
expressions when $L_{\rm out}$ is comparable to $L$), where
\begin{equation}
    \epsilon_{\rm GR} \equiv \p{\Omega_{\rm GR} t_{\rm LK}}_{e = 0}
        = \frac{3Gm_{12}^2 \tilde{a}_{\rm out}^3}{c^2m_3a^4}.
\end{equation}
The conservation laws can be combined to obtain the maximum eccentricity
$e_{\max}$ as a function of the initial $I_0$ (and initial $e_0 \ll 1$).
The largest value of $e_{\max}$ occurs at $I_0 = 90^\circ$ and is given
by
\begin{equation}
    j(e_{\max})_{I_0 = 90^\circ} = (8/9) \epsilon_{\rm GR}.
        \label{eq:jemaxepsgr}
\end{equation}
Eccentricity excitation then requires $\epsilon_{\rm GR} < 9/8$. Our fiducial
examples in Figs.~\ref{fig:4sim_90_350} and~\ref{fig:qslscan} satisfy
$\epsilon_{\rm GR} \ll 1$ at $a = a_0$, leading to $e_{\max} \sim 1$ within a
narrow inclination window around $I_0 = 90^\circ$.

Eqs.~\eqref{eq:const1} and~\eqref{eq:const2} imply that $e$ is a function of
$\sin^2\omega$ alone \citep[see][for the exact form]{kinoshita, storch}, so an
eccentricity maximum occurs every half period of $\omega$. We define the LK
period of eccentricity oscillation $P_{\rm LK}$ and its corresponding angular
frequency $\Omega_{\rm LK}$ via
\begin{align}
    \pi &= \int\limits_0^{P_{\rm LK}} \rd{\omega}{t}\;\mathrm{d}t,&
    \Omega_{\rm LK} &\equiv \frac{2\pi}{P_{\rm LK}}.\label{eq:PLK_def}
\end{align}

In LK cycles, the inner binary oscillates between the eccentricity minimum
$e_{\min}$ and maximum $e_{\max}$. The oscillation is ``uneven'': when
$e_{\min} \ll e_{\max}$, the binary spends a fraction $\sim j(e_{\max})$ of the
LK cycle, or time $\Delta t \sim t_{\rm LK} j\p{e_{\max}}$, near $e \simeq
e_{\max}$ [see Eq.~\eqref{eq:dwdt}].

\subsection{Behavior with GW Radiation}

Including the effect of GW radiation, orbital decay predominantly occurs at $e
\simeq e_{\max}$ with the timescale of $t_{\rm GW}\p{e_{\max}}$ [see
Eq.~\eqref{eq:t_gw}]. On the other hand, Eq.~\eqref{eq:dwdt} implies that, when
$\epsilon_{\rm GR} \ll 1$, the binary spends only a small fraction ($\sim
j(e_{\max})$) of the time near $e \simeq e_{\max}$. Thus, we expect two
qualitatively different merger behaviors:
\begin{itemize}
    \item ``Rapid mergers'': When $t_{\rm GW}\p{e_{\max}} \lesssim t_{\rm
        LK}j(e_{\max})$, the binary is ``pushed'' into high eccentricity and
        exhibits a ``one shot merger'' without any $e$-oscillations.

    \item ``Smooth mergers'': When $t_{\rm GW}\p{e_{\max}} \gtrsim t_{\rm
        LK}j(e_{\max})$, the binary goes through a phase of eccentricity
        oscillations while the orbit gradually decays. In this case, the
        LK-averaged orbital decay rate is $\sim j(e_{\max})t_{\rm
        GW}^{-1}(e_{\max})$. As $a$ decreases, $e_{\max}$ decreases slightly
        while the minimum eccentricity increases, approaching $e_{\max}$ (see
        Fig.~\ref{fig:4sim_90_350}). This eccentricity oscillation ``freeze''
        ($e_{\min} \sim e_{\max}$) is due to GR-induced apsidal precession
        ($\epsilon_{\rm GR}$ increases as $a$ decreases), and occurs when
        $\epsilon_{\rm GR}(a) \gg j(e_{\max})$. After the eccentricity is
        frozen, the binary circularizes and decays on the timescale $t_{\rm
        GW}\p{e_{\max}}$.
\end{itemize}

\section{Spin Dynamics: Equations}\label{s:setup_spin}

We are interested in the spin orientations of the inner BHs at merger as a
function of initial conditions. Since they evolve independently to leading
post-Newtonian order, we focus on the dynamics of $\uv{S}_1 = \uv{S}$, the unit
spin vector of $m_1$. Since the spin magnitude does not enter into the dynamics,
we write $\bv{S} \equiv \uv{S}$ for brevity (i.e.\ $\bv{S}$ is a unit vector).
Neglecting spin-spin interactions, $\bv{S}$ undergoes de Sitter precession about
$\bv{L}$ as
\begin{align}
    \rd{\bv{S}}{t} &= \Omega_{\rm SL}\hat{\bv{L}} \times \bv{S},
            \label{eq:dsdt}
\end{align}
with
\begin{align}
    \Omega_{\rm SL} &= \frac{3Gn\p{m_2 + \mu/3}}{2c^2aj^2(e)}.
\end{align}

In the presence of a tertiary companion, the orbital axis $\uv{L}$ of the inner
binary precesses around $\uv{L}_{\rm out}$ with rate $\rdil{\ascnode}{t}$ and
nutates with varying $I$ [see Eqs.~\eqref{eq:dWdt} and~\eqref{eq:dIdt}]. To
analyze the dynamics of the spin vector, we go to the co-rotating frame with
$\uv{L}$ about $\uv{L}_{\rm out}$, in which Eq.~\eqref{eq:dsdt} becomes
\begin{align}
    \p{\rd{\bv{S}}{t}}_{\rm rot}
        &= \bv{\Omega}_{\rm e} \times \bv{S}\label{eq:dsdt_weff},
\end{align}
where we have defined an effective rotation vector
\begin{align}
    \bv{\Omega}_{\rm e} &\equiv \Omega_{\rm L}\uv{L}_{\rm out} + \Omega_{\rm SL}
            \uv{L},\label{eq:weff_def}
\end{align}
with [see Eq.~\eqref{eq:dWdt}]
\begin{align}
    \Omega_{\rm L} &\equiv -\rd{\ascnode}{t}.\label{eq:Wldef}
\end{align}
In this rotating frame, the plane spanned by $\uv{L}_{\rm out}$ and $\uv{L}$ is constant
in time, only the inclination angle $I$ can vary.

\subsection{Nondissipative Spin Dynamics}\label{ss:monodromy}

We first consider the limit where dissipation via GW radiation is completely
neglected ($t_{\rm GW}(e) \to \infty$). Then $\bv{\Omega}_{\rm e}$ is exactly
periodic with period $P_{\rm LK}$ [see Eq.~\eqref{eq:PLK_def}] We can rewrite
Eq.~\eqref{eq:dsdt_weff} in Fourier components
\begin{equation}
    \p{\rd{\bv{S}}{t}}_{\rm rot}
        = \s{\overline{\bv{\Omega}}_{\rm e} + \sum\limits_{N = 1}^\infty
            \bv{\Omega}_{\rm eN}\cos \p{N\Omega_{\rm LK}t}}
            \times \bv{S}.\label{eq:dsdt_fullft}
\end{equation}
Note that $\overline{\bv{\Omega}}_{\rm e}$ is the zeroth Fourier component,
where the bar denotes an average over a LK cycle. We have adopted the convention
where $t = 0$ is the time of maximum eccentricity of the LK cycle, so that
Eq.~\eqref{eq:dsdt_fullft} does not have $\sin\p{N\Omega_{\rm LK}t}$ terms.

This system superficially resembles that considered in \citet{storch} (SL15),
who studied the dynamics of the spin axis of a star when driven by a giant
planet undergoing LK oscillations \citep[see also][]{storch2014chaotic,
storch2017dynamics}. In their system, the spin-orbit coupling arises from
Newtonian interaction between the planet ($M_{\rm p}$) and the rotation-induced
stellar quadrupole ($I_{\rm 3}-I_1$), and the spin precession frequency is
\begin{equation}
    \Omega_{\rm SL}^{\rm (Newtonian)} = -\frac{3GM_{\rm p} \p{I_{\rm 3} -
        I_1}}{ 2a^3j^3(e)}\frac{\cos \theta_{\rm sl}}{I_3 \Omega_{\rm s}},
\end{equation}
where $I_3 \Omega_{\rm s}$ is the spin angular momentum of the star. SL15 showed
that under some conditions that depend on a dimensionless adiabaticity parameter
(roughly the ratio between the magnitudes of $\Omega_{\rm SL}^{\rm (Newtonian)}$
and $\Omega_{\rm L}$ when factoring out the eccentricity and obliquity
dependence), the stellar spin axis can vary chaotically. One strong indicator of
chaos in their study is the presence of irregular, fine structure in a
bifurcation diagram [Fig.~1 of \citet{storch}] that shows the values of the
spin-orbit misalignment angle $\theta_{\rm sl}$ when varying system parameters
in the ``transadiabatic'' regime, where the adiabaticity parameter crosses
unity.

\begin{figure}
    \centering
    \includegraphics[width=\colummwidth]{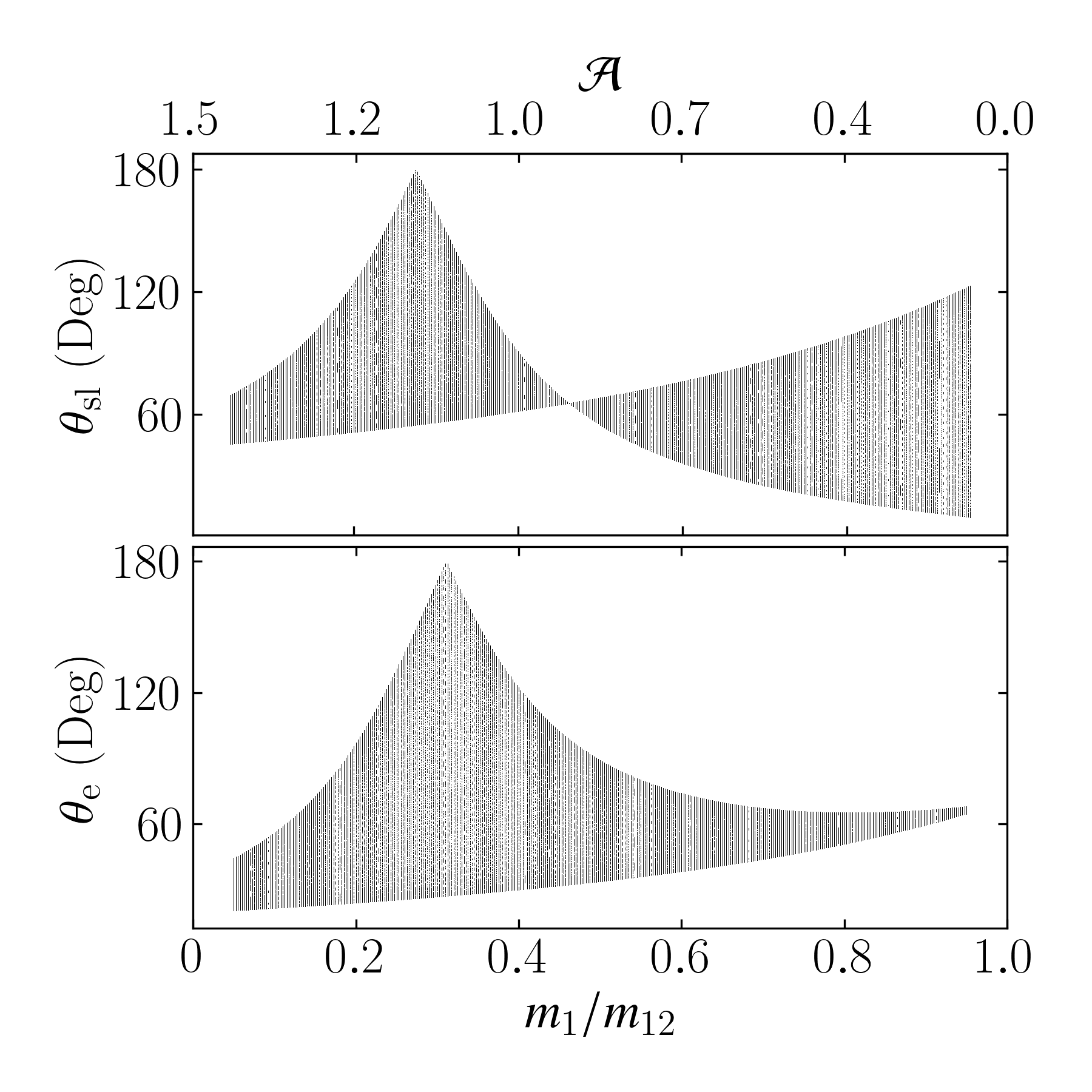}
    \caption{Bifurcation diagram for the BH spin orientation during LK
    oscillations. The physical parameters are $m_{12} = 60M_{\odot}$, $m_3 = 3
    \times 10^7 M_{\odot}$, $a = 0.1\;\mathrm{AU}$, $e_0 = 10^{-3}$,
    $I_0 = 70^\circ$, $a_{\rm out} = 300\;\mathrm{AU}$, $e_{\rm out} = 0$,
    and initial condition $\theta_{\rm sl, 0} = 0$. For each mass ratio $m_1 /
    m_{12}$, the orbit-spin system is solved over $500$ LK cycles, and both
    $\theta_{\rm sl}$ (the angle between $\bv{S}$ and $\uv{L}$) and $\theta_{\rm
    e}$ [defined by Eq.~\eqref{eq:q_eff_inst}] are sampled at every eccentricity
    maximum and are plotted. The top axis shows the adiabaticity parameter
    $\mathcal{A}$ as defined by Eq.~\eqref{eq:abar_def}. Note that for a given
    $m_{12}$, changing the mass ratio $m_1 / m_{12}$ only changes the spin
    evolution and not the orbital evolution.}\label{fig:bifurcation_70}
\end{figure}

To generate an analogous bifurcation diagram for our problem, we consider a
sample system with $m_{12} = 60M_{\odot}$, $m_3 = 3 \times 10^7 M_{\odot}$, $a =
0.1\;\mathrm{AU}$, $e_0 = 10^{-3}$, $I_0 = 70^\circ$, $a_{\rm out} =
300\;\mathrm{AU}$, $e_{\rm out} = 0$, and initial $\theta_{\rm sl} = 0$
(note that these parameters are different from those in
Fig.~\ref{fig:4sim_90_350}). We then
evolve Eq.~\eqref{eq:dsdt} together with the orbital evolution equations
[Eqs.~(\ref{eq:dadt}--\ref{eq:dwdt}) without the GW terms] while sampling both
$\theta_{\rm sl}$ and $\theta_{\rm e}$ at eccentricity maxima, where
$\theta_{\rm e}$ is the angle between
$\overline{\bv{\Omega}}_{\rm e}$ and $\bv{S}$, i.e.
\begin{equation}
    \cos \theta_{\rm e} = \frac{\overline{\bv{\Omega}}_{\rm e}}{
        \overline{\Omega}_{\rm e}}\cdot \bv{S},\label{eq:q_eff_inst}
\end{equation}
where $\overline{\Omega}_{\rm e} \equiv \abs{\overline{\bv{\Omega}}_{\rm e}}$.
We repeat this procedure with different mass ratios $m_1 / m_{12}$ of the inner
binary, which only changes $\Omega_{\rm SL}$ without changing the orbital
evolution (note that the LK oscillation depends only on $m_{12}$ and not on
individual masses of the inner binary). Analogous to SL15, we consider systems
with a range of the adibaticity parameter $\mathcal{A}$ [to be defined later in
Eq.~\eqref{eq:abar_def}] that crosses order unity. Note that
the fiducial system of Fig.~\ref{fig:4sim_90_350} does not serve this purpose
because the initial $\Omega_{\rm SL}$ is too small. Our result is depicted in
Fig.~\ref{fig:bifurcation_70}.

While our bifurcation diagram has interesting structure, the features are all
regular. This is in contrast to the star-planet system studied by SL15 (see
their Fig.~1). A key difference is that in our system, $\Omega_{\rm SL}$ does
not depend on $\theta_{\rm sl}$, while for the planet-star system, $\Omega_{\rm
SL}^{\rm (Newtonian)}$ does, and this latter feature introduces nonlinearity to
the dynamics.

A more formal understanding of the dynamical behavior of our spin-orbit system
comes from Floquet theory\citep{floquet1883equations, chicone2006ordinary}, as
Eq.~\eqref{eq:dsdt_weff} is a linear system with periodic coefficients (the
system studied in SL15 is nonlinear). By Floquet's theorem, when a linear system
with periodic coefficients is integrated over a period, the evolution can be
described by the linear transformation
\begin{equation}
    \bv{S}\p{t + P_{\rm LK}} = \bv{\tilde{M}} \bv{S}(t),
\end{equation}
where $\bv{\tilde{M}}$ is called the \emph{monodromy matrix} and is independent
of $\bv{S}$.

For our system, while $\bv{\tilde{M}}$ can be easily defined, it cannot be
evaluated in closed form. Instead, we can reason directly about the general
properties of $\bv{\tilde{M}}$: it must be a proper orthogonal matrix, or a
rotation matrix, as it represents the effect of many infinitesimal rotations,
each about the instantaneous $\bv{\Omega}_{\rm e}$\footnote{More formally,
$\bv{\tilde{M}} = \bv{\tilde{\Phi}}(P_{\rm LK})$ where $\bv{\tilde{\Phi}}(t)$ is
the \emph{principal fundamental matrix solution}: the columns of
$\bv{\tilde{\Phi}}$ are solutions to Eq.~\eqref{eq:dsdt_weff} and
$\bv{\tilde{\Phi}}(0)$ is the identity. By linearity, the columns of
$\bv{\tilde{\Phi}}(t)$ remain orthonormal, while its determinant does not
change, so $\bv{\tilde{M}}$ is a proper orthogonal matrix, or a rotation
matrix.}. Therefore, over each period $P_{\rm LK}$, the dynamics of $\bv{S}$ are
equivalent to a rotation about a fixed axis, prohibiting chaotic behavior.

Another traditional indicator of chaos is a positive Lyapunov exponent, obtained
when the separation between nearby trajectories diverges \emph{exponentially} in
time. In Floquet theory, the Lyapunov exponent is the logarithm of the largest
eigenvalue of the monodromy matrix. Since $\bv{\tilde{M}}$ is a rotation matrix
in our problem, the Lyapunov exponent must be $0$, indicating no chaos. We have
verified this numerically.

\subsection{Spin Dynamics With GW Dissipation}

When $t_{\rm GW}$ is finite, the coefficients $\bv{\Omega}_{\rm eN}$, including
$\overline{\bv{\Omega}}_{\rm e}$ [see Eq.~\eqref{eq:dsdt_fullft}], are no longer
constant, but change over time. For ``smooth'' mergers (satisfying $t_{\rm
GW}\p{e_{\max}} \gg t_{\rm LK} j(e_{\max})$; see Section~\ref{s:setup_orbital}),
the binary goes through a sequence of LK cycles, and the coefficients vary on the
LK-averaged orbital decay time $t_{\rm GW}\p{e_{\max}} / j\p{e_{\max}}$.
After the LK oscillation freezes, we have $\bv{\Omega}_{\rm e}
\simeq \overline{\bv{\Omega}}_{\rm e}$ (and $\bv{\Omega}_{\rm eN} \simeq 0$ for
$N \geq 1$), which evolves on timesale $t_{\rm GW}(e)$ as the orbit decays and
circularizes.

Once $a$ is sufficiently small that $\Omega_{\rm SL} \gg \Omega_{\rm L}$, it can
be seen from Eqs.~(\ref{eq:dsdt_weff}--\ref{eq:weff_def}) that $\theta_{\rm e} =
\theta_{\rm sl}$ is constant, i.e.\ the spin-orbit misalignment angle is frozen
(see bottom right panel of Fig.~\ref{fig:4sim_90_350}). This is the ``final''
spin-orbit misalignment, although the binary may still be far from the final
merger. Note that at such separations, $\epsilon_{\rm GR} \gg 1$ as well since
$\Omega_{\rm SL} \sim \Omega_{\rm GR}$, and so LK eccentricity excitation is
suppressed. For the fiducial examples depicted in
Figs.~\ref{fig:4sim_90_350}--\ref{fig:qslscan}, we stop the simulation at $a =
0.5\;\mathrm{AU}$, as $\theta_{\rm sl}$ has converged to its final value.

\subsection{Spin Dynamics Equation in Component Form}

\begin{figure}
    \centering
    \includegraphics[width=0.5\colummwidth]{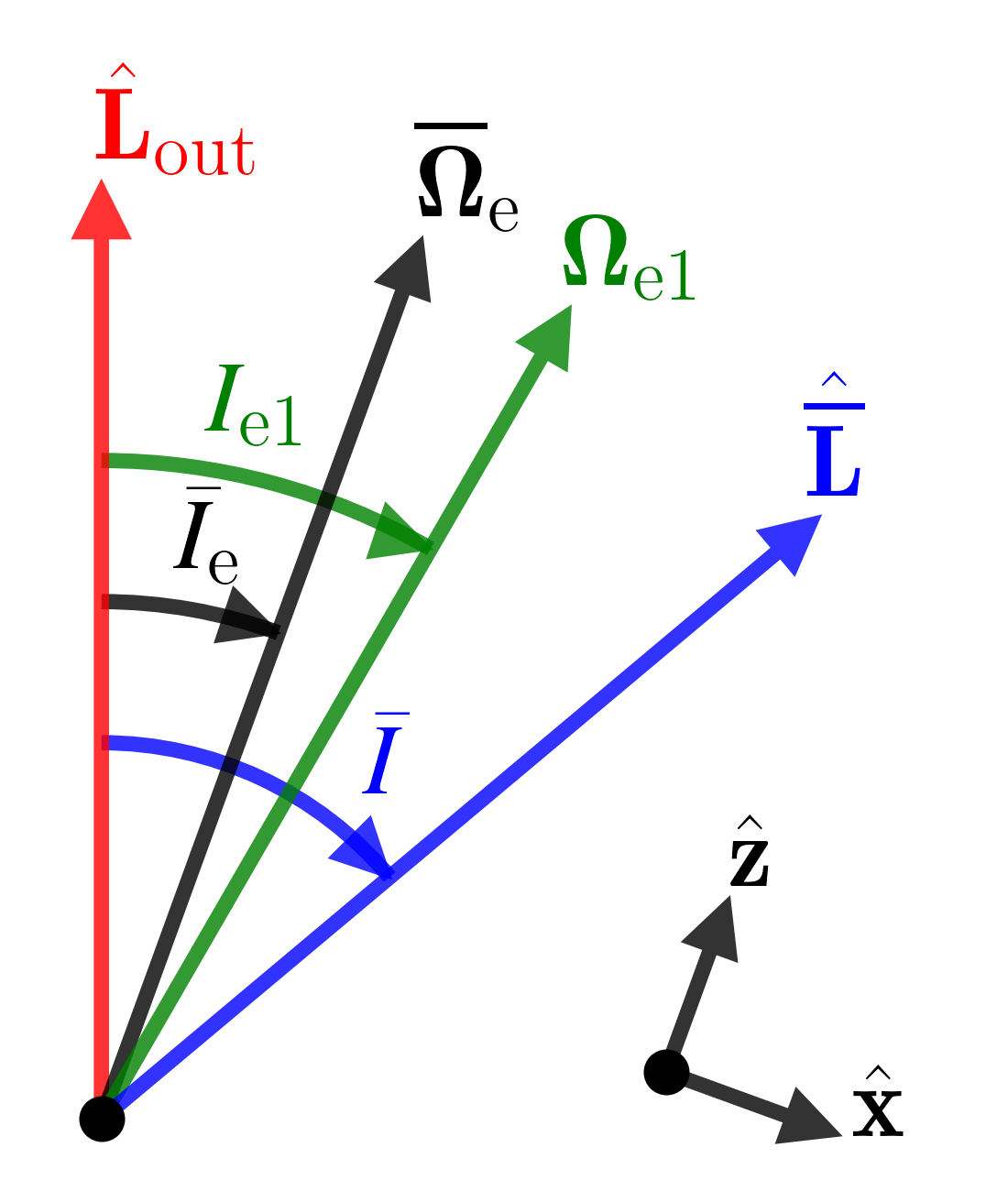}
    \caption{Definition of angles in the problem, shown in plane of the two
    angular momenta $\bv{L}_{\rm out}$ and $\bv{L}$. Here,
    $\overline{\bv{\Omega}}_{\rm e}$ is the LK-averaged $\bv{\Omega}_{\rm e}$,
    and $\bv{\Omega}_{\rm e1}$ is the first harmonic component (see
    Eqs.~\eqref{eq:weff_def} and~\eqref{eq:dsdt_fullft}). Note that for $I_0 >
    90^\circ$, we choose $\bar{I}_{\rm e} \in (90^ \circ, 180^\circ)$ so that
    $\overline{\Omega}_{\rm e} > 0$ (since $\Omega_{\rm L} < 0$). The bottom
    right shows our choice of coordinate axes with $\uv{z} \propto
    \overline{\Omega}_{\rm e}$.
    }\label{fig:3vec}
\end{figure}

For later analysis, it is useful to write Eq.~\eqref{eq:dsdt_fullft} in
component form. To do so, we define the inclination angle
$\bar{I}_{\rm e}$ as the angle between $\overline{\bv{\Omega}}_{\rm e}$ and
$\bv{L}_{\rm out}$ as shown in Fig.~\ref{fig:3vec}. To express $\bar{I}_{\rm
e}$ algebraically, we define the LK-averaged quantities
\begin{align}
    \overline{\Omega_{\rm SL} \sin I} &\equiv
            \overline{\Omega}_{\rm SL} \sin \bar{I},&
    \overline{\Omega_{\rm SL} \cos I} &\equiv
            \overline{\Omega}_{\rm SL} \cos \bar{I}.\label{eq:barI}
\end{align}
It then follows from Eq.~\eqref{eq:weff_def} that
\begin{equation}
    \tan \bar{I}_{\rm e} = \frac{\mathcal{A}\sin \bar{I}}{
        1 + \mathcal{A}\cos \bar{I}},\label{eq:ie_def}
\end{equation}
where $\mathcal{A}$ is the adiabaticity parameter, given by
\begin{equation}
    \mathcal{A} \equiv \frac{\overline{\Omega}_{\rm SL}}{
        \overline{\Omega}_{\rm L}}.\label{eq:abar_def}
\end{equation}
Note that in Eq.~\eqref{eq:ie_def}, $\bar{I}_{\rm e}$ is defined in the domain
$[0^\circ, 180^\circ]$, i.e.\ $\bar{I}_{\rm e} \in (0, 90)$ when $\tan
\bar{I}_{\rm e} > 0$ and $\bar{I}_{\rm e} \in (90,180)$ when $\tan \bar{I}_{\rm
e} < 0$.

We now choose a non-inertial coordinate system where $\uv{z} \propto
\overline{\bv{\Omega}}_{\rm e}$ and $\uv{x}$ lies in the plane of $\bv{L}_{\rm
out}$ and $\bv{L}$ (see Fig.~\ref{fig:3vec}). In this reference frame, the
spin orientation is specified by the polar angle $\theta_{\rm
e}$ as defined above in Eq.~\eqref{eq:q_eff_inst}, and the
spin evolution equation becomes
\begin{align}
    \p{\rd{\bv{S}}{t}}_{\rm xyz} &= \s{\overline{\Omega}_{\rm e}\uv{z}
         + \sum\limits_{N = 1}^\infty
            \bv{\Omega}_{\rm eN}\cos \p{N\Omega_{\rm LK}t }}
        \times \bv{S}
        - \dot{\bar{I}}_{\rm e} \uv{y} \times \bv{S}\label{eq:eom_prime}.
\end{align}
One further simplification lets us cast this vector equation of motion into a
scalar form. Break $\bv{S}$ into components $\bv{S} = S_x\uv{x} + S_y \uv{y} +
\cos \theta_{\rm e} \uv{z}$ and define complex variable
\begin{equation}
    S_\perp \equiv S_x + iS_y.
\end{equation}
Then, we can rewrite Eq.~\eqref{eq:eom_prime} as
\begin{align}
    \rd{S_{\perp}}{t} ={}& i\overline{\Omega}_{\rm e}S_\perp
            - \dot{\bar{I}}_{\rm e} \cos \theta_{\rm e}
        + \sum\limits_{N = 1}^\infty\Big[
            \cos \p{\Delta I_{\rm eN}}S_\perp \nonumber\\
        &- i\cos \theta_{\rm e} \sin \p{\Delta I_{\rm eN}}\Big]
            \Omega_{\rm eN}\cos \p{N\Omega_{\rm LK} t},
            \label{eq:formal_eom_allgen}
\end{align}
where $\dot{\bar{I}}_{\rm e} = \rdil{\bar{I}_{\rm e}}{t}$ and
$\Omega_{\rm eN}$ is the magnitude of the vector $\bv{\Omega}_{\rm eN}$ [see
Eq.~\eqref{eq:dsdt_fullft}] and $\Delta I_{\rm eN} = I_{\rm eN} - \bar{I}_{\rm
e}$, with $I_{\rm eN}$ the angle between $\bv{\Omega}_{\rm
eN}$ and $\bv{L}_{\rm out}$ (see Fig.~\ref{fig:3vec}). Since
$\cos \theta_{\rm e} = \pm \sqrt{1 - \abs{S_{\perp}}^2}$,
Eq.~\eqref{eq:formal_eom_allgen} is generally nonlinear in $S_{\perp}$, but
becomes approximately linear when $\abs{\theta_{\rm e}} \ll 1$.

\section{Analysis: Approximate Adiabatic Invariant}\label{s:fast_merger}

In general, Eqs.~\eqref{eq:dsdt_fullft} and~\eqref{eq:formal_eom_allgen} are
difficult to study analytically. In this section, we neglect the harmonic terms
and focus on how the varying $\overline{\bv{\Omega}}_{\rm e}$ affects the
evolution of the BH spin axis. The effect of the harmonic terms is studied in
Section~\ref{s:harmonic}.

\subsection{The Adiabatic Invariant}

When neglecting the $N \geq 1$ harmonic terms, Eq.~\eqref{eq:dsdt_fullft}
reduces to
\begin{equation}
    \p{\rd{\overline{\bv{S}}}{t}}_{\rm rot}
        = \overline{\bv{\Omega}}_{\rm e}
            \times \overline{\bv{S}}.\label{eq:dsdt_0only}
\end{equation}
It is not obvious to what extent the analysis of Eq.~\eqref{eq:dsdt_0only} is
applicable to Eq.~\eqref{eq:dsdt_fullft}. From our numerical calculations, we
find that the LK-average of $\bv{S}$ often evolves following
Eq.~\eqref{eq:dsdt_0only}, motivating our notation $\overline{\bv{S}}$. Over
timescales shorter than the LK period $P_{\rm LK}$, Eq.~\eqref{eq:dsdt_0only}
loses accuracy as the evolution of $\bv{S}$ itself is dominated by the $N \geq
1$ harmonics we have neglected. An intuitive interpretation of this result is
that the $N \geq 1$ harmonics vanish when integrating Eq.~\eqref{eq:dsdt_fullft}
over a LK cycle.

Eq.~\eqref{eq:dsdt_0only} has one desirable property: $\bar{\theta}_{\rm e}$,
defined by
\begin{equation}
    \cos \bar{\theta}_{\rm e} \equiv
        \frac{\overline{\bv{\Omega}}_{\rm e}}{\overline{\Omega}_{\rm e}}
            \cdot \overline{\bv{S}},
        \label{eq:q_eff}
\end{equation}
is an adiabatic invariant. This follows from the fact that
$\overline{\bv{S}}_z$, the projection of $\overline{\bv{S}}$ on the
$\overline{\bv{\Omega}}_{\rm e}$ axis, and $\varphi$, the precessional angle of
$\overline{\bv{S}}$ around $\overline{\bv{\Omega}}_{\rm e}$, form a pair of
action-angle variables. The adiabaticity condition requires
that the precession axis $\uv{z} =
\overline{\bv{\Omega}}_{\rm e} / \overline{\Omega}_{\rm e}$ evolve slowly
compared to the precession frequency at all times, i.e.
\begin{equation}
    \abs{\rd{\bar{I}_{\rm e}}{t}} \ll \overline{\Omega}_{\rm e}.\label{eq:ad_constr}
\end{equation}
For our fiducial example depicted in Fig.~\ref{fig:4sim_90_350}, the values of
$\dot{\bar{I}}_{\rm e}$ and $\overline{\Omega}_{\rm e}$ are shown in the top
panel of Fig.~\ref{fig:4sim_90_350_supp}, and the evolution of
$\bar{\theta}_{\rm e}$ in the bottom panel. The net change in $\bar{\theta}_{\rm
e}$ in this simulation is $0.01^\circ$, small as expected since
$|\dot{\bar{I}}_{\rm e}| \ll \overline{\Omega}_{\rm e}$ at all times.

\subsection{Deviation from Adiabaticity}\label{ss:eom_0}

The extent to which $\bar{\theta}_{\rm e}$ is conserved depends on how well
Eq.~\eqref{eq:ad_constr} is satisfied. In this subsection, we derive a bound on
the total non-conservation of $\bar{\theta}_{\rm e}$, then in the next
subsection we show how this bound can be estimated from the
initial conditions.

\begin{figure}
    \centering
    \includegraphics[width=\colummwidth]{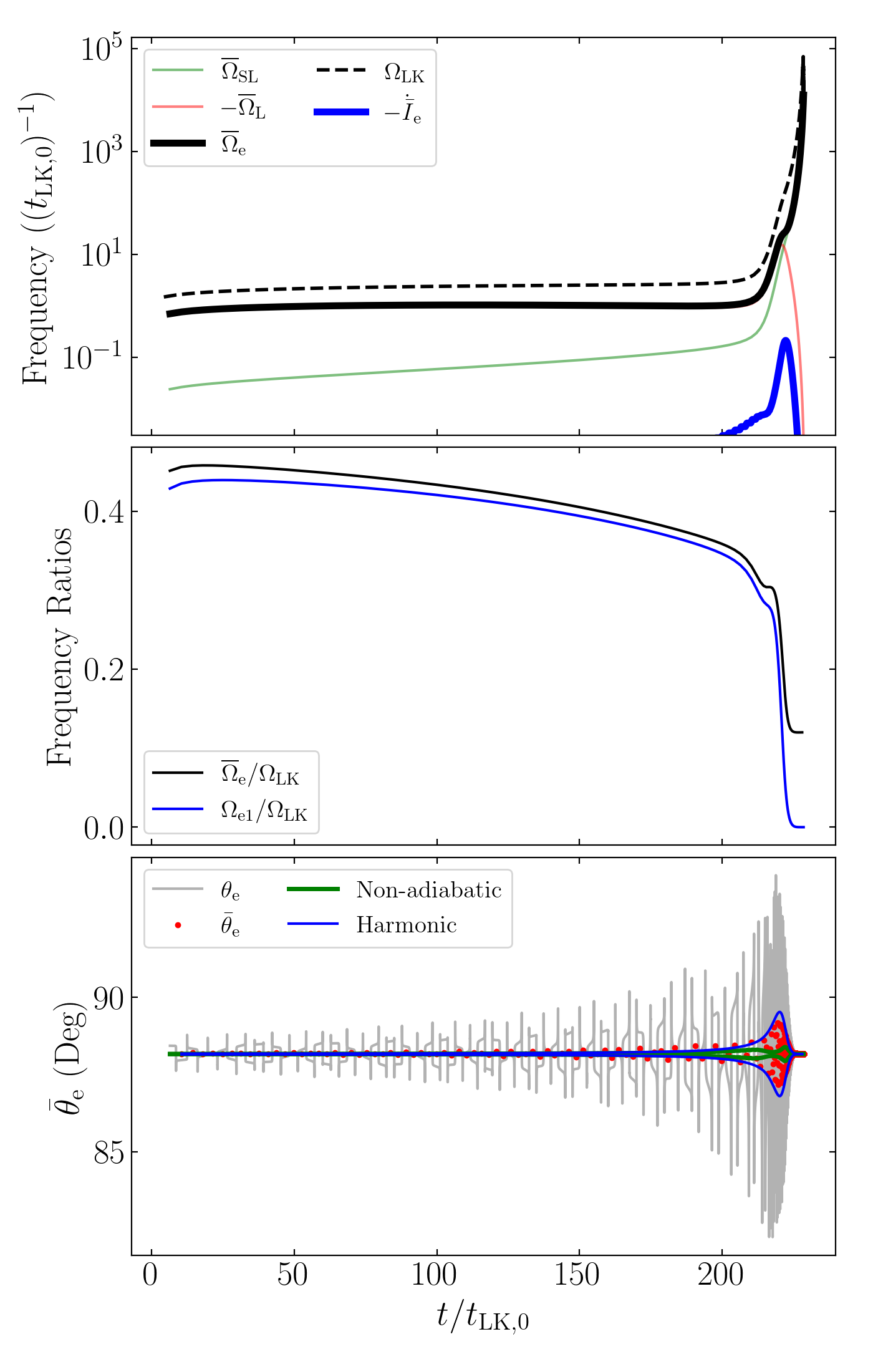}
    \caption{The same simulation as depicted in Fig.~\ref{fig:4sim_90_350} but
    showing several calculated quantities relevant to the
    theory of the spin evolution. Top: the four characteristic frequencies of
    the system and $\rdil{\bar{I}_{\rm e}}{t}$. Middle: the frequency ratios
    between the zeroth and first Fourier components of $\bv{\Omega}_{\rm e}$ to
    the LK frequency $\Omega_{\rm LK}$. Bottom: Time evolution of $\theta_{\rm
    e}$ [grey line; Eq.~\eqref{eq:q_eff_inst}], $\bar{\theta}_{\rm e}$ [red
    dots; Eq.~\eqref{eq:q_eff}], as well as estimates of the deviations from
    perfect conservation of $\bar{\theta}_{\rm e}$ due to nonadiabaticity
    [green, Eq.~\eqref{eq:nonad_dqeff}] and due to the resonance
    $\overline{\Omega}_{\rm e} \approx \Omega_{\rm LK}$ [blue,
    Eq.~\eqref{eq:harmonic_dqeff}].}\label{fig:4sim_90_350_supp}
\end{figure}

When neglecting harmonic terms, the scalar evolution equation
Eq.~\eqref{eq:formal_eom_allgen} becomes
\begin{align}
    \rd{S_{\perp}}{t} &= i\overline{\Omega}_{\rm e}S_\perp
            - \dot{\bar{I}}_{\rm e} \cos \bar{\theta}_{\rm e}.
\end{align}
This can be solved in closed form. Defining
\begin{equation}
    \Phi(t) \equiv \int\limits^t \overline{\Omega}_{\rm e}\;\mathrm{d}t,
        \label{eq:Phi_t}
\end{equation}
we obtain the solution at time $t$:
\begin{equation}
    e^{-i\Phi}S_{\perp}\bigg|_{0}^{t}
        = -\int\limits_{0}^{t}
            e^{-i\Phi(\tau)}\dot{\bar{I}}_{\rm e} \cos \bar{\theta}_{\rm e}
                \;\mathrm{d}\tau.\label{eq:formal_sol_0}
\end{equation}
Recalling $\abs{S_{\perp}} = \sin \bar{\theta}_{\rm e}$ and analyzing
Eq.~\eqref{eq:formal_sol_0}, we see that $\bar{\theta}_{\rm e}$ oscillates about
its initial value with semi-amplitude
\begin{equation}
    \abs{\Delta \bar{\theta}_{\rm e}} \sim
        \abs{\frac{\dot{\bar{I}}_{\rm e}}{\overline{\Omega}_{\rm
        e}}}.\label{eq:nonad_dqeff}
\end{equation}
In the adiabatic limit [Eq.~\eqref{eq:ad_constr}], $\bar{\theta}_{\rm e}$ is
indeed conserved, as the right-hand side of Eq.~\eqref{eq:nonad_dqeff} goes to
zero. The bottom panel of Fig.~\ref{fig:4sim_90_350_supp} shows $\Delta
\bar{\theta}_{\rm e}$ for the fiducial example. Note that $\bar{\theta}_{\rm
e}$ is indeed mostly constant where Eq.~\eqref{eq:nonad_dqeff} predicts small
oscillations.

If we denote $\abs{\Delta \bar{\theta}_{\rm e}}_{\rm f}$ to be the net change
in $\bar{\theta}_{\rm e}$ over $t \in [0, t_{\rm f}]$, we can give a
loose bound
\begin{equation}
    \abs{\Delta \bar{\theta}_{\rm e}}_{\rm f} \lesssim
        \abs{\frac{\dot{\bar{I}}_{\rm e}}{\overline{\Omega}_{\rm e}}}_{\max}.
        \label{eq:nonad_dqeff_tot}
\end{equation}

\begin{figure*}
    \centering
    \includegraphics[width=\textwidth]{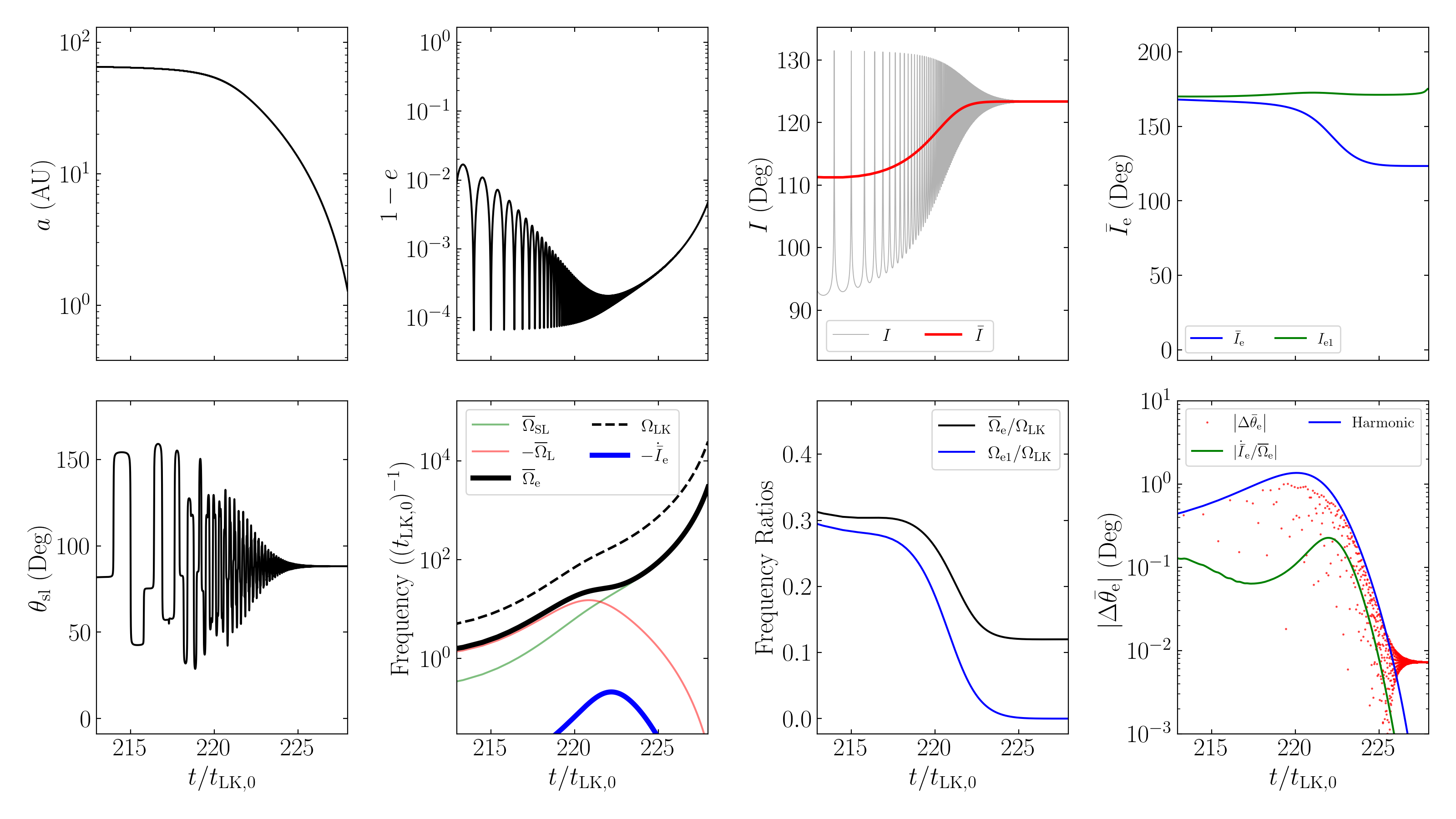}
    \caption{The same simulation as Fig.~\ref{fig:4sim_90_350} but zoomed in on
    the region around $\mathcal{A} \equiv \overline{\Omega}_{\rm SL} /
    \overline{\Omega}_{\rm L} \simeq 1$ and showing a wide range of relevant
    quantities. The first three panels in the upper row depict $a$, $e$, $I$ and
    $\bar{I}$ as in Fig.~\ref{fig:4sim_90_350}, while the fourth shows
    $\bar{I}_{\rm e}$ [Eq.~\eqref{eq:ie_def}] and $I_{\rm e1}$. The bottom four
    panels depict $\theta_{\rm sl}$, the four characteristic frequencies of the
    system and $\rdil{\bar{I}_{\rm e}}{t}$ [Eqs.~\ref{eq:weff_def}
    and~\eqref{eq:Wldef}] (as in the top panel of
    Fig.~\ref{fig:4sim_90_350_supp}), the relevant frequency ratios (as in the
    middle panel of Fig.~\ref{fig:4sim_90_350_supp}), and the deviation of
    $\bar{\theta}_{\rm e}$ from its initial value compared to the predictions of
    Eqs.~\eqref{eq:nonad_dqeff}
    and~\eqref{eq:harmonic_dqeff}.}\label{fig:4sim_90_350_zoom}
\end{figure*}

Inspection of Fig.~\ref{fig:4sim_90_350_supp} indicates that the spin dynamics
are mostly uninteresting except near the peak of $|\dot{\bar{I}}_{\rm e}|$,
which occurs where $\bar{\Omega}_{\rm SL} \simeq \abs{\Omega_{\rm L}}$. We
present a zoomed-in view of dynamical quantities near the peak of
$|\dot{\bar{I}}_{\rm e}|$ in
Fig.~\ref{fig:4sim_90_350_zoom}. In particular, in the bottom-rightmost panel,
we see that the fluctuations in $\bar{\theta}_{\rm e}$ are dominated by a second
contribution, the subject of the discussion in Section~\ref{s:harmonic}.

\begin{figure*}
    \centering
    \includegraphics[width=\textwidth]{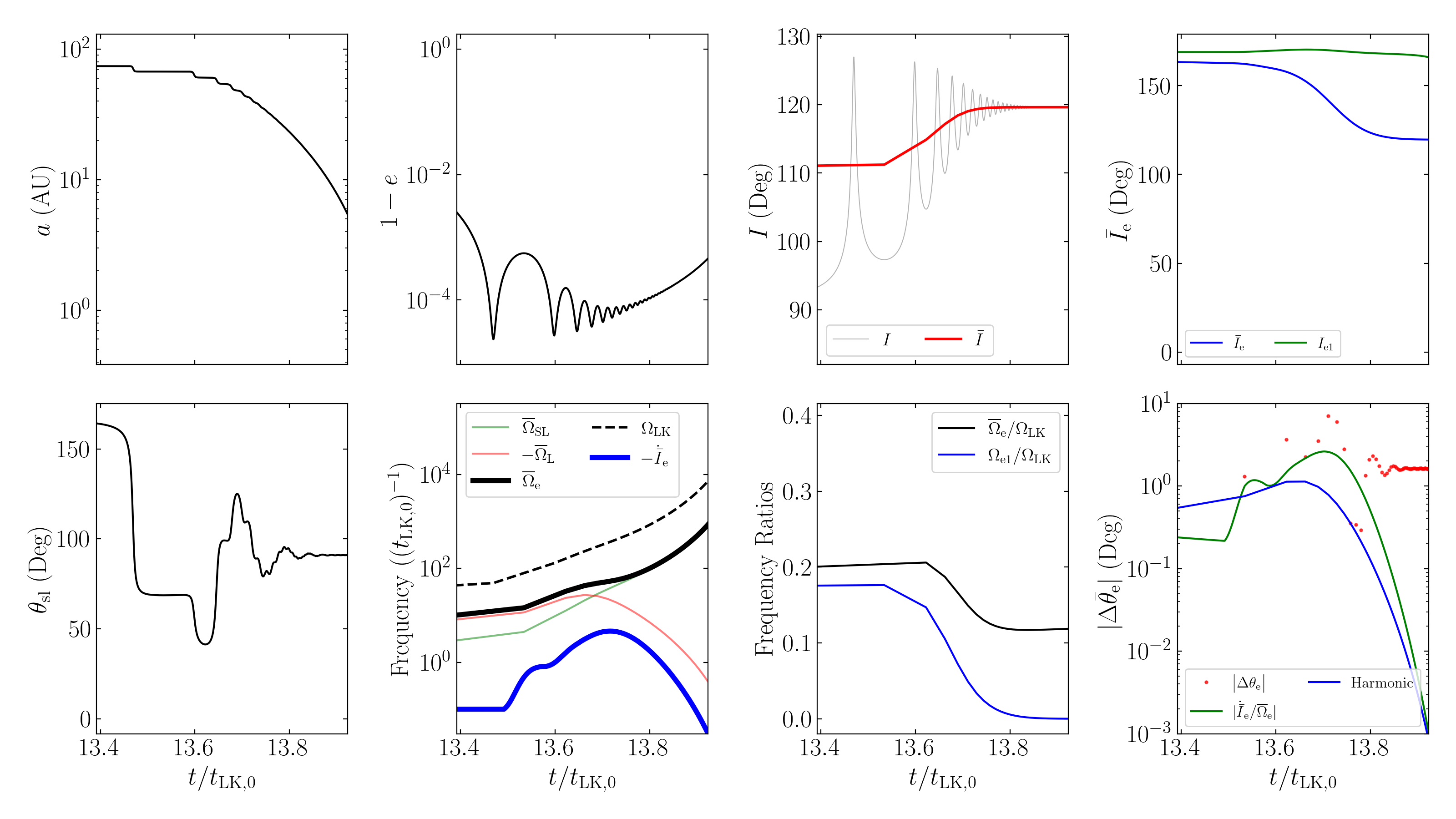}
    \caption{Same as Fig.~\ref{fig:4sim_90_350_zoom} except for $I_0 =
    90.2^\circ$ (and all other parameters are the same as in
    Fig.~\ref{fig:4sim_90_350}), corresponding to a faster coalescence. The
    total change in $\bar{\theta}_{\rm e}$ for this simulation is $\approx
    2^\circ$.}\label{fig:4sim_90_200_zoom}
\end{figure*}

For comparison, we show in Fig.~\ref{fig:4sim_90_200_zoom} a more rapid
binary merger starting with $I_0 = 90.2^\circ$, for which $\abs{\Delta
\theta_{\rm e}}_{\rm f} \approx 2^\circ$. If we again examine the
bottom-rightmost panel, we see that the net $\abs{\Delta \bar{\theta}_{\rm
e}}_{\rm f}$ obeys Eq.~\eqref{eq:nonad_dqeff_tot}.


\subsection{Estimate of Deviation from Adiabaticity from Initial Conditions}

To estimate Eq.~\eqref{eq:nonad_dqeff_tot} as a function of initial conditions,
we first differentiate Eq.~\eqref{eq:ie_def},
\begin{equation}
    \dot{\bar{I}}_{\rm e} = \p{\frac{\dot{\mathcal{A}}}{
            \mathcal{A}}}
        \frac{\mathcal{A} \sin \bar{I}}{
            1 + 2\mathcal{A}\cos \bar{I}
                + \mathcal{A}^2}.
\end{equation}
It also follows from Eq.~\eqref{eq:weff_def} that
\begin{equation}
    \overline{\Omega}_{\rm e} = \abs{\overline{\Omega}_{\rm L}}
        \p{1 + 2\mathcal{A}\cos \bar{I}
            + \mathcal{A}^2}^{1/2},
\end{equation}
from which we obtain
\begin{equation}
    \frac{\abs{\dot{\bar{I}}_{\rm e}}}{\overline{\Omega}_{\rm e}}
        = \p{\frac{\dot{\mathcal{A}}}{
            \mathcal{A}}}
        \frac{1}{\abs{\overline{\Omega}_{\rm L}}}
        \frac{\mathcal{A} \sin \bar{I}}{
            \p{1 + 2\mathcal{A}\cos \bar{I}
                + \mathcal{A}^2}^{3/2}}.\label{eq:idot_over_W_med}
\end{equation}
Moreover, if we assume the eccentricity is frozen around $e \simeq 1$ and use
$\overline{\cos^2 \omega} \simeq 1/2$ in $\abs{\Omega_{\rm L}} =
\abs{\rdil{\ascnode}{t}}$, we obtain the estimate
\begin{align}
    \mathcal{A} = \frac{\overline{\Omega}_{\rm SL}}{\overline{\Omega}_{\rm L}}
        &\simeq \frac{3Gn\p{m_2 + \mu/3}}{
            2c^2a j^2(e)}
                \s{\frac{15\cos \bar{I}}{
                    8t_{\rm LK}j(e)}}^{-1}\nonumber\\
        &\simeq \frac{4}{5}\frac{G(m_2 + \mu/3) m_{12}\tilde{a}_{\rm out}^3}{
            c^2m_3a^4 j(e) \cos \bar{I}},\label{eq:abar_eq1}\\
    \frac{\dot{\mathcal{A}}}{\mathcal{A}}
        &= -4\p{\frac{\dot{a}}{a}}_{\rm GW}
            + \frac{e}{j^2(e)}\p{\rd{e}{t}}_{\rm GW}.\label{eq:abar_dot}
\end{align}
With these, we see that Eq.~\eqref{eq:idot_over_W_med} is largest around
$\mathcal{A} \simeq 1$, and so we find that the maximum
$|\dot{\bar{I}}_{\rm e}| / \overline{\Omega}_{\rm e}$ is given
by
\begin{equation}
    \abs{\frac{\dot{\bar{I}}_{\rm e}}{\overline{\Omega}_{\rm e}}}_{\max}
        \simeq \abs{\frac{\dot{\mathcal{A}}}{\mathcal{A}}}
            \frac{1}{\abs{\overline{\Omega}_{\rm L}}}
            \frac{\sin \bar{I}}{\p{2 + 2\cos \bar{I}}^{3/2}}.
            \label{eq:idot_over_W}
\end{equation}

To evaluate this, we make two assumptions: (i) $\bar{I}$ is approximately
constant (see the third panels of Figs.~\ref{fig:4sim_90_350_zoom}
and~\ref{fig:4sim_90_200_zoom}), and (ii) $j(e)$ evaluated at $\mathcal{A}
\simeq 1$ can be approximated as a constant multiple of the initial
$j(e_{\max})$, i.e.
\begin{equation}
    j_{\star} \equiv j(e_{\star}) = f
        \sqrt{\frac{5}{3}\cos^2 I_0},\label{eq:jstar_ansatz}
\end{equation}
where the star subscript denotes evaluation at $\mathcal{A} \simeq 1$ and $f>1$
is a constant. Eq.~\eqref{eq:jstar_ansatz} assumes that $I_0$ far enough from
$90^\circ$ that the GR effect is unimportant in determining $e_{\max}$. The
value of f turns out to be relatively insensitive to $I_0$.

\begin{figure}
    \centering
    \includegraphics[width=\colummwidth]{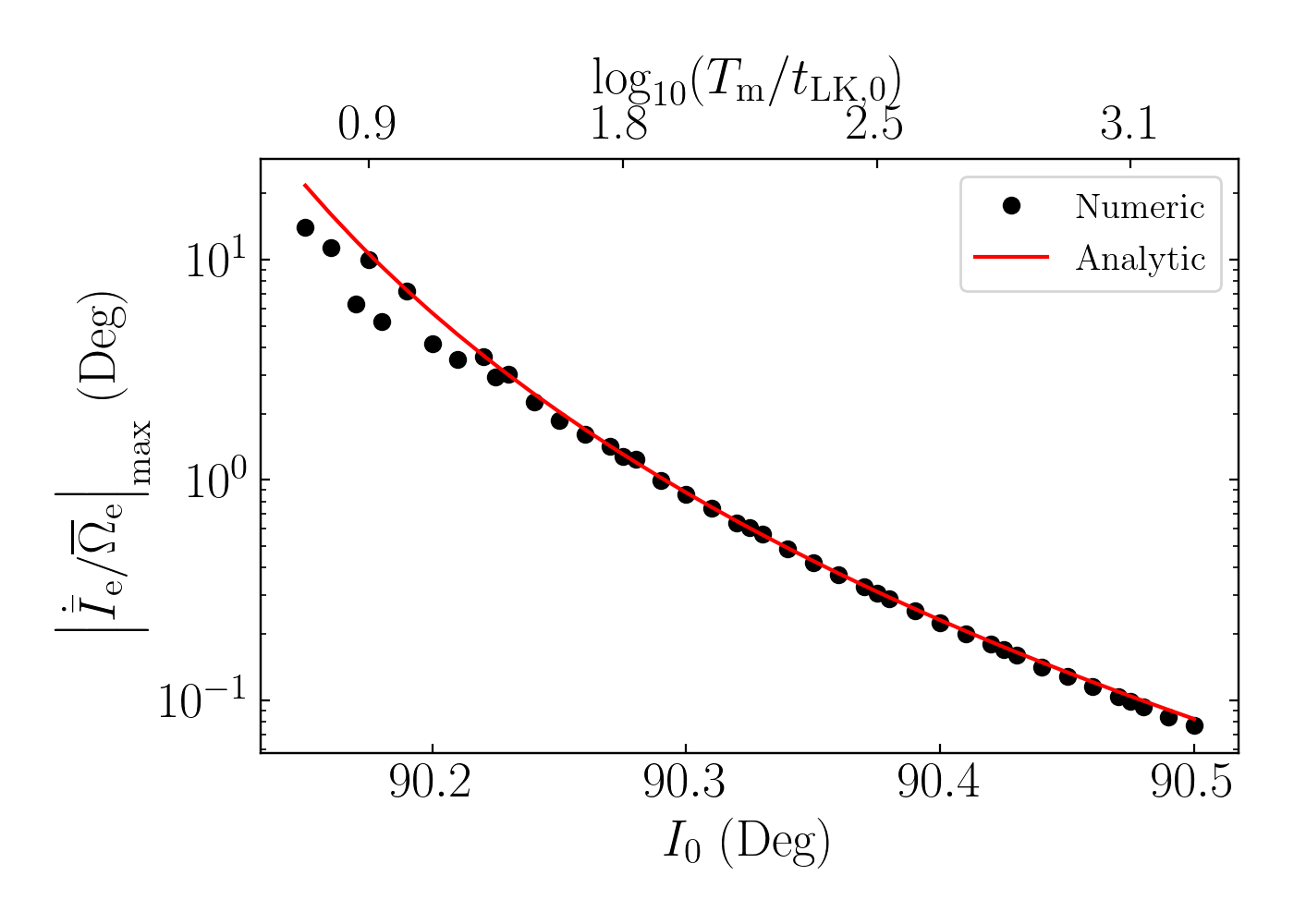}
    \caption{Comparison of $\abs{\dot{\bar{I}}_{\rm e} / \overline{\Omega}_{\rm
    e}}_{\max}$ obtained from simulations and from the analytical expression
    Eq.~\eqref{eq:prediction}, where we take $f = 2.72$ in
    Eq.~\eqref{eq:jstar_ansatz}. The merger time $T_{\rm m}$ is shown along
    the top axis of the plot in units of the initial LK
    timescale $t_{\rm LK, 0}$. The agreement between the analytical and
    numerical results is excellent for $T_{\rm m} \gg t_{\rm LK,
    0}$.}\label{fig:good_quants}
\end{figure}

Using Eq.~\eqref{eq:abar_dot} and approximating $e_\star \approx 1$ in
Eqs.~\eqref{eq:dadt_gw} and~\eqref{eq:dedt_gw} give
\begin{equation}
    \s{\frac{\dot{\mathcal{A}}}{\mathcal{A}}}_{\star}
        \simeq \frac{G^3 \mu m_{12}^2}{c^5a_\star^4j_\star^7} \frac{595}{3}.
        \label{eq:adot_star}
\end{equation}
To determine $a_\star$, we require Eq.~\eqref{eq:abar_eq1} to give $\mathcal{A}
= 1$ for $a_\star$ and $j_\star$. Taking this and Eq.~\eqref{eq:adot_star}, we
rewrite Eq.~\eqref{eq:idot_over_W_med} as
\begin{align}
    \abs{\frac{\dot{\bar{I}}_{\rm e}}{\overline{\Omega}_{\rm e}}}_{\max}
        \approx
            \frac{595 \sin\bar{I} \abs{\cos\bar{I}}^{3/8}}{36
                \left(\cos\bar{I} + 1\right)^{\frac{3}{2}}}
        \left[\frac{8000 G^{9} m_{12}^{9} m_3^{3}
            \mu^{8}}{\tilde{a}_{\rm out}^{9} j_\star^{37}c^{18} (m_2 + \mu /
            3)^{11}}\right]^{1/8}.
\end{align}
We can also calculate $|\dot{\bar{I}}_{\rm e}| / \overline{\Omega}_{\rm e}$ from
numerical simulations. Taking characteristic $\bar{I} \approx 120^\circ$
(Figs.~\ref{fig:4sim_90_350_zoom} and~\ref{fig:4sim_90_200_zoom} show that this
holds across a range of $I_0$), we fit the last remaining free parameter $f$
[Eq.~\eqref{eq:jstar_ansatz}] to the data from numerical simulations. This
yields $f \approx 2.72$, leading to
\begin{align}
    \abs{\frac{\dot{\bar{I}}_{\rm e}}{\overline{\Omega}_{\rm e}}}_{\max}
        \simeq{}& 0.98^\circ \p{\frac{\cos I_0}{\cos (90.3^\circ)}}^{-37/8}
            \p{\frac{\tilde{a}_{\rm out}}{2.2\;\mathrm{pc}}}^{-9/8}\nonumber\\
        & \times \p{\frac{m_3}{3 \times 10^7\;M_{\odot}}}^{3/8}
            \p{\frac{m_{12}^9 \mu^8 / (m_2 + \mu/3)^{11}}{\p{28.64M_{\odot}}^6}}
                ^{1/8}.
            \label{eq:prediction}
\end{align}
Figure~\ref{fig:good_quants} shows that when the merger time $T_{\rm m}$ is much
larger than the initial LK timescale, Eq.~\eqref{eq:prediction} provides an
accurate estimate for $|\dot{\bar{I}}_{\rm e} / \bar{\Omega}_{\rm e}|_{\max}$
when compared with numerical results.

In the above, we have assumed that the system evolves through $\mathcal{A}
\simeq 1$ when the eccentricity is mostly frozen (see
Fig.~\ref{fig:4sim_90_350} for an indication of how accurate this is for the
parameter space explored in Fig.~\ref{fig:good_quants}). It is also possible
that $\mathcal{A} \simeq 1$ occurs when the eccentricity is still undergoing
substantial oscillations. In fact, Eq.~\eqref{eq:prediction} remains accurate
in this case when replacing $e$ with $e_{\max}$, due to the following
analysis. Recall that when $e_{\min} \ll e_{\max}$, the binary spends a fraction
$\sim j(e_{\max})$ of the LK cycle near $e \simeq e_{\max}$. This fraction of
the LK cycle dominates both GW dissipation and $\overline{\Omega}_{\rm e}$
precession. Thus, both $\dot{\bar{I}}_{\rm e}$ and $\overline{\Omega}_{\rm e}$
in the eccentricity-oscillating regime can be evaluated by
setting $e \approx e_{\max}$ and multiplying by a prefactor of $j(e_{\max})$.
This factor cancels when computing the quotient $|\dot{\bar{I}}_{\rm e}| /
\overline{\Omega}_{\rm e}$.

\begin{figure}
    \centering
    \includegraphics[width=\colummwidth]{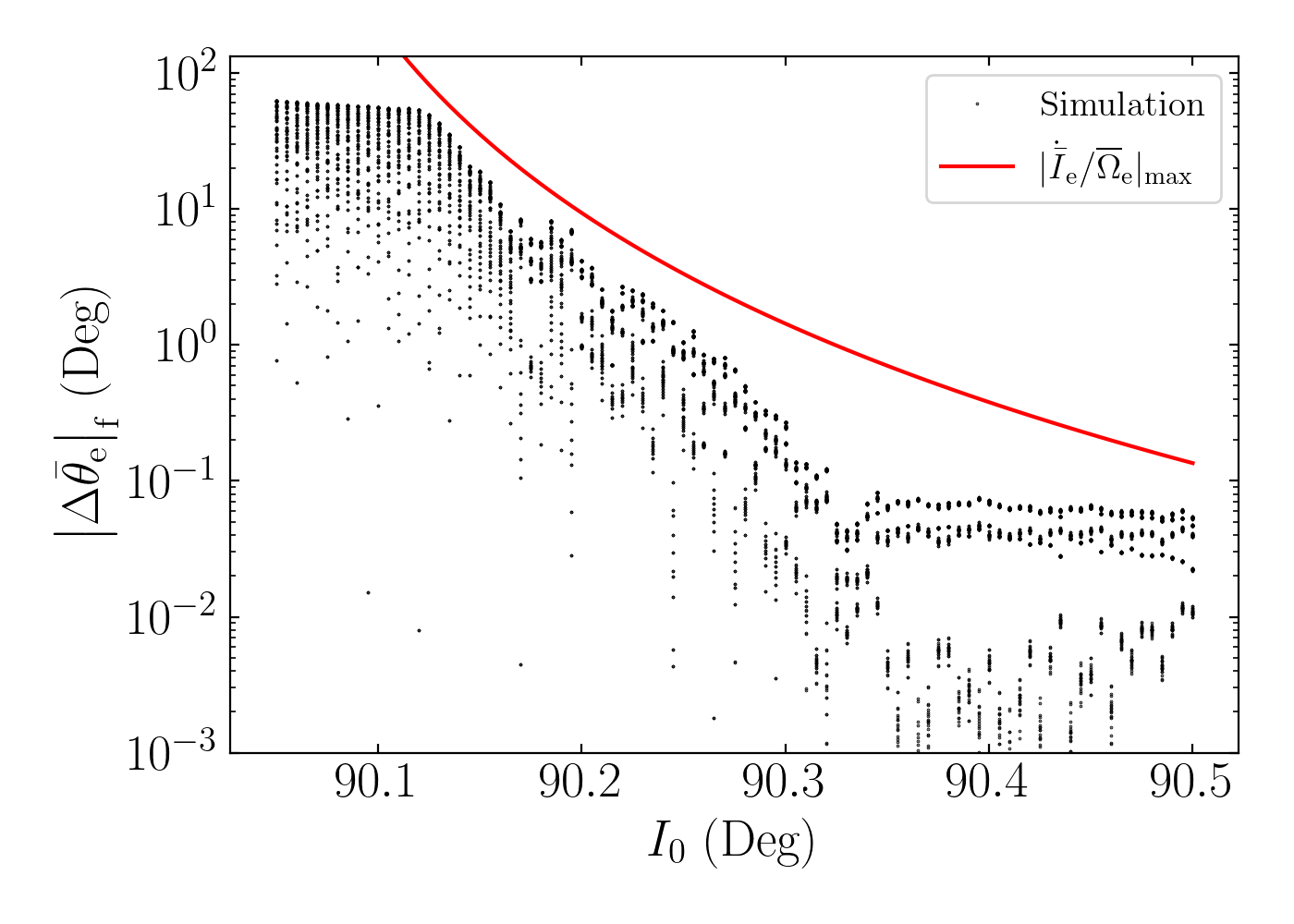}
    \caption{Net change in $\bar{\theta}_{\rm e}$ over the binary inspiral as a
    function of initial inclination $I_0$. For each $I_0$, $100$ simulations are
    run for $\bv{S}$ on a uniform, isotropic grid. Plotted for comparison is the
    bound $\abs{\Delta \bar{\theta}_{\rm e}}_{\rm f} \lesssim
    \abs{\dot{\bar{I}}_{\rm e} / \overline{\Omega}_{\rm e}}_{\max}$, using the
    analytical expression given by Eq.~\eqref{eq:prediction}. It is clear that the
    expression provides a robust upper bound for the non-conservation of
    $\bar{\theta}_{\rm e}$ due to nonadiabatic effects. Note that at the right of the
    plot, the numerical $\abs{\Delta \bar{\theta}_{\rm e}}_{\rm f}$ saturates;
    this is because we compute the initial $\bar{\Omega}_e$ (in order to
    evaluate the initial $\bar\theta_e$) without GW dissipation, and such a
    procedure inevitably introduces fuzziness in $\bar{\theta}_{\rm e}$.
    }\label{fig:deviations}
\end{figure}

The accuracy of Eq.~\eqref{eq:prediction} in bounding $\abs{\Delta
\bar{\theta}_{\rm e}}_{\rm f}$ is shown in Fig.~\ref{fig:deviations}, where we
carry out simulations for a range of $I_0$, and for each $I_0$ we consider $100$
different, isotropically distributed initial orientations for $\bv{S}$ (thus
sampling a wide range of initial initial $\bar{\theta}_{\rm e}$). Note that
conservation of $\bar{\theta}_{\rm e}$ is generally much better than
Eq.~\eqref{eq:prediction} predicts. This is because cancellation of phases in
Eq.~\eqref{eq:formal_sol_0} is generally more efficient than
Eq.~\eqref{eq:prediction} assumes (recall that Eq.~\eqref{eq:nonad_dqeff} only
provides an estimate for the amplitude of ``local'' oscillations of
$\bar{\theta}_{\rm e}$). Nevertheless, it is clear that
Eq.~\eqref{eq:prediction} provides a robust upper bound of
$\abs{\Delta\bar{\theta}_{\rm e}}_{\rm f}$, and serves as a good indicator for
the breakdown of adiabatic invariance.

\subsection{Origin of the $\theta_{\rm sl, f} = 90^\circ$ Attractor
}\label{ss:effect}

Using the approximate adiabatic invariant, we can now understand the origin of
the $\theta_{\rm sl, f} = 90^\circ$ attractor as shown in
Fig.~\ref{fig:qslscan}.

Recall from Eq.~\eqref{eq:weff_def}
\begin{align}
    \overline{\bv{\Omega}}_{\rm e} &=
            \overline{\Omega}_{\rm L} \uv{L}_{\rm out}
            + \overline{\Omega_{\rm SL}\uv{L}},\nonumber\\
        &= \p{\overline{\Omega}_{\rm L}
            + \overline{\Omega}_{\rm SL} \cos \bar{I}}\uv{Z}
            + \overline{\Omega}_{\rm SL} \sin \bar{I} \uv{X},
\end{align}
where $\uv{Z} = \uv{L}_{\rm out}$ and $\uv{X}$ is perpendicular to $\uv{Z}$ in
the $\uv{L}_{\rm out}$--$\uv{L}$ plane. Note that
\begin{align}
    \overline{\Omega}_{\rm L} &\propto \frac{\cos \bar{I}}{t_{\rm LK}}
        \propto a^{3/2}\cos \bar{I},\\
    \overline{\Omega}_{\rm SL} &\propto \frac{1}{a^{5/2}}.
\end{align}
Adiabatic invariance implies that $\bar{\theta}_{\rm e}$, the angle between
$\bv{S}$ and $\overline{\bv{\Omega}}_{\rm e}$ is conserved between $t = 0$ and
$t = t_{\rm f}$, i.e.
\begin{equation}
    \bar{\theta}_{\rm e, f} \simeq \bar{\theta}_{\rm e, 0}.
        \label{eq:4_4_qeffcons}
\end{equation}
At $t = t_{\rm f}$, $\overline{\Omega}_{\rm SL} \gg \abs{\overline{\Omega}_{\rm
L}}$ and the spin-orbit misalignment angle $\theta_{\rm sl}$ is ``frozen'',
implying $\overline{\bv{\Omega}}_{\rm e}$ is parallel to $\bv{L}$, and so
$\bar{\theta}_{\rm e, f} = \theta_{\rm sl, f}$. Eq.~\eqref{eq:4_4_qeffcons} then
gives
\begin{equation}
    \theta_{\rm sl, f} \simeq \bar{\theta}_{\rm e, 0},
        \label{eq:4_4_general}
\end{equation}
i.e.\ the final $\theta_{\rm sl}$ is determined by the initial angle between
$\bv{S}$ and $\overline{\bv{\Omega}}_{\rm e}$.

Now, first consider the case where the initial spin $\bv{S}_0$ is aligned
with the initial $\bv{L}_0$. This initial spin is inclined with respect to
$\overline{\bv{\Omega}}_{\rm e}$ by $\bar{\theta}_{\rm e, 0} = \abs{I_0 -
\bar{I}_{\rm e, 0}}$, where $I_0$ is the initial inclination angle
between $\bv{L}$ and $\bv{L}_{\rm out}$ and $\bar{I}_{\rm e, 0}$ is the
initial value of $\bar{I}_{\rm e}$. Thus, adiabatic invariance implies
\begin{equation}
    \theta_{\rm sl, f} \simeq \abs{I_0 - \bar{I}_{\rm e, 0}}.
\end{equation}
In the special case where the binary initially satisfies $\overline{\Omega}_{\rm
SL} \ll \abs{\overline{\Omega}_{\rm L}}$ or $\abs{\mathcal{A}_0} \ll 1$,
we find that $\overline{\bv{\Omega}}_{\rm e}$ is nearly parallel to $\bv{L}_{\rm
out}$ (for $I_0 < 90^\circ$) or antiparallel to $\bv{L}_{\rm out}$ (for
$I_0 > 90^\circ$). Thus,
\begin{equation}
    \theta_{\rm sl, f} =
    \begin{cases}
        I_0 & I_0 < 90^\circ,\\
        180^\circ - I_0 & I_0 > 90^\circ.
    \end{cases}
\end{equation}
Since LK-induced mergers necessarily require $I_0$ close to $90^\circ$, we
find that $\theta_{\rm sl, f}$ is ``attracted'' to $90^\circ$.

Eq.~\eqref{eq:4_4_general} can be applied to more general initial spin
orientations. For initial $\abs{\mathcal{A}_0} \ll 1$ (as required for
LK-induced mergers), $\overline{\Omega}_{\rm e, 0}$ is parallel to $\pm
\bv{L}_{\rm out}$. Suppose the initial inclination between $\bv{S}$ and
$\bv{L}_{\rm out}$ is $\theta_{\rm s,out, 0}$, then $\bar{\theta}_{\rm
e, 0} = \theta_{\rm s,out, 0}$ or $180^\circ - \theta_{\rm s,
out, 0}$ (depending on whether $I_0 < 90^\circ$ or $I_0 >
90^\circ$). Thus,
\begin{equation}
    \theta_{\rm sl, f} \simeq
    \begin{cases}
        \theta_{\rm s,out, 0} & I_0 < 90^\circ,\\
        180^\circ - \theta_{\rm s,out, 0} & I_0 > 90^\circ,
    \end{cases}
\end{equation}
\citep[see also][]{yu2020spin}.

So far, we have analyzed the $\theta_{\rm sl, f}$ distribution for smooth
mergers. Next, we can consider rapid mergers, for which $\bar{\theta}_{\rm e}$
conservation is imperfect. We expect
\begin{equation}
    \abs{\theta_{\rm sl, f} - \bar{\theta}_{\rm e, 0}}
        \lesssim \abs{\Delta \bar{\theta}_{\rm e}}_{\rm f}.\label{eq:qslf_plot_black}
\end{equation}
where $\abs{\Delta \bar{\theta}_{\rm e}}_{\rm f}$ is given by
Eq.~\eqref{eq:prediction}. This is shown as the black dotted line in
Fig.~\ref{fig:qslscan}, and we see it predicts the maximum deviation of
$\theta_{\rm sl, f}$ from $\sim 90^\circ$ except very near $I_0 =
90^\circ$. This is expected, as Eq.~\eqref{eq:prediction} is not very accurate
very near $I_0 = 90^\circ$, where it diverges (see Fig.~\ref{fig:deviations}).

When $I_0 = 90^\circ$ exactly, Fig.~\ref{fig:qslscan} shows that $\theta_{\rm
sl, f} = 0^\circ$. This can be understood: $I_0 = 90^\circ$ gives
$\rdil{I}{t} = 0$ by Eq.~\eqref{eq:dIdt}, so $I = 90^\circ$
for all time. This then yields $\rdil{\ascnode}{t} = 0$
[Eq.~\eqref{eq:dWdt}], implying that $\bv{L}$ is constant. Thus, $\bv{L}$ is
fixed as $\bv{S}$ precesses around it, and $\theta_{\rm sl}$ can never change.
In Fig.~\ref{fig:qslscan}, we take $\theta_{\rm sl, 0} = 0$, so $\theta_{\rm
sl, f} = 0$.

Finally, Fig.~\ref{fig:qslscan} shows that the actual $\theta_{\rm sl, f}$ are
oscillatory within the envelope bounded by Eq.~\eqref{eq:qslf_plot_black} above.
This can also be understood: Eq.~\eqref{eq:nonad_dqeff_tot} only bounds the
maximum of the absolute value of the change in $\bar{\theta}_{\rm e}$, while the
actual change depends on the initial and final complex phases of $S_{\perp}$ in
Eq.~\eqref{eq:formal_sol_0}, denoted $\Phi(0)$ and $\Phi(t_{\rm f})$. When
$\theta_{\rm sl, 0} = 0$, we have $\Phi(0) = 0$, as $\bv{S}$ starts in the
$\uv{x}$-$\uv{z}$ plane. Then, as $I_0$ is smoothly varied, the final
phase $\Phi\p{t_{\rm f}}$ must also vary smoothly [since $\overline{\Omega}_{\rm
e}$ in Eq.~\eqref{eq:Phi_t} is a continuous function, $\Phi\p{t}$ must be as
well], so the total phase difference between the initial and final values of
$S_{\perp}$ varies smoothly. This means the total change in $\bar{\theta}_{\rm
e}$ will fluctuate smoothly between $\pm \abs{\Delta \bar{\theta}_{\rm e}}_{\rm
f}$ as $I_0$ is changed, giving rise to the sinusoidal shape seen in
Fig.~\ref{fig:qslscan}.

%

\section{Analysis: Effect of Resonances}\label{s:harmonic}

In the previous section, we have shown that the $\theta_{\rm
sl, f}$ distribution in Fig.~\ref{fig:qslscan} and the
``$90^\circ$ attractor'' can be understood when neglecting the $N \geq 1$
Fourier harmonics in Eq.~\eqref{eq:dsdt_fullft}. In this section, we study the
effects of these neglected terms and show that they have a negligible effect in
LK-induced mergers. Separately, we will also consider the
LK-enhanced regime and show that these Fourier harmonics play a dominant role in
shaping the $\theta_{\rm sl, f}$ distribution.

For simplicity, we ignore the effects of GW dissipation in this section and
assume the system is exactly periodic (so $\dot{\bar{I}}_{\rm e} = 0$). The
scalar spin evolution equation
Eq.~\eqref{eq:formal_eom_allgen} is then
\begin{align}
    \rd{S_{\perp}}{t} ={}& i\overline{\Omega}_{\rm e}S_\perp
        + \sum\limits_{N = 1}^\infty\Big[
            \cos \p{\Delta I_{\rm eN}}S_\perp \nonumber\\
        &- i\cos \theta_{\rm e} \sin \p{\Delta I_{\rm eN}}\Big]
            \Omega_{\rm eN}\cos (N\Omega_{\rm LK} t).\label{eq:formal_sol_gen}
\end{align}

\subsection{Intuitive Analysis}\label{ss:intuitive_resonance}

We first restrict our attention to the effect of just the
$N$-th Fourier harmonic, and Eq.~\eqref{eq:formal_sol_gen}
further simplifies to
\begin{align}
    \rd{S_{\perp}}{t} ={}& \s{i\overline{\Omega}_{\rm e}
        + \Omega_{\rm eN}\cos \p{\Delta I_{\rm eN}} \cos\p{N \Omega_{\rm LK}t}
    } S_\perp\nonumber\\
        &- i\Omega_{\rm eN}\cos \bar{\theta}_{\rm e} \sin \p{\Delta I_{\rm eN}}
            \cos (N\Omega_{\rm LK} t).\label{eq:formal_sol_gen1}
\end{align}
There are two time-dependent perturbations, a modulation of the oscillation
frequency (the term proportional to $S_\perp$ on the right-hand
side) and a driving term. We can begin to understand the dynamics of
$S_{\perp}$ by considering the effect of each of these two terms
separately.

First, we consider the effect of frequency modulation alone. The equation of
motion is
\begin{equation}
    \rd{S_\perp}{t} \approx \s{i\overline{\Omega}_{\rm e}
        + \Omega_{\rm eN}\cos\p{\Delta I_{\rm eN}}\cos\p{N\Omega_{\rm LK}t}}
            S_\perp.
\end{equation}
The exact solution is
\begin{align}
    S_\perp(t) &= S_\perp(0)
        \exp\s{i\overline{\Omega}_{\rm e}t + \frac{\Omega_{\rm eN}
            \cos\p{\Delta I_{\rm eN}} }{N\Omega_{\rm LK}}
            \sin\p{N\Omega_{\rm LK}t}}.
\end{align}
There is no combination of parameters for which the magnitude of $S_{\perp}$
diverges, so the modulation of the oscillation frequency does not cause any
resonant behavior.

When considering only the time-dependent driving term instead, the equation of
motion is
\begin{align}
    \rd{S_{\perp}}{t} &\approx i\overline{\Omega}_{\rm e}S_\perp
        - i \Omega_{\rm eN}\cos \theta_{\rm e} \sin \p{\Delta I_{\rm eN}}
            \cos \p{N\Omega_{\rm LK} t}.
\end{align}
We can approximate $\cos \p{N\Omega_{\rm LK}t} \approx e^{iN\Omega_{\rm LK} t} /
2$, as the $e^{-iN\Omega_{\rm LK} t}$ component is always farther from resonance
(as $\Omega_{\rm LK}, \overline{\Omega}_{\rm e} > 0$). Then the equation of
motion has the solution
\begin{align}
    e^{-i\overline{\Omega}_{\rm e}t}S_{\perp}\Bigg|_{0}^{t}
        &= -\int\limits_{0}^{t}
            \frac{i \Omega_{\rm eN}\sin\p{\Delta I_{\rm eN}}}{2}
                e^{-i\overline{\Omega}_{\rm e}t + iN\Omega_{\rm LK} t} \cos
                \theta_{\rm e}
            \;\mathrm{d}t.\label{eq:harmonic_dS}
\end{align}
Since $\abs{S_{\perp}} = \sin \theta_{\rm e}$, the instantaneous
oscillation amplitude $\abs{\Delta \theta_{\rm e}}$ can be bound by
\begin{equation}
    \abs{\Delta \theta_{\rm e}} \sim \frac{1}{2}
        \abs{\frac{\Omega_{\rm eN} \sin \p{\Delta I_{\rm eN}}}{
             \overline{\Omega}_{\rm e} - N\Omega_{\rm LK}}}.
        \label{eq:harmonic_dqeff1}
\end{equation}
Thus, we see that large oscillation amplitudes in $\bar{\theta}_{\rm e}$ occur
when $\overline{\Omega}_{\rm e} \approx N\Omega_{\rm LK}$, the frequency of the
$N$-th Fourier harmonic.

\subsection{General Solution}\label{ss:resonance}

\subsubsection{Single Fourier Harmonic}

Above, we began the analysis of Eq.~\eqref{eq:formal_sol_gen1} by considering
the two time-dependent perturbations separately. However, this is not necessary.
Eq.~\eqref{eq:formal_sol_gen1} can be solved exactly:
\begin{align}
    e^{-i\Phi} S_\perp\Bigg|_{0}^{t} &=
        -i\Omega_{\rm eN} \cos \bar{\theta}_{\rm e} \sin \Delta I_{\rm eN}
        \int\limits_{0}^{t}
            \cos\p{N \Omega \tau}e^{-i\Phi(\tau)} \;\mathrm{d}\tau,\nonumber\\
        &= iA\s{e^{-i\Phi}\Bigg|_{0}^{t}
            + i\overline{\Omega}_{\rm e}\int\limits_{0}^t
                e^{-i\Phi(\tau)}\;\mathrm{d}\tau}.\label{eq:formal_sol_terms}
\end{align}
where $A = -\tan \Delta I_{\rm eN} \cos \bar{\theta}_{\rm e}$, and
\begin{align}
    i\Phi(t) &\equiv \int\limits^t
        i \overline{\Omega}_{\rm e} + \Omega_{\rm eN}\cos (N\Omega_{\rm LK}
            \tau) \cos \p{\Delta I_{\rm eN}}\;\mathrm{d}\tau,\nonumber\\
        &\equiv i\overline{\Omega}_{\rm e}t + \eta \sin\p{N\Omega_{\rm LK} t},
\end{align}
where $\eta \equiv \p{\Omega_{\rm eN} \cos \Delta I_{\rm eN}}/\p{N\Omega_{\rm
LK}}$. Eq.~\eqref{eq:formal_sol_terms} shows that $\abs{S_{\perp}}$ is bounded
for all $t$ unless the integral $\mathcal{I}(x)$, given by
\begin{equation}
    \mathcal{I}(x) = \int\limits_0^x e^{-i\xi - \eta \sin\p{\beta
        \xi}}\;\mathrm{d}\xi,
\end{equation}
grows without bound as $x \to \infty$, where $x = \overline{\Omega}_{\rm e}t$
and $\beta = N\Omega_{\rm LK} / \overline{\Omega}_{\rm e}$.

To see where $\mathcal{I}(x)$ grows without bound, we rewrite
\begin{equation}
    \mathcal{I}(x) = \sum\limits_{k = 0}^\infty \int\limits_{0}^x
        \p{\cos \xi + i\sin \xi} \frac{\p{-\eta \sin \p{\beta \xi}}^k}{k!}
        \;\mathrm{d}\xi.\label{eq:I_sum}
\end{equation}
These $\sin^{k}\p{\beta \xi}$ terms can be expanded using the general
trigonometric power-reduction identities \citep{zwillinger2002crc}:
\begin{align}
    \sin^{2n}y
        &= \frac{1}{2^{2n}} \binom{2n}{n}
            + \frac{\p{-1}^n}{2^{2n - 1}}\sum\limits_{l = 0}^{n - 1}
                \p{-1}^l\binom{2n}{l}
                \cos\s{2\p{n - l}y},\label{eq:identity1}\\
    \sin^{2n + 1}y &= \frac{\p{-1}^n}{4^n}\sum\limits_{l = 0}^n
        \p{-1}^l \binom{2n + 1}{l}\sin\s{\p{2n + 1 - 2l}y}.\label{eq:identity2}
\end{align}
Due to the orthogonality relations among the trigonometric functions,
$\mathcal{I}(x)$ only grows without bound if $\sin^k(\beta \xi)$ contains a
$\cos \xi$ or $\sin \xi$ term. Eqs.~\eqref{eq:identity1}
and~\eqref{eq:identity2} show that $\sin^{k}\p{\beta \xi}$ only contains a term
with unit frequency if $\beta = 1 / q$ for some integer $q \geq 1$.
When this is the case, all terms with $k \geq q$ in Eq.~\eqref{eq:I_sum} contain
a term with unit frequency. Among these terms, the $\sin^q\p{\beta \xi}$ term
has the largest prefactor. Neglecting the other terms with $k > q$, we can
evaluate $\mathcal{I}$ for any integer multiple $m$ of its
period $2\pi q$ to be
\begin{equation}
    \abs{\mathcal{I}\p{2\pi mq}} \approx 2\pi mq \p{\frac{\eta^q}{2^q q!}}.
        \label{eq:I_growth}
\end{equation}
Within each period, $\mathcal{I}(x)$ has additional oscillatory behavior due to
the other, off-resonance terms in Eq.~\eqref{eq:I_sum}. However, these
oscillations are periodic and vanish at every $x = 2\pi mq$, so they are bounded
and do not affect the divergence in Eq.~\eqref{eq:I_growth}. We conclude that
$\mathcal{I}(x)$ grows without bound when $\beta = 1 / q$, or
\begin{equation}
    \overline{\Omega}_{\rm e} = Nq\Omega_{\rm LK}.\label{eq:resonance}
\end{equation}
We see that this differs from the result of the intuitive
analysis [Eq.~\eqref{eq:harmonic_dqeff1}], as the $N$-th Fourier harmonic
generates infinitely many resonances indexed by $q \geq 1$.

Instead, if $\beta$ is near but not on a resonance, i.e.\ $0 < \abs{1 - q\beta}
\ll 1$, the amplitude of oscillation of $\mathcal{I}(x)$ is large but bounded.
If we take $q$ to be even, then the maximum value of $\abs{\mathcal{I}(x)}$ is
dominated by the near-resonance term in Eq.~\eqref{eq:I_sum},
\begin{align}
    \abs{\mathcal{I}\p{x}} &\simeq \abs{\int\limits_0^{x}
        \cos \xi \frac{\eta^q}{q!}\frac{1}{2^{q - 1}}
            \cos \p{q \beta \xi}\;\mathrm{d}\xi}\nonumber\\
        &\leq \frac{\abs{\eta}^q}{2^qq!}\frac{1}{\abs{1 -
            q\beta}}.\label{eq:scaling}
\end{align}
If $q$ is instead odd, we use Eq.~\eqref{eq:identity2} instead of
Eq.~\eqref{eq:identity1} and integrate against $i\sin \xi$ instead of $\cos \xi$
in Eq.~\eqref{eq:I_sum}, which results in the same bound on
the oscillation amplitude. Returning to Eq.~\eqref{eq:formal_sol_terms}, we
neglect the first, bounded term ($\lesssim e^\eta \simeq 1$) on the right-hand
side and obtain the total oscillation amplitude due to a $q$-th order resonance
with the $N$-th Fourier harmonic
\begin{equation}
    \abs{\Delta \bar{\theta}_{\rm e}}_{Nq} \sim \frac{1}{2^qq!}
        \abs{\tan \Delta I_{\rm eN}\s{\frac{
            \Omega_{\rm eN}\cos\p{\Delta I_{\rm eN}}}{
            N \Omega_{\rm LK}}}^q
            \p{\frac{\overline{\Omega}_{\rm e}}{
            \overline{\Omega}_{\rm e} - qN\Omega_{\rm LK}}}}.\label{eq:dq_Nq}
\end{equation}
Since $\overline{\Omega}_{\rm e} / N \Omega_{\rm LK} \approx q$, this reduces to
Eq.~\eqref{eq:harmonic_dqeff1} when $q = 1$, as expected.

\subsubsection{Generalization to Multiple Fourier Harmonics}

After having understood the effect of a single Fourier harmonic, we now return
to the spin evolution equation containing all of the Fourier
harmonics [Eq.~\eqref{eq:formal_sol_gen}]. If $\overline{\Omega}_{\rm e} /
\Omega_{\rm LK} \approx M$, there can now be multiple $N$ and $q$ satisfying $Nq
= M$. By linearity, the total $\abs{\Delta \bar{\theta}_{\rm e}}$ is given by
the sum over all of the resonances, so that \begin{equation} \abs{\Delta
\bar{\theta}_{\rm e}} \approx \sum\limits_{N, q \mid Nq = M} \abs{\Delta
\bar{\theta}_{\rm e}}_{Nq},\label{eq:dq_sum_res} \end{equation} where
$\abs{\Delta \bar{\theta}_{\rm e}}_{Nq}$ is given by Eq.~\eqref{eq:dq_Nq}.

We next attempt to understand whether particular combinations of $N$ and $q$
dominate this sum. We make a few simplifying assumptions: (i)
all $N$ harmonics are approximately equal\footnote{This approximation is
suitable for the problem studied in the main text because the only
characteristic frequency scale is $j_{\min}^{-1} \gg 1$, so all Fourier
harmonics $\bv{\Omega}_{\rm eN}$ for $N \lesssim j_{\min}^{-1}$ are similar.},
(ii) $\Omega_{\rm eN} \sim \overline{\Omega}_{\rm e}$ and $\cos \Delta I_{\rm
eN} \simeq 1$ (e.g.\ Figs.~\ref{fig:4sim_90_350_supp}
and~\ref{fig:4sim_90_350_zoom}). Under these assumptions, $\abs{\Delta
\bar{\theta}_{\rm e}}_{Nq}$ with fixed $Nq = M$ scales with respect to $q$ as
\begin{equation}
    \abs{\Delta \bar{\theta}_{\rm e}}_{Nq} \propto \frac{\cos^q\p{\Delta I_{\rm
        eN}} q^q}{2^qq!}.
\end{equation}
Stirling's formula then suggests that $\abs{\Delta
\bar{\theta}_{\rm e}}_{Nq} \propto \p{\cos\p{\Delta I_{\rm eN}} e / 2}^q /
\sqrt{q} \sim 1$. Thus, we conclude that all combinations of
$N$ and $q$ satisfying $Nq = M$ result in comparable oscillation amplitudes. For
simplicity, we evaluate this amplitude for $q = 1$ and $N = M$. If we denote the
number of pairs of $N$ and $q$ satisfying $Nq = M$ by $d(M)$ (the number of
positive divisors of $M$), we can approximate Eq.~\eqref{eq:dq_sum_res} by:
\begin{align}
    \abs{\Delta \bar{\theta}_{\rm e}}
        &\approx d(M) \abs{\Delta \bar{\theta}_{\rm e}}_{M1}\nonumber\\
        &\sim \frac{d(M)}{2}
            \abs{\frac{\Omega_{\rm eM}\sin \p{\Delta I_{\rm eM}}}
            {\overline{\Omega}_{\rm e} - M\Omega_{\rm LK}}},
            \label{eq:harmonic_dqeff}
\end{align}
Note that this agrees with Eq.~\eqref{eq:harmonic_dqeff1} except for the factor
of $d(M)$. Appendix~\ref{app:numeric} demonstrates that
Eq.~\eqref{eq:harmonic_dqeff} is in good agreement with detailed numerical
simulations when $M = 1$ or $M = 2$, the two most relevant
cases for our study.

\subsection{Effect of Resonances in LK-Induced Mergers}

\begin{figure}
    \centering
    \includegraphics[width=\colummwidth]{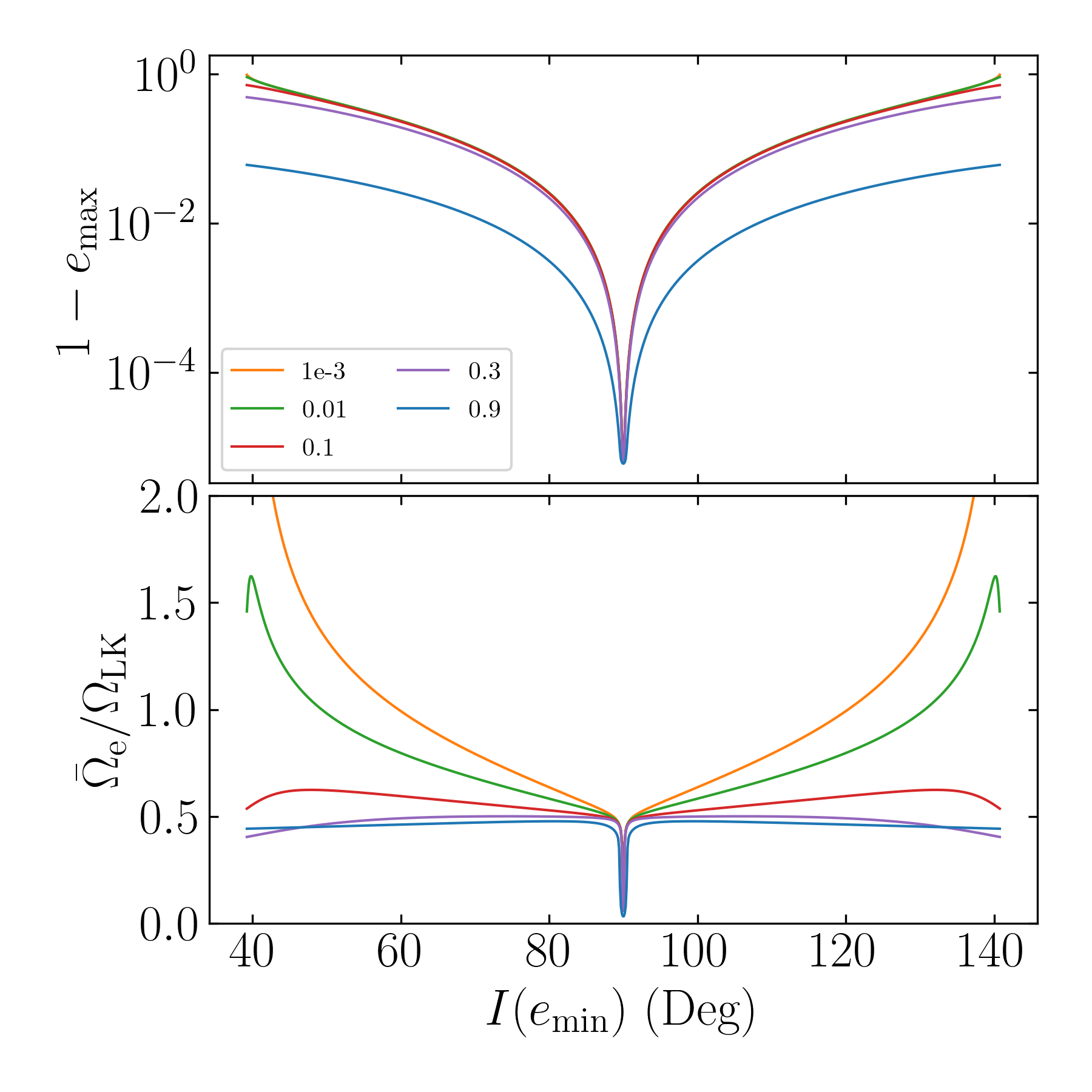}
    \caption{$e_{\max}$ and $\overline{\Omega}_{\rm e} / \Omega_{\rm LK}$ as a
    function of $I(e_{\min})$, the inclination of the inner binary at
    eccentricity minimum, for varying values of $e_{\min}$ (different colors
    as labeled) for a LK-induced merger (with
    the parameters the same as in Figs.~\ref{fig:4sim_90_350}
    and~\ref{fig:qslscan}). In this case, only systems with
    $I_0$ close to $90^\circ$ will merge within a Hubble time,
    $I(e_{\min}) \sim 90^\circ$ for most of the evolution (see
    Fig.~\ref{fig:4sim_90_350}) until $e_{\min} \approx 1$ is satisfied. This
    plot shows that $\overline{\Omega}_{\rm e} \lesssim 0.5 \Omega_{\rm LK}$ is
    a general feature of LK-induced mergers, as is the case for the fiducial
    simulation (see Fig.~\ref{fig:4sim_90_350_supp}).}\label{fig:dWs}
\end{figure}

We first consider the effect of these resonances
($\bar{\Omega}_{\rm e} = M\Omega_{\rm LK}$) on the LK-induced
regime, using the fiducial parameters (as in
Figs.~\ref{fig:4sim_90_350} and~\ref{fig:qslscan}). Numerically, we find that
$\overline{\Omega}_{\rm e} < \Omega_{\rm LK}$ for the region of parameter space
relevant to LK-induced mergers (see Fig.~\ref{fig:dWs}), so we focus on the
effect of the $M = 1$ resonance. If, for the entire inspiral,
$\overline{\Omega}_{\rm e} < \Omega_{\rm LK}$ by a sufficient margin that
Eq.~\eqref{eq:harmonic_dqeff} remains small, then the conservation of
$\bar{\theta}_{\rm e}$ cannot be significantly affected by this resonance. For
the fiducial simulation, Fig.~\ref{fig:4sim_90_350_supp} shows the ratio
$\overline{\Omega}_{\rm e} / \Omega_{\rm LK}$ in the middle panel (black) and
the amplitude of oscillation of $\bar{\theta}_{\rm e}$ due to the $M = 1$
resonance [Eq.~\eqref{eq:harmonic_dqeff}] in the bottom panel (blue). We see
that the system is never close to the resonant condition $\overline{\Omega}_{\rm
e} / \Omega_{\rm LK} = 1$, and as a result the net effect of
the resonance never exceeds a few degrees.

For a more precise comparison, the bottom-rightmost panel of
Fig.~\ref{fig:4sim_90_350_zoom} compares $\abs{\Delta \bar{\theta}_{\rm e}}$
in the fiducial simulation to the expected contributions from nonadiabatic
[Eq.~\eqref{eq:nonad_dqeff}] and resonant [Eq.~\eqref{eq:harmonic_dqeff} for
$M = 1$] effects in the regime where $\mathcal{A} \simeq 1$.
We see that Eq.~\eqref{eq:harmonic_dqeff} for $M = 1$ describes the oscillations
in $\bar{\theta}_{\rm e}$ very well. The agreement is poorer in the
bottom-rightmost panel of Fig.~\ref{fig:4sim_90_200_zoom}, as the nonadiabatic
effect is comparatively stronger.

It is somewhat surprising that the contribution of the resonances to the
instantaneous $\abs{\Delta \bar{\theta}_{\rm e}}$ when $\mathcal{A} \simeq 1$ is
dominant over that of the nonadiabatic contribution. In
Section~\ref{s:fast_merger}, we have shown
that neglecting resonant terms still allows for an accurate prediction of the
final $\bar{\theta}_{\rm e}$ deviation [Eq.~\eqref{eq:prediction}]. This implies
that, while the resonances have a larger contribution to $\abs{\Delta
\bar{\theta}_{\rm e}}$, the nonadiabatic effect is more important in determining
$\abs{\Delta \bar{\theta}_{\rm e}}_{\rm f}$. This also requires that a
$\abs{\Delta \bar{\theta}_{\rm e}}$ of up to a few degrees due to resonant
effects not affect $\abs{\Delta \bar{\theta}_{\rm e}}_{\rm f}$ by more than
$\sim 0.01^\circ$. This differs from the nonadiabatic case, where we find
$\abs{\Delta \bar{\theta}_{\rm e}}_{\rm f} \sim \max \abs{\Delta
\bar{\theta}_{\rm e}}$. The origin of these differences in behaviors may be due
to the complex phases cancelling differently in Eqs.~\eqref{eq:formal_sol_0}
and~\eqref{eq:formal_sol_terms} as the BH binary coalesces.

\subsection{Effect of Resonances in LK-Enhanced Mergers}\label{ss:lk_enhanced}

We turn now to the case of LK-enhanced mergers, as was studied in LL17, where
the inner binary is sufficiently compact ($\sim 0.1\;\mathrm{AU}$) that it can
merge in isolation via GW radiation. We consider a set of parameters that has
the same $t_{\rm LK, 0}$ as the system studied in LL17 but has a tertiary
SMBH\@: $m_1 = m_2 = 30M_{\odot}$, $a_0 = 0.1\;\mathrm{AU}$, $e_0 =
10^{-3}$, $m_3 = 3 \times 10^7 M_{\odot}$, $\tilde{a}_{\rm out} =
300\;\mathrm{AU}$, and $e_{\rm out} = 0$. We show that the resonances studied
above play an important role in shaping the $\theta_{\rm sl, f}$ distribution
for this regime.

\begin{figure}
    \centering
    \includegraphics[width=0.97\colummwidth]{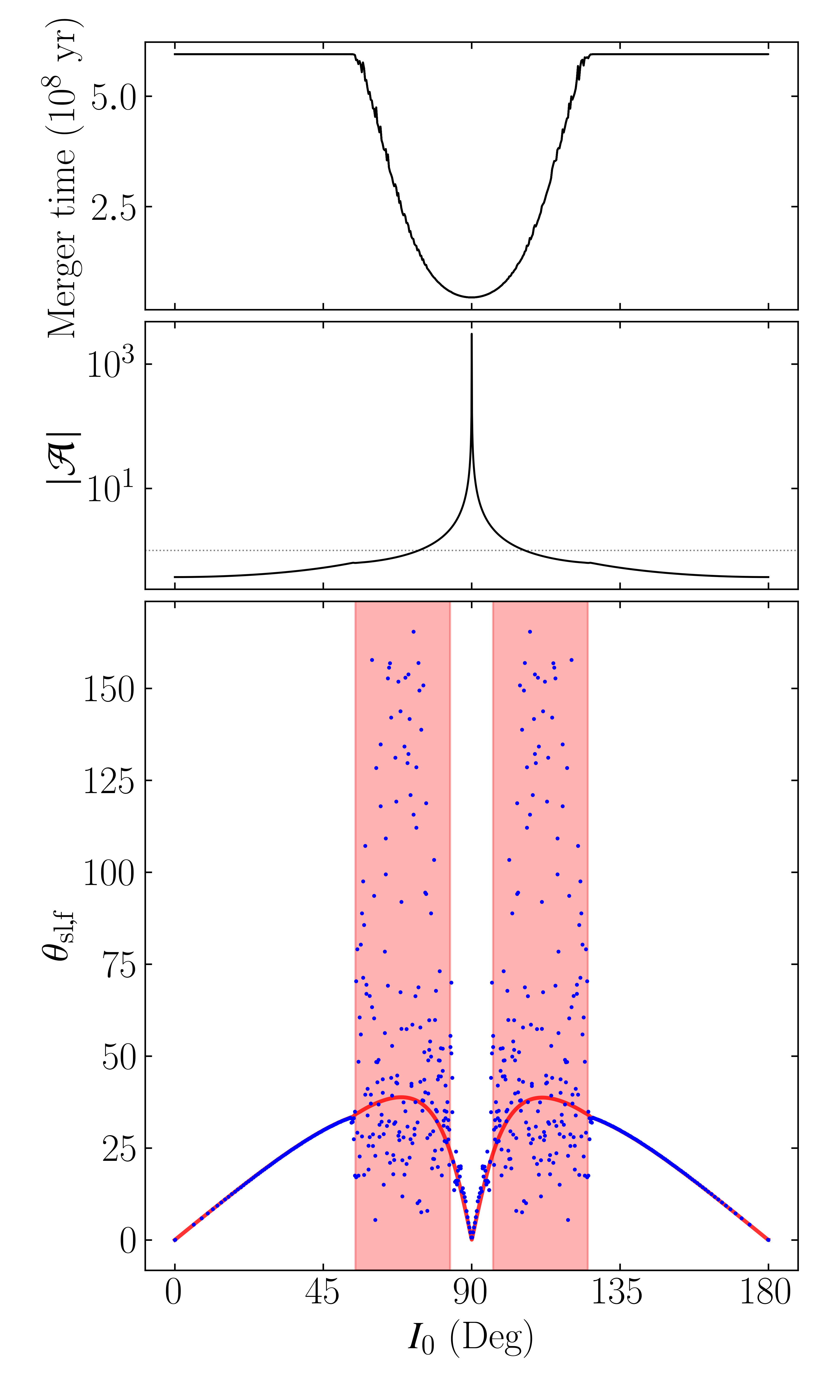}
    \caption{The merger time (top), the magnitude of the
    initial adiabaticity parameter $\abs{\mathcal{A}} \equiv
    \overline{\Omega}_{\rm SL} / |\overline{\Omega}_{\rm L}|$ (middle), and
    the final spin-orbit misalignment angle $\theta_{\rm sl, f}$ (bottom) for
    LK-enhanced mergers, with $m_1 = m_2 = 30M_{\odot}$, $m_3
    = 3 \times 10^{7}M_{\odot}$, $a_0 = 0.1\;\mathrm{AU}$, $e_0 = 10^{-3}$,
    $\tilde{a}_{\rm out} = 300\;\mathrm{AU}$, and $e_{\rm out} = 0$. In the
    middle panel, the horizontal dashed line indicates $\abs{\mathcal{A}} = 1$.
    In the bottom panel, the blue dots denote results from numerical simulations
    with $\theta_{\rm sl, 0} = 0$ [these are symmetric about $I_0 = 90^\circ$,
    as the equations of motion (\ref{eq:dadt}--\ref{eq:dwdt}) are as well]. The
    prediction for $\theta_{\rm sl, f}$ assuming conservation of
    $\bar{\theta}_{\rm e}$ is shown as the red line, which agrees well with the
    data both when there is no eccentricity excitation ($I_0 \lesssim 50^\circ$
    and $I_0 \gtrsim 130^\circ$) and when $\abs{\mathcal{A}} \gg 1$. For a
    substantial range of intermediate inclinations ($I_0 \in [50^\circ,
    80^\circ]$ and $I_0 \in [100^\circ, 130^\circ]$), $\theta_{\rm sl, f}$ is
    significantly affected by the resonances as they evolve  (see
    Fig.~\ref{fig:dWs_inner}). As such, these initial inclinations are expected
    to give rise to a wide range of $\theta_{\rm sl, f}$, and
    we denote this with broad red shaded regions.}\label{fig:bin_comp}
\end{figure}

First, we illustrate the $\theta_{\rm sl, f}$ distribution obtained via
numerical simulation, shown as the blue dots in Fig.~\ref{fig:bin_comp}. The
prediction assuming adiabatic invariance (i.e.\ the
conservation of $\bar{\theta}_{\rm e}$) is shown in the red solid line. Good
agreement is observed both when no eccentricity excitation occurs ($I_0 \lesssim
50^\circ$ and $I_0 \gtrsim 130^\circ$) and when $\abs{\mathcal{A}} \gg 1$
($80^\circ \lesssim I_0 \lesssim 100^\circ$). However, we see in
Fig.~\ref{fig:bin_comp} that for intermediate inclinations, $I_0 \in [50, 80]$
and $I_0 \in [100, 130]$, $\theta_{\rm sl, f}$ varies over a large range and
does not agree with the prediction of $\bar{\theta}_{\rm e}$ conservation. Note
that these inclinations correspond to neither the fastest nor slowest merging
systems.

\begin{figure}
    \centering
    \includegraphics[width=\colummwidth]{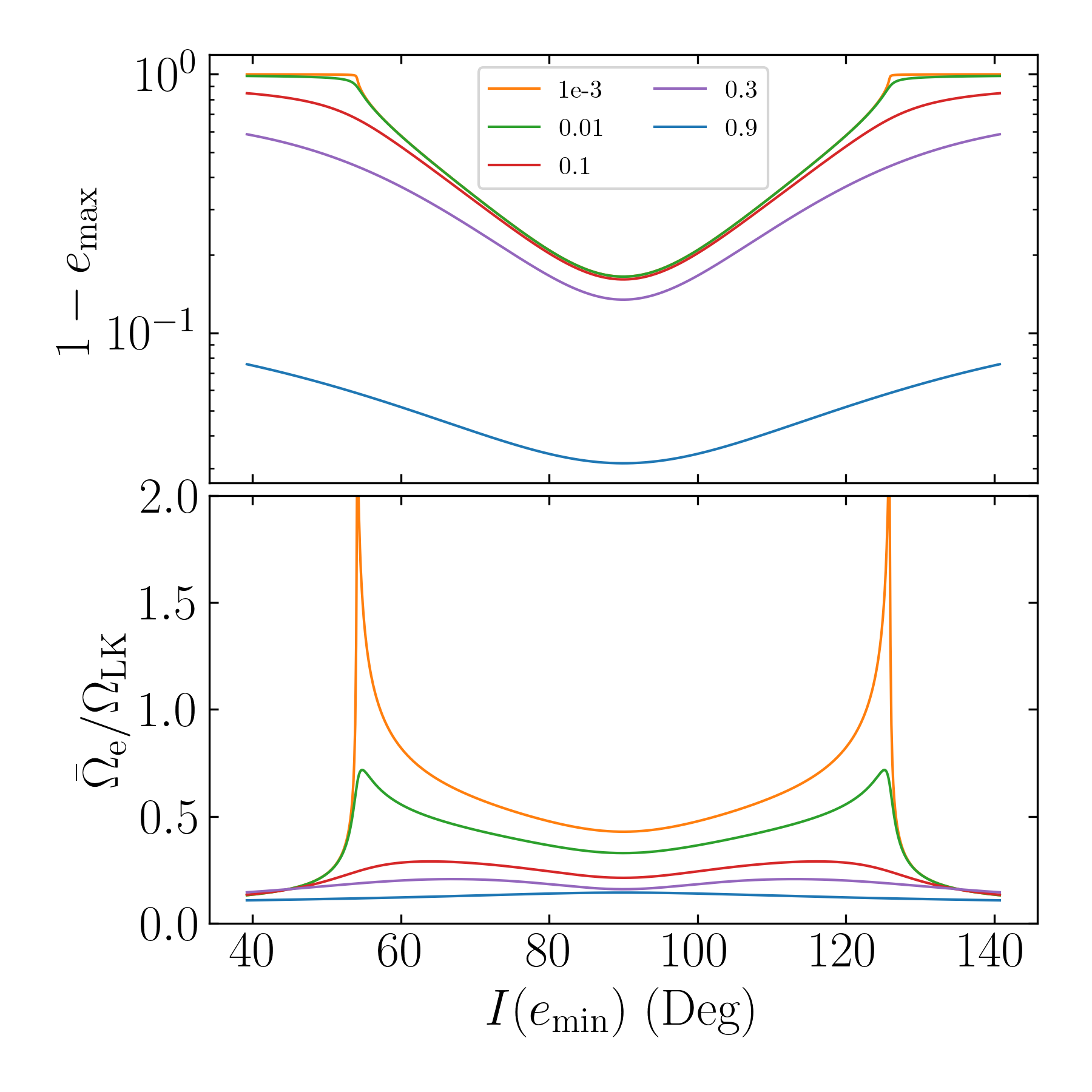}
    \caption{Same as Fig.~\ref{fig:dWs}, but for a LK-enhanced merger (with the
    parameters of Fig.~\ref{fig:bin_comp}). In the LK-enhanced regime, all
    initial inclinations merge within a Hubble time, and it is clear that
    $\overline{\Omega}_{\rm e} \approx \Omega_{\rm LK}$ can be satisfied for a
    wide range of initial inclinations when the initial
    eccentricity is sufficiently small.}\label{fig:dWs_inner}
\end{figure}

We attribute the origin of this wide scatter to resonance interactions.
Figure~\ref{fig:dWs_inner} illustrates that for these
intermediate inclinations, the condition
$\overline{\Omega}_{\rm e} \sim \Omega_{\rm LK}$ is satisfied.
In addition, as the inner binary coalesces under GW radiation, $e_{\min}$
becomes larger (e.g.\ see Fig.~\ref{fig:4sim_90_350}). This causes the locations
of the resonances at inclinations less than (greater than) $I =
90^\circ$ to evolve to smaller (larger) inclinations. As the location of the
resonances is a sensitive function of $e_{\min}$, resonance passage is
nonadiabatic. Thus, we expect that all systems that encounter the resonance,
i.e.\ all systems with intermediate initial inclinations, will experience an
impulsive kick to $\bar{\theta}_{\rm e}$, resulting in poor $\bar{\theta}_{\rm
e}$ conservation. This result is denoted by the two red shaded
regions in Fig.~\ref{fig:bin_comp}.

While the outer edges of the red shaded regions described above are located at
the critical $I_0$ required for nonzero eccentricity excitation, the inner edges
are harder to characterize. Systems with initial inclinations close to
$90^\circ$ start at the edge of the $M = 1$ resonance and quickly evolve
away from it (as $e_{\min}$ increases and $a$ decreases). As
such, they only interact briefly and weakly with the resonances, and the
cumulative effect of the resonance interaction can be estimated by evaluating
Eq.~\eqref{eq:harmonic_dqeff} for $M = 1$ at the initial conditions. We
empirically choose the transition between such ``weakly'' and ``strongly''
resonant systems, i.e.\ the inner edge of the broad red shaded region in
Fig.~\ref{fig:bin_comp}, when the oscillation semi-amplitude $\abs{\Delta
\bar{\theta}_{\rm e}}$ predicted by Eq.~\eqref{eq:harmonic_dqeff} exceeds
$3^\circ$.

To understand the general characteristics of systems that interact strongly with
resonances, we examine the quantities in Eq.~\eqref{eq:harmonic_dqeff}:
\begin{itemize}
    \item $\sin \p{\Delta I_{\rm eN}}$ is small unless $\mathcal{A} \simeq
        1$. Otherwise, $\bv{\Omega}_{\rm e}$ does not nutate substantially
        within a LK cycle, and all the $\bv{\Omega}_{\rm eN}$ are aligned with
        $\overline{\bv{\Omega}}_{\rm e}$ which implies that the $\Delta I_{\rm
        eN} \approx 0$ for all $N \geq 1$.

    \item Smaller values of $e_{\min}$ increase $\overline{\Omega}_{\rm e} /
        \Omega_{\rm LK}$, as shown in Fig.~\ref{fig:dWs_inner}.
\end{itemize}
However, the timescales over which $\mathcal{A}$ increases and $e_{\min}$
decreases are comparable (see Fig.~\ref{fig:4sim_90_350}). This implies that, if
$\mathcal{A} \ll 1$ initially, which is the case for LK-induced mergers, then
$e_{\min}$ will be very close to unity when $\mathcal{A}$ grows to be $\simeq
1$, and the contribution predicted by Eq.~\eqref{eq:harmonic_dqeff} will remain
small throughout the entire evolution. On the other hand, only if $\mathcal{A}
\simeq 1$ and $e_{\min} \ll 1$ initially, as is the case for the intermediate
inclinations in the LK-enhanced regime, are resonant interactions likely to be
significant.

\section{Stellar Mass Black Hole Triples}\label{s:stellar}

\begin{figure}
    \centering
    \includegraphics[width=0.7\colummwidth]{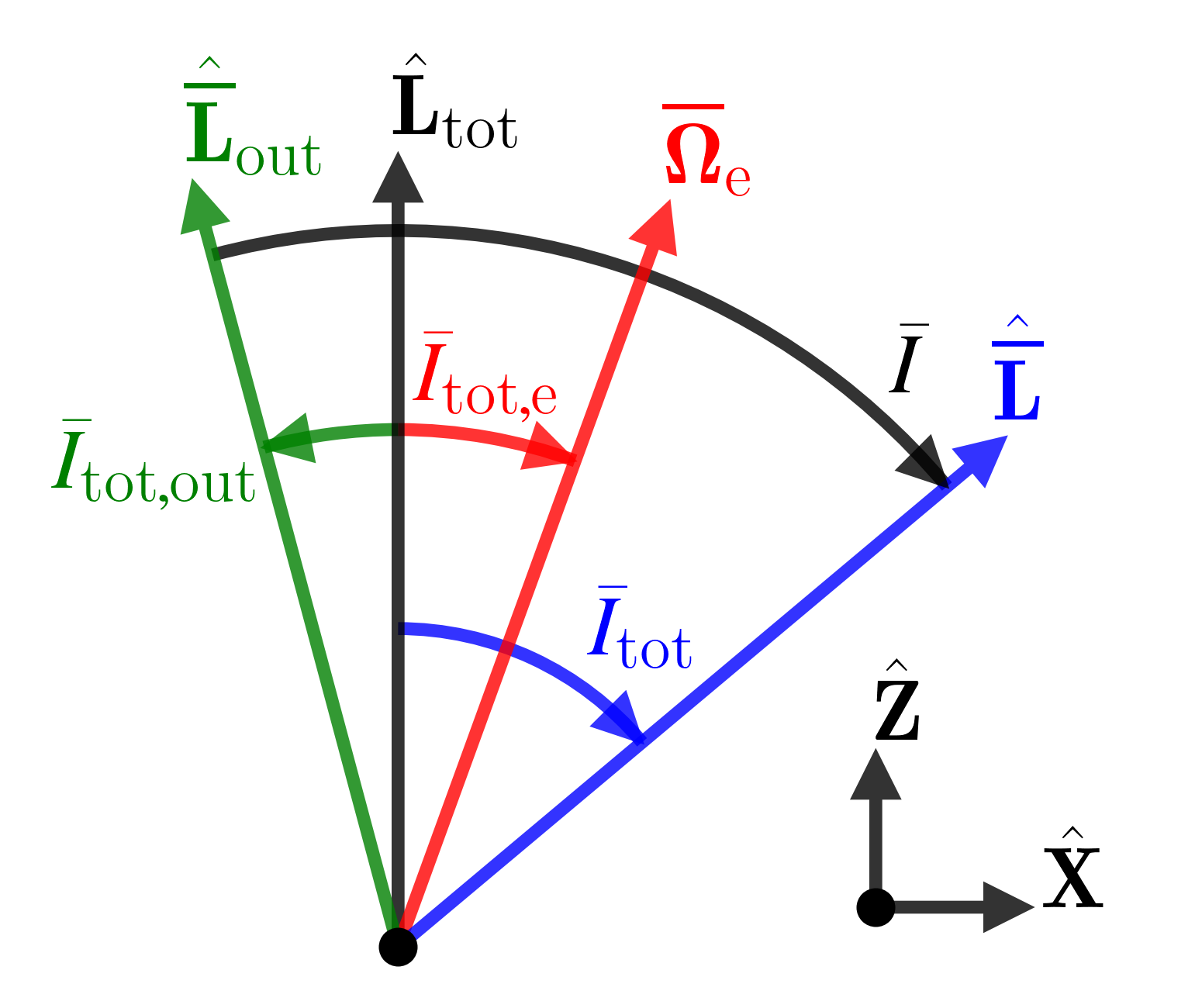}
    \caption{Definition of angles in the case where $L / L_{\rm out}$ is
    nonzero. Similar to before, we choose the convention where
    $I_{\rm tot,e} \in [0^\circ, 90^\circ]$ when
    $\overline{\Omega}_{\rm L} > 0$ and $I_{\rm tot,
    e} \in [90^\circ, 180^\circ]$ when $\overline{\Omega}_{\rm L} < 0$. Here,
    $\bv{L}_{\rm out}$ is not fixed, but $\bv{L}_{\rm tot} \equiv \bv{L} +
    \bv{L}_{\rm out}$ is. Note that the coordinate system is now oriented with
    $\uv{Z} \propto \bv{L}_{\rm tot}$.}\label{fig:3vec_eta}
\end{figure}

In this section, we extend our predictions for the final spin-orbit misalignment
angle $\theta_{\rm sl, f}$ to systems where all three masses are comparable
and the ratio of the angular momenta of the two binaries, given by
\begin{equation}
    \eta \equiv \at{\frac{L}{L_{\rm out}}}_{e = e_{\rm out} = 0} =
        \frac{\mu}{\mu_{\rm out}}
            \s{\frac{m_{12}a}{m_{123}a_{\rm out}}}^{1/2},
\end{equation}
where $m_{123} = m_{12} + m_3$ and $\mu_{\rm out} = m_{12}m_3 / m_{123}$, is not
negligible. When $\eta \neq 0$, $\bv{L}_{\rm out}$ is no longer fixed, but the
total angular momentum $\bv{L}_{\rm tot} \equiv \bv{L} + \bv{L}_{\rm out}$ is
fixed. We choose the coordinate system with $\uv{Z} = \uv{L}_{\rm tot}$, shown
in Fig.~\ref{fig:3vec_eta}.

To analyze this system, we still assume $e_{\rm out} \ll 1$ so that the
octupole-order effects are negligible. To calculate the evolution of $\bv{L}$,
it is only necessary to evolve the orbital elements of the inner binary [$a$,
$e$, $\ascnode$, $I_{\rm tot}$ (its inclination relative to $\uv{L}_{\rm tot}$),
and $\omega$] and a single orbital element for the outer binary, its inclination
$I_{\rm tot, out}$ relative to $\uv{L}_{\rm tot}$. The equations of motion are
given by \citep{bin_diego}:
\begin{align}
    \rd{a}{t} ={}& \p{\rd{a}{t}}_{\rm GW},\\
    \rd{e}{t} ={}& \frac{15}{8t_{\rm LK}} e\,j(e)\sin 2\omega
        \sin^2 I + \p{\rd{e}{t}}_{\rm GW},\\
    \rd{\ascnode}{t} ={}& \frac{L_{\rm tot}}{L_{\rm out}}\frac{3}{4t_{\rm LK}}
        \frac{\cos I\p{5e^2 \cos^2\omega - 4e^2 - 1}}{j(e)}
            \label{eq:dWdt_eta},\\
    \rd{I_{\rm tot}}{t} ={}& -\frac{15}{16t_{\rm LK}}\frac{e^2\sin 2\omega \sin
        2I}{j(e)},\\
    \rd{I_{\rm tot,out}}{t} ={}& -\eta\frac{15}{8t_{\rm LK}}
        \p{e^2\sin 2\omega \sin I},\\
    \rd{\omega}{t} ={}& \frac{3}{t_{\rm LK}} \left\{
        \frac{4 \cos^2 I + (5 \cos(2\omega) - 1)(1 - e^2 - \cos^2 I)} {8j(e)}
        \right.\nonumber\\
        &+ \left.\frac{\eta \cos I}{8} \s{2 + e^2 (3 - 5\cos(2\omega))}\right\}
        + \Omega_{\rm GR},
\end{align}
where $I = I_{\rm tot} + I_{\rm tot, out}$ is the relative inclination between
the two angular momenta. The spin evolution of one of the inner BHs is then
described in the frame corotating with $\bv{L}$ about $\bv{L}_{\rm tot}$ by the
equation of motion
\begin{equation}
    \p{\rd{\bv{S}}{t}}_{\rm rot} = \bv{\Omega}_{\rm e} \times \bv{S},
\end{equation}
where
\begin{align}
    \bv{\Omega}_{\rm e} &\equiv \Omega_{\rm SL}\uv{L}
        + \Omega_{\rm L} \uv{L}_{\rm tot}.
\end{align}
where $\Omega_{\rm L} = -\rdil{\ascnode}{t}$ [Eq.~\eqref{eq:dWdt_eta}] is the
rate of precession of $\bv{L}$ about $\bv{L}_{\rm tot}$.
As in Section~\ref{s:fast_merger}, we consider the LK-averaged
$\bv{\Omega}_{\rm e}$ and neglect the harmonic terms:
\begin{equation}
    \p{\rd{\overline{\bv{S}}}{t}}_{\rm rot}
        = \overline{\bv{\Omega}}_{\rm e} \times
        \overline{\bv{S}},\label{eq:dsdt_eta}
\end{equation}
where
\begin{align}
    \overline{\bv{\Omega}}_{\rm e} &= \overline{\Omega_{\rm SL} \sin I_{\rm
        tot}}\uv{X} + \p{\overline{\Omega}_{\rm L} + \overline{\Omega_{\rm
        SL}\cos I_{\rm tot}}}\uv{Z}\nonumber\\
        &\equiv \overline{\Omega}_{\rm SL} \sin \bar{I}_{\rm
        tot}\uv{X} + \p{\overline{\Omega}_{\rm L} + \overline{\Omega}_{\rm
        SL}\cos \bar{I}_{\rm tot}}\uv{Z}.\label{eq:We_component_finiteeta}
\end{align}

The results of Section~\ref{s:fast_merger} suggest that the angle
$\bar{\theta}_{\rm e}$ is an adiabatic invariant, where $\bar{\theta}_{\rm e}$
is given by
\begin{equation}
    \cos \bar{\theta}_{\rm e} \equiv
        \frac{\overline{\bv{\Omega}}_{\rm e}}{\overline{\Omega}_{\rm e}}
            \cdot \overline{\bv{S}},
\end{equation}
where $\overline{\bv{S}}$ is the spin vector averaged over a LK cycle. The
orientation of $\overline{\bv{\Omega}}_{\rm e}$ is described by the inclination
angle $\bar{I}_{\rm tot,e}$ (Fig.~\ref{fig:3vec_eta}), which can be expressed
using Eq.~\eqref{eq:We_component_finiteeta}
\begin{equation}
    \tan \bar{I}_{\rm tot,e} =
        \frac{\mathcal{A}\sin \bar{I}_{\rm tot}}{1 + \mathcal{A}\cos
        \bar{I}_{\rm tot}},
\end{equation}
where $\mathcal{A} \equiv \overline{\Omega}_{\rm SL} / \overline{\Omega}_{\rm
L}$ is the adiabaticity parameter.

At $t = t_{\rm f}$, the inner binary is sufficiently compact that $\theta_{\rm
sl}$ is frozen (see bottom right panel of Fig.~\ref{fig:4sim_90_350}), and the
system satisfies $\mathcal{A} \gg 1$ ($\overline{\Omega}_{\rm SL} \propto
a^{-5/2}$ while $\overline{\Omega}_{\rm L} \propto a^{3/2}$). When this is the
case, $\overline{\bv{\Omega}}_{\rm e} \parallel \bv{L}$, and so
$\bar{\theta}_{\rm e, f} = \theta_{\rm sl, f}$. Then, since adiabatic invariance
implies $\bar{\theta}_{\rm e, f} = \bar{\theta}_{\rm e, 0}$,
\begin{equation}
    \theta_{\rm sl, f} = \bar{\theta}_{\rm e, 0}.
\end{equation}

We first consider the case where $\bv{S}_0 \propto \bv{L}_{\rm
0}$. Then $\bar{\theta}_{\rm e, 0} = \abs{I_{\rm tot, 0} - \bar{I}_{\rm tot,e,
0}}$ (see Fig.~\ref{fig:3vec_eta}), and so
\begin{equation}
    \theta_{\rm sl, f} = \abs{I_{\rm tot, 0} - \bar{I}_{\rm e, 0}}.
        \label{eq:qslf_eta_pred}
\end{equation}
Suppose additionally that the binary initially satisfies
$|\overline{\Omega}_{\rm L}| \gg \overline{\Omega}_{\rm SL}$, then
$\overline{\bv{\Omega}}_{\rm e}$ is either parallel or anti-parallel to
$\bv{L}_{\rm tot}$ depending on whether $\overline{\Omega}_{\rm L}$ is positive
or negative. We denote the starting mutual inclination for which
$\overline{\Omega}_{\rm L}$ changes sign by $I_{\rm c}$. Note that $I_{\rm c} >
90^\circ$: even though $\Omega_{\rm L}$ changes sign at $I_0 =
90^\circ$, the inclination decreases over a LK cycle for $I < I_{\lim}$
(where $I_{\lim} > 90^\circ$ is the starting mutual inclination that maximizes
$e_{\max}$ \citep{bin2}), so the sign of $\Omega_{\rm L}$ changes over a LK
cycle for some $I_{\rm c} \in (90^\circ, I_{\lim})$. We then obtain that
\begin{equation}
    \theta_{\rm sl, f} =
    \begin{cases}
        I_{\rm tot, 0}, & I_0 < I_{\rm c},\\
        180^\circ - I_{\rm tot, 0}, & I_0 > I_{\rm c}.
    \end{cases}
\end{equation}

More generally, so long as $\mathcal{A} \ll 1$ initially, we can specify the
initial spin orientation by $\theta_{\rm s,tot, 0}$, the initial angle
between $\bv{S}$ and $\bv{L}_{\rm tot}$, giving
\begin{equation}
    \theta_{\rm sl, f} =
    \begin{cases}
        \theta_{\rm s,tot, 0}, & I < I_{\rm c},\\
        180^\circ - \theta_{\rm s,tot, 0}, & I > I_{\rm c}.
    \end{cases}
\end{equation}

\begin{figure}
    \centering
    \includegraphics[width=\colummwidth]{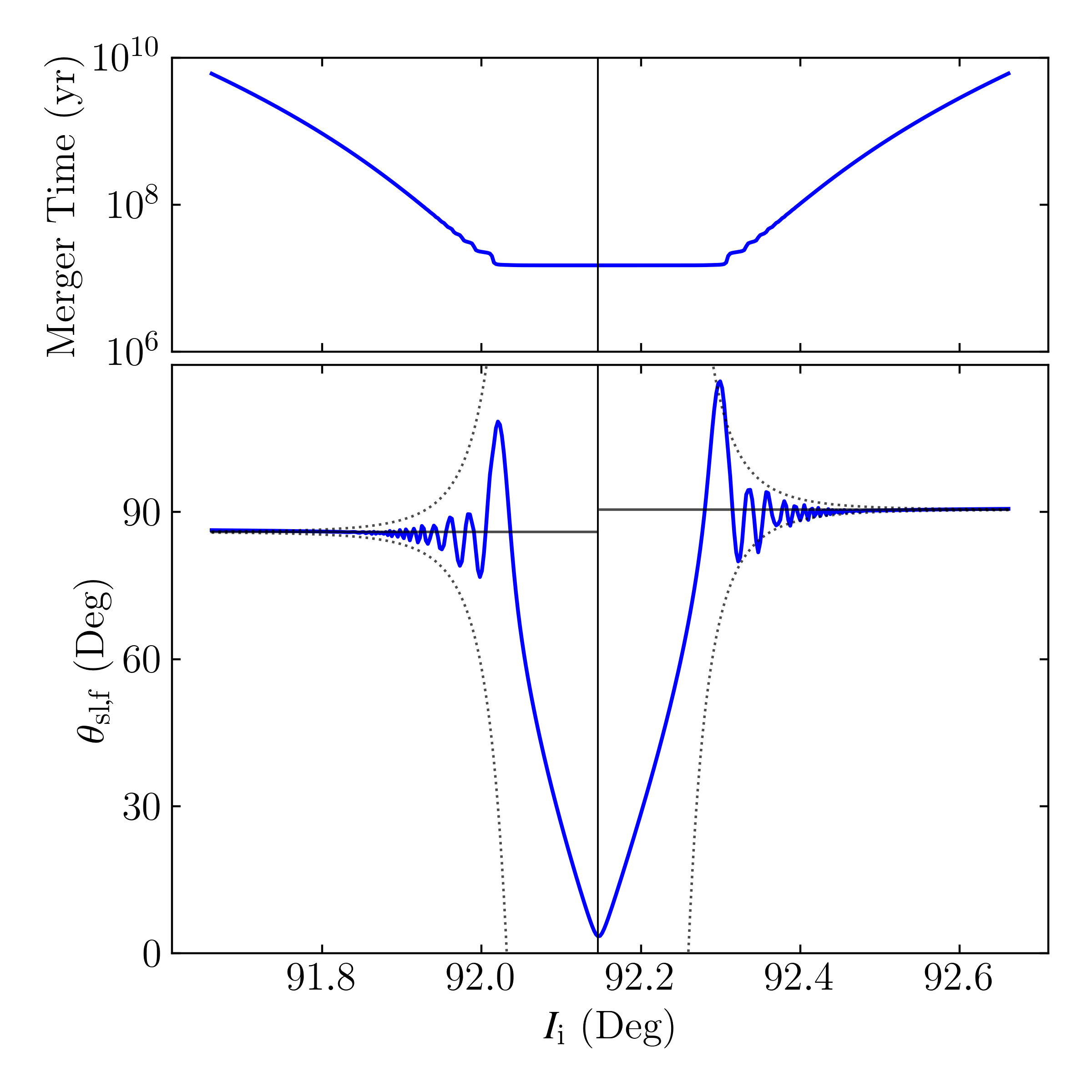}
    \caption{Similar to Fig.~\ref{fig:qslscan} but for stellar-mass tertiary
    $m_3 = 30M_{\odot}$ and $\tilde{a}_3 = 4500\;\mathrm{AU}$. The vertical
    black line denotes $I_{\rm c} \approx 92.14^\circ$, the
    initial inclination for which $\overline{\Omega}_{\rm L}$
    changes signs, and the two horizontal lines denote the predictions of
    Eq.~\eqref{eq:qslf_eta_pred}. The dotted black lines bound the deviation due
    to non-adiabatic evolution, given by Eq.~\eqref{eq:qslf_plot_black}.
    }\label{fig:finite_qslfscan}
\end{figure}

We first compare these results to numerical simulations by considering a
stellar-mass BH triple that is in the LK-induced regime: we use the same inner
binary parameters as the example of Fig.~\ref{fig:4sim_90_350},
but for a tertiary companion with $m_3 = 30M_{\odot}$, $a_{\rm out} =
4500\;\mathrm{AU}$, and $e_{\rm out} = 0$. Figure~\ref{fig:finite_qslfscan} shows
that Eq.~\eqref{eq:qslf_eta_pred} accurately predicts $\theta_{\rm sl, f}$ when
$\theta_{\rm sl, 0} = 0^\circ$ for this parameter regime when conservation of
$\bar{\theta}_{\rm e}$ is good. Furthermore, deviations from exact
$\bar{\theta}_{\rm e}$ conservation are well described by
Eq.~\eqref{eq:qslf_plot_black}, the prediction of the theory in
Section~\ref{s:fast_merger}. Unlike the $\eta = 0$ case
(Fig.~\ref{fig:qslscan}), $\theta_{\rm sl, f}$ is not symmetric about $I_{\rm c}
\approx 92.14^\circ$. This is because
$|\overline{\Omega}_{\rm L}|$ is not exactly equal on
either side of $I_{\rm c}$. Additionally, unlike in the $\eta = 0$ case, the
minimum $\theta_{\rm sl,f}$ is not exactly zero. We showed in
Section~\ref{ss:effect} that when $\eta = 0$, $\bv{L}$ is fixed when $I_0 =
90^\circ$, as $\rdil{\ascnode}{t} = \rdil{I}{t} = 0$. For nonzero $\eta$,
neither $\rdil{\ascnode}{t}$ nor $\rdil{I_{\rm tot}}{t}$ is
zero at $I_{\rm 0} = I_{\rm c}$.

\begin{figure}
    \centering
    \includegraphics[width=\colummwidth]{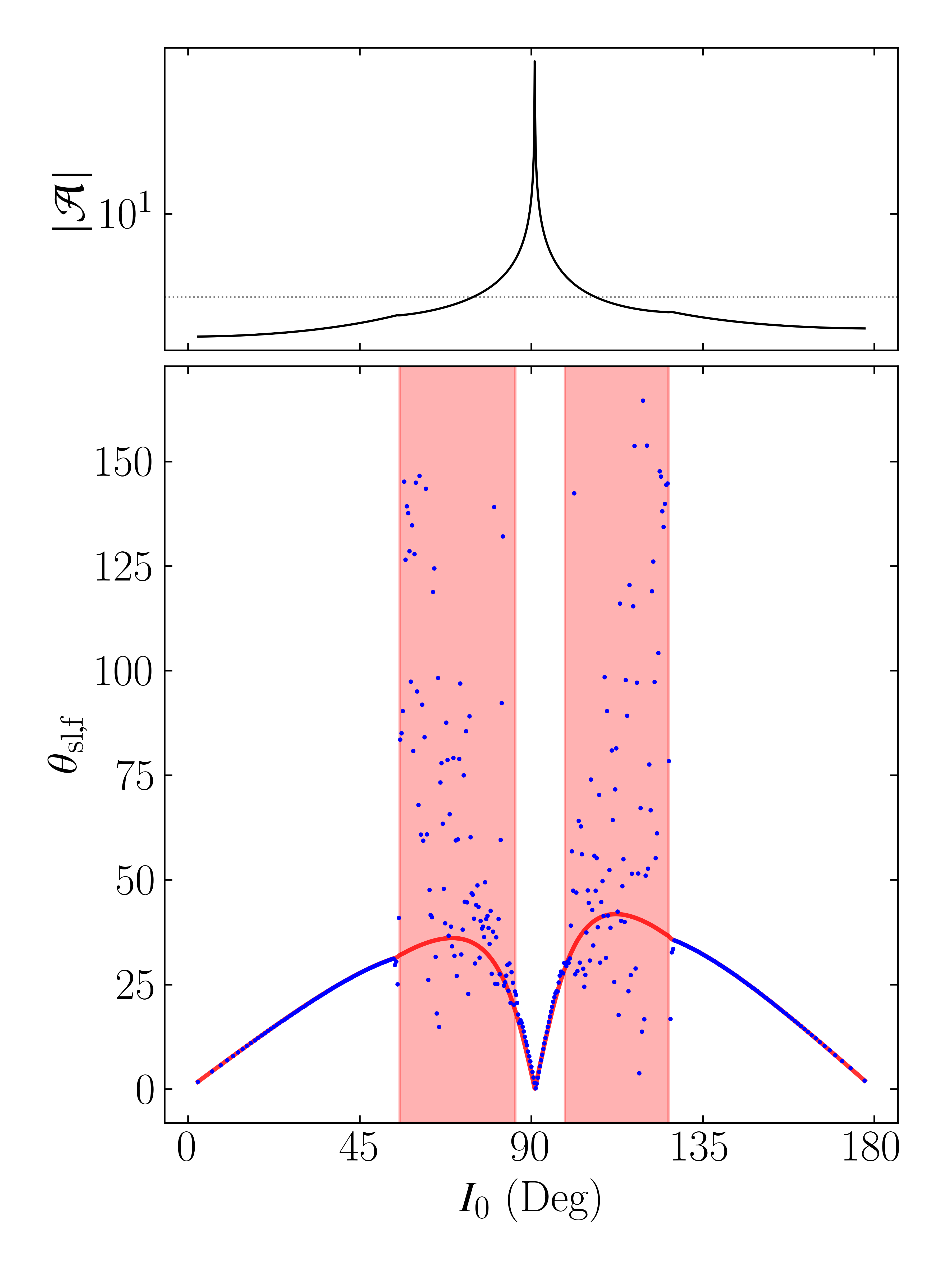}
    \caption{Similar to Fig.~\ref{fig:bin_comp} except for a
    stellar mass tertiary $m_3 = 30M_{\odot}$ and $\tilde{a}_{\rm out} =
    3\;\mathrm{AU}$.}\label{fig:finite_bin_comp}
\end{figure}

Eq.~\eqref{eq:qslf_eta_pred} also gives good agreement in the
LK-enhanced merger regime. We consider the same inner binary parameters as in
Fig.~\ref{fig:bin_comp} but use a tertiary companion with $m_3 = 30M_{\odot}$,
$a_{\rm out} = 3\;\mathrm{AU}$, and $e_{\rm out} = 0$. The results are shown in
Fig.~\ref{fig:finite_bin_comp}.

\section{Conclusion and Discussion}\label{s:discussion}

In this paper, we have carried out a theoretical study on the
evolution of spin-orbit misalignments in tertiary-induced black-hole (BH) binary
mergers. Recent numerical works \citep{bin1, bin2, bin3,
antonini2018precessional, yu2020spin} have revealed that when binary BHs undergo
mergers due to Lidov-Kozai (LK) oscillations driven by a tertiary companion, the
BH spin may evolve toward a perpendicular state where the final spin-orbit
misalignment angle $\theta_{\rm sl}$ is close to $90^\circ$. Our theoretical
analysis in this paper provides an understanding of this ``$90^\circ$
attractor'' and characterizes its regime of validity and various spin evolution
behaviors during such LK-induced mergers. We focus on
hierachical triple systems where the inner BH binary experiences the
``standard'' quadupole LK oscillations and eventually merges, with the octuple
effects playing a negigible role [$\epsilon_{\rm oct}\ll 1$; see
Eq.~\eqref{eq:epsoct}]. For such systems, the spin vectors of the inner BHs obey
a simple evolution equation, Eq.~\eqref{eq:dsdt_weff} or
Eq.~\eqref{eq:dsdt_eta}, where the ``effective'' precssion rate
$\bv{\Omega}_{\rm e}$ varies quasi-periodically due to the combined effects of
LK oscillations and gravitational radiation. Analysis of this equation yields
the following conclusions:

\begin{itemize}
    \item For BH binaries that have too large initial separations to merge
        in isolation, LK-induced mergers require large/extreme
        eccentricity excitations in the binary driven by a highly inclined
        tertiary companion. For such systems, the BH spin evolution behavior can
        be generally captured by replacing $\bv{\Omega}_{\rm e}$ with its
        LK-average $\overline{\bv{\Omega}}_{\rm e}$ [thus neglecting the
        Fourier harmonic terms in Eq.~\eqref{eq:dsdt_fullft}]. If the orbital decay is
        sufficiently gradual, the angle $\bar{\theta}_{\rm e}$
        [Eq.~\eqref{eq:q_eff}] between the spin axis and
        $\overline{\bv{\Omega}}_{\rm e}$ is an adiabatic invariant. This
        naturally explains the ``$90^\circ$ attractor'' for the final spin-orbit
        misalignment angle when the initial tertiary inclination $I_0$ is not
        too close to $90^\circ$ and the initial BH spin axis is aligned with the
        orbital angular momentum axis (see Fig.~\ref{fig:qslscan}). We show that
        the deviation from perfect adiabaticity can be predicted from initial
        conditions [see Eq.~\ref{eq:prediction} and Fig.~\ref{fig:deviations}].

    \item When the resonant condition $\overline{\Omega}_{\rm e} \approx
        M\Omega_{\rm LK}$ for integer $M$ is satisfied, significant variations
        in $\bar{\theta}_{\rm e}$ can arise. We derive an analytic estimate of
        this variation amplitude [Eq.~\eqref{eq:harmonic_dqeff}]. This estimate
        demonstrates that the resonances are unimportant for ``LK-induced''
        mergers (as depicted in Figs.~\ref{fig:4sim_90_350}--\ref{fig:qslscan}
        and~\ref{fig:finite_qslfscan}), but become important for ``LK-enhanced''
        mergers, where the BH binaries exhibit only modest eccentricity
        excitations. Our analysis of the resonance effects qualitatively explain
        the behavior in $\theta_{\rm sl, f}$ as seen in LK-enhanced mergers (see
        Figs.~\ref{fig:bin_comp} and~\ref{fig:finite_bin_comp}).

    \item For LK-induced mergers of BH binaries with general tertiary
        companions, we provide an analytic prescription for calculating the
        final spin-orbit misalignment angle for arbitrary initial spin
        orientations (Section~\ref{s:stellar}). This prescirption is based on
        the approximate adiabatic invariance of $\bar{\theta}_{\rm e}$, and
        produces results that are in agreement with numerical simulations (see
        Fig.~\ref{fig:finite_qslfscan}) in the appropriate regime.
\end{itemize}

There are several simplificatons in our theoretical analysis that are worth
mentioning. (i) We have neglected the octupole effects in the LK oscillations.
This is appropriate for systems where the tertiary orbit is circular and/or the
semi-major axis $a_{\rm out}$ is much larger than the inner binary (as in the
case when the tertiary is a SMBH) and/or the inner binary BHs have nearly equal
masses [see Eq.~\eqref{eq:epsoct}]. The octupole effects are known to
significantly broaden the inclination window for extreme eccentricity
excitations, and therefore enhance the efficiency of LK-induced mergers
\citep{bin2}. When the octupole effects are significant, the LK orbital
evolution is not integrable, and the eccentricity excitations are no longer
regular. As a result, $\bv{\Omega}_{\rm e}$ has neither consistent direction nor
magnitude, and our theory cannot be applied. In fact, the resulting $\theta_{\rm
sl, f}$ distribution is largely unrelated to the initial $\bar{\theta}_{\rm e,
0}$ and the ``$90^\circ$ attractor'' is significantly ``erased'' \citep{bin2}.
(ii) If the system is not sufficiently hierarchical ($a_{\rm out}$ is too
small), the double averaging approximation for the dynamics of the triple breaks
down \citep{bin2}. In this case, there is little reason to expect any relation
between $\theta_{\rm sl, f}$ and $\bar{\theta}_{\rm e, 0}$. However,
\citet{bin2} found that the double averaged orbital equations predict the
correct merger window and merger fractions even beyond their regime of validity
if the octupole effect is weak, so it is possible that our results concerning
$\theta_{\rm sl, f}$ are also somewhat robust even when the double averaged
equations formally break down. (iii) In this work, we only consider spin-orbit
coupling, and follow the evolution of $\theta_{\rm sl}$ until the orbital
separation is sufficiently small such that the inner binary is gravitationally
decoupled from the tertiary and the spin-orbit misalignment angle is frozen. To
leading post-Newtonian (PN) order, $\theta_{\rm sl}$ is constant for small
separations until the spin-spin interaction (2 PN) becomes important. This
interaction is non-negligible only when binary enters the LIGO band and when the
spin magnitude of each BH is appreciable \citep{bin3, yu2020spin}.

As noted in Section~\ref{s:intro}, the merging BH binaries detected by
LIGO/VIRGO in O1 and O2 have $\chi_{\rm eff} \sim 0$
\citep{Abbott:2016blz,abbott2019binary}. One possible explanation for this is
that BHs are born slowly rotating \citep[e.g.][]{fuller2019most}. But our
``$90^\circ$ attractor'' provides an alternative explanation with no assumptions
on the BH spin magnitudes if the mergers are ``LK-induced'' and $\theta_{\rm sl,
0} \approx 0^\circ$. In the O3 event GW190521, each BH has a significant spin
magnitude and a large spin-orbit misalignment angle \citep{190521}. If the
evolution history of the system resembled our LK-induced scenario
\citep[see][]{bin_misc1}, a primordial $\theta_{\rm sl, 0} \approx 0^\circ$
would be consistent with the observed outcome of $\theta_{\rm sl, f} \sim
90^\circ$.

\section{Acknowledgements}

This work has been supported in part by the NSF grant AST-17152. YS is supported
by the NASA FINESST grant 19-ASTRO19-0041.

\bibliographystyle{apsrev4-2} 
\bibliography{Su_LK90}

\begin{thebibliography}{59}%
\makeatletter
\providecommand \@ifxundefined [1]{%
 \@ifx{#1\undefined}
}%
\providecommand \@ifnum [1]{%
 \ifnum #1\expandafter \@firstoftwo
 \else \expandafter \@secondoftwo
 \fi
}%
\providecommand \@ifx [1]{%
 \ifx #1\expandafter \@firstoftwo
 \else \expandafter \@secondoftwo
 \fi
}%
\providecommand \natexlab [1]{#1}%
\providecommand \enquote  [1]{``#1''}%
\providecommand \bibnamefont  [1]{#1}%
\providecommand \bibfnamefont [1]{#1}%
\providecommand \citenamefont [1]{#1}%
\providecommand \href@noop [0]{\@secondoftwo}%
\providecommand \href [0]{\begingroup \@sanitize@url \@href}%
\providecommand \@href[1]{\@@startlink{#1}\@@href}%
\providecommand \@@href[1]{\endgroup#1\@@endlink}%
\providecommand \@sanitize@url [0]{\catcode `\\12\catcode `\$12\catcode
  `\&12\catcode `\#12\catcode `\^12\catcode `\_12\catcode `\%12\relax}%
\providecommand \@@startlink[1]{}%
\providecommand \@@endlink[0]{}%
\providecommand \url  [0]{\begingroup\@sanitize@url \@url }%
\providecommand \@url [1]{\endgroup\@href {#1}{\urlprefix }}%
\providecommand \urlprefix  [0]{URL }%
\providecommand \Eprint [0]{\href }%
\providecommand \doibase [0]{https://doi.org/}%
\providecommand \selectlanguage [0]{\@gobble}%
\providecommand \bibinfo  [0]{\@secondoftwo}%
\providecommand \bibfield  [0]{\@secondoftwo}%
\providecommand \translation [1]{[#1]}%
\providecommand \BibitemOpen [0]{}%
\providecommand \bibitemStop [0]{}%
\providecommand \bibitemNoStop [0]{.\EOS\space}%
\providecommand \EOS [0]{\spacefactor3000\relax}%
\providecommand \BibitemShut  [1]{\csname bibitem#1\endcsname}%
\let\auto@bib@innerbib\@empty
\bibitem [{\citenamefont {Abbott}\ \emph {et~al.}(2016)\citenamefont {Abbott},
  \citenamefont {Abbott}, \citenamefont {Abbott}, \citenamefont {Abernathy},
  \citenamefont {Acernese}, \citenamefont {Ackley}, \citenamefont {Adams},
  \citenamefont {Adams}, \citenamefont {Addesso}, \citenamefont {Adhikari}
  \emph {et~al.}}]{Abbott:2016blz}%
  \BibitemOpen
  \bibfield  {author} {\bibinfo {author} {\bibfnamefont {B.~P.}\ \bibnamefont
  {Abbott}}, \bibinfo {author} {\bibfnamefont {R.}~\bibnamefont {Abbott}},
  \bibinfo {author} {\bibfnamefont {T.}~\bibnamefont {Abbott}}, \bibinfo
  {author} {\bibfnamefont {M.}~\bibnamefont {Abernathy}}, \bibinfo {author}
  {\bibfnamefont {F.}~\bibnamefont {Acernese}}, \bibinfo {author}
  {\bibfnamefont {K.}~\bibnamefont {Ackley}}, \bibinfo {author} {\bibfnamefont
  {C.}~\bibnamefont {Adams}}, \bibinfo {author} {\bibfnamefont
  {T.}~\bibnamefont {Adams}}, \bibinfo {author} {\bibfnamefont
  {P.}~\bibnamefont {Addesso}}, \bibinfo {author} {\bibfnamefont
  {R.}~\bibnamefont {Adhikari}}, \emph {et~al.},\ }\href@noop {} {\bibfield
  {journal} {\bibinfo  {journal} {Physical review letters}\ }\textbf {\bibinfo
  {volume} {116}},\ \bibinfo {pages} {131102} (\bibinfo {year}
  {2016})}\BibitemShut {NoStop}%
\bibitem [{\citenamefont {Abbott}\ \emph {et~al.}(2019)\citenamefont {Abbott},
  \citenamefont {Abbott}, \citenamefont {Abbott}, \citenamefont {Abraham},
  \citenamefont {Acernese}, \citenamefont {Ackley}, \citenamefont {Adams},
  \citenamefont {Adhikari}, \citenamefont {Adya}, \citenamefont {Affeldt} \emph
  {et~al.}}]{abbott2019binary}%
  \BibitemOpen
  \bibfield  {author} {\bibinfo {author} {\bibfnamefont {B.}~\bibnamefont
  {Abbott}}, \bibinfo {author} {\bibfnamefont {R.}~\bibnamefont {Abbott}},
  \bibinfo {author} {\bibfnamefont {T.}~\bibnamefont {Abbott}}, \bibinfo
  {author} {\bibfnamefont {S.}~\bibnamefont {Abraham}}, \bibinfo {author}
  {\bibfnamefont {F.}~\bibnamefont {Acernese}}, \bibinfo {author}
  {\bibfnamefont {K.}~\bibnamefont {Ackley}}, \bibinfo {author} {\bibfnamefont
  {C.}~\bibnamefont {Adams}}, \bibinfo {author} {\bibfnamefont
  {R.}~\bibnamefont {Adhikari}}, \bibinfo {author} {\bibfnamefont
  {V.}~\bibnamefont {Adya}}, \bibinfo {author} {\bibfnamefont {C.}~\bibnamefont
  {Affeldt}}, \emph {et~al.},\ }\href@noop {} {\bibfield  {journal} {\bibinfo
  {journal} {The Astrophysical Journal Letters}\ }\textbf {\bibinfo {volume}
  {882}},\ \bibinfo {pages} {L24} (\bibinfo {year} {2019})}\BibitemShut
  {NoStop}%
\bibitem [{\citenamefont {Lipunov}\ \emph {et~al.}(1997)\citenamefont
  {Lipunov}, \citenamefont {Postnov},\ and\ \citenamefont
  {Prokhorov}}]{lipunov1997black}%
  \BibitemOpen
  \bibfield  {author} {\bibinfo {author} {\bibfnamefont {V.}~\bibnamefont
  {Lipunov}}, \bibinfo {author} {\bibfnamefont {K.}~\bibnamefont {Postnov}},\
  and\ \bibinfo {author} {\bibfnamefont {M.}~\bibnamefont {Prokhorov}},\
  }\href@noop {} {\bibfield  {journal} {\bibinfo  {journal} {Astronomy
  Letters}\ }\textbf {\bibinfo {volume} {23}},\ \bibinfo {pages} {492}
  (\bibinfo {year} {1997})}\BibitemShut {NoStop}%
\bibitem [{\citenamefont {Lipunov}\ \emph {et~al.}(2017)\citenamefont
  {Lipunov}, \citenamefont {Kornilov}, \citenamefont {Gorbovskoy},
  \citenamefont {Buckley}, \citenamefont {Tiurina}, \citenamefont {Balanutsa},
  \citenamefont {Kuznetsov}, \citenamefont {Greiner}, \citenamefont
  {Vladimirov}, \citenamefont {Vlasenko} \emph {et~al.}}]{lipunov2017first}%
  \BibitemOpen
  \bibfield  {author} {\bibinfo {author} {\bibfnamefont {V.}~\bibnamefont
  {Lipunov}}, \bibinfo {author} {\bibfnamefont {V.}~\bibnamefont {Kornilov}},
  \bibinfo {author} {\bibfnamefont {E.}~\bibnamefont {Gorbovskoy}}, \bibinfo
  {author} {\bibfnamefont {D.}~\bibnamefont {Buckley}}, \bibinfo {author}
  {\bibfnamefont {N.}~\bibnamefont {Tiurina}}, \bibinfo {author} {\bibfnamefont
  {P.}~\bibnamefont {Balanutsa}}, \bibinfo {author} {\bibfnamefont
  {A.}~\bibnamefont {Kuznetsov}}, \bibinfo {author} {\bibfnamefont
  {J.}~\bibnamefont {Greiner}}, \bibinfo {author} {\bibfnamefont
  {V.}~\bibnamefont {Vladimirov}}, \bibinfo {author} {\bibfnamefont
  {D.}~\bibnamefont {Vlasenko}}, \emph {et~al.},\ }\href@noop {} {\bibfield
  {journal} {\bibinfo  {journal} {Monthly Notices of the Royal Astronomical
  Society}\ }\textbf {\bibinfo {volume} {465}},\ \bibinfo {pages} {3656}
  (\bibinfo {year} {2017})}\BibitemShut {NoStop}%
\bibitem [{\citenamefont {Podsiadlowski}\ \emph {et~al.}(2003)\citenamefont
  {Podsiadlowski}, \citenamefont {Rappaport},\ and\ \citenamefont
  {Han}}]{podsiadlowski2003formation}%
  \BibitemOpen
  \bibfield  {author} {\bibinfo {author} {\bibfnamefont {P.}~\bibnamefont
  {Podsiadlowski}}, \bibinfo {author} {\bibfnamefont {S.}~\bibnamefont
  {Rappaport}},\ and\ \bibinfo {author} {\bibfnamefont {Z.}~\bibnamefont
  {Han}},\ }\href@noop {} {\bibfield  {journal} {\bibinfo  {journal} {Monthly
  Notices of the Royal Astronomical Society}\ }\textbf {\bibinfo {volume}
  {341}},\ \bibinfo {pages} {385} (\bibinfo {year} {2003})}\BibitemShut
  {NoStop}%
\bibitem [{\citenamefont {Belczynski}\ \emph {et~al.}(2010)\citenamefont
  {Belczynski}, \citenamefont {Dominik}, \citenamefont {Bulik}, \citenamefont
  {O'Shaughnessy}, \citenamefont {Fryer},\ and\ \citenamefont
  {Holz}}]{belczynski2010effect}%
  \BibitemOpen
  \bibfield  {author} {\bibinfo {author} {\bibfnamefont {K.}~\bibnamefont
  {Belczynski}}, \bibinfo {author} {\bibfnamefont {M.}~\bibnamefont {Dominik}},
  \bibinfo {author} {\bibfnamefont {T.}~\bibnamefont {Bulik}}, \bibinfo
  {author} {\bibfnamefont {R.}~\bibnamefont {O'Shaughnessy}}, \bibinfo {author}
  {\bibfnamefont {C.}~\bibnamefont {Fryer}},\ and\ \bibinfo {author}
  {\bibfnamefont {D.~E.}\ \bibnamefont {Holz}},\ }\href@noop {} {\bibfield
  {journal} {\bibinfo  {journal} {The Astrophysical Journal Letters}\ }\textbf
  {\bibinfo {volume} {715}},\ \bibinfo {pages} {L138} (\bibinfo {year}
  {2010})}\BibitemShut {NoStop}%
\bibitem [{\citenamefont {Belczynski}\ \emph {et~al.}(2016)\citenamefont
  {Belczynski}, \citenamefont {Holz}, \citenamefont {Bulik},\ and\
  \citenamefont {O'Shaughnessy}}]{belczynski2016first}%
  \BibitemOpen
  \bibfield  {author} {\bibinfo {author} {\bibfnamefont {K.}~\bibnamefont
  {Belczynski}}, \bibinfo {author} {\bibfnamefont {D.~E.}\ \bibnamefont
  {Holz}}, \bibinfo {author} {\bibfnamefont {T.}~\bibnamefont {Bulik}},\ and\
  \bibinfo {author} {\bibfnamefont {R.}~\bibnamefont {O'Shaughnessy}},\
  }\href@noop {} {\bibfield  {journal} {\bibinfo  {journal} {Nature}\ }\textbf
  {\bibinfo {volume} {534}},\ \bibinfo {pages} {512} (\bibinfo {year}
  {2016})}\BibitemShut {NoStop}%
\bibitem [{\citenamefont {Dominik}\ \emph {et~al.}(2012)\citenamefont
  {Dominik}, \citenamefont {Belczynski}, \citenamefont {Fryer}, \citenamefont
  {Holz}, \citenamefont {Berti}, \citenamefont {Bulik}, \citenamefont
  {Mandel},\ and\ \citenamefont {O'Shaughnessy}}]{dominik2012double}%
  \BibitemOpen
  \bibfield  {author} {\bibinfo {author} {\bibfnamefont {M.}~\bibnamefont
  {Dominik}}, \bibinfo {author} {\bibfnamefont {K.}~\bibnamefont {Belczynski}},
  \bibinfo {author} {\bibfnamefont {C.}~\bibnamefont {Fryer}}, \bibinfo
  {author} {\bibfnamefont {D.~E.}\ \bibnamefont {Holz}}, \bibinfo {author}
  {\bibfnamefont {E.}~\bibnamefont {Berti}}, \bibinfo {author} {\bibfnamefont
  {T.}~\bibnamefont {Bulik}}, \bibinfo {author} {\bibfnamefont
  {I.}~\bibnamefont {Mandel}},\ and\ \bibinfo {author} {\bibfnamefont
  {R.}~\bibnamefont {O'Shaughnessy}},\ }\href@noop {} {\bibfield  {journal}
  {\bibinfo  {journal} {The Astrophysical Journal}\ }\textbf {\bibinfo {volume}
  {759}},\ \bibinfo {pages} {52} (\bibinfo {year} {2012})}\BibitemShut
  {NoStop}%
\bibitem [{\citenamefont {Dominik}\ \emph {et~al.}(2013)\citenamefont
  {Dominik}, \citenamefont {Belczynski}, \citenamefont {Fryer}, \citenamefont
  {Holz}, \citenamefont {Berti}, \citenamefont {Bulik}, \citenamefont
  {Mandel},\ and\ \citenamefont {O'Shaughnessy}}]{dominik2013double}%
  \BibitemOpen
  \bibfield  {author} {\bibinfo {author} {\bibfnamefont {M.}~\bibnamefont
  {Dominik}}, \bibinfo {author} {\bibfnamefont {K.}~\bibnamefont {Belczynski}},
  \bibinfo {author} {\bibfnamefont {C.}~\bibnamefont {Fryer}}, \bibinfo
  {author} {\bibfnamefont {D.~E.}\ \bibnamefont {Holz}}, \bibinfo {author}
  {\bibfnamefont {E.}~\bibnamefont {Berti}}, \bibinfo {author} {\bibfnamefont
  {T.}~\bibnamefont {Bulik}}, \bibinfo {author} {\bibfnamefont
  {I.}~\bibnamefont {Mandel}},\ and\ \bibinfo {author} {\bibfnamefont
  {R.}~\bibnamefont {O'Shaughnessy}},\ }\href@noop {} {\bibfield  {journal}
  {\bibinfo  {journal} {The Astrophysical Journal}\ }\textbf {\bibinfo {volume}
  {779}},\ \bibinfo {pages} {72} (\bibinfo {year} {2013})}\BibitemShut
  {NoStop}%
\bibitem [{\citenamefont {Dominik}\ \emph {et~al.}(2015)\citenamefont
  {Dominik}, \citenamefont {Berti}, \citenamefont {O'Shaughnessy},
  \citenamefont {Mandel}, \citenamefont {Belczynski}, \citenamefont {Fryer},
  \citenamefont {Holz}, \citenamefont {Bulik},\ and\ \citenamefont
  {Pannarale}}]{dominik2015double}%
  \BibitemOpen
  \bibfield  {author} {\bibinfo {author} {\bibfnamefont {M.}~\bibnamefont
  {Dominik}}, \bibinfo {author} {\bibfnamefont {E.}~\bibnamefont {Berti}},
  \bibinfo {author} {\bibfnamefont {R.}~\bibnamefont {O'Shaughnessy}}, \bibinfo
  {author} {\bibfnamefont {I.}~\bibnamefont {Mandel}}, \bibinfo {author}
  {\bibfnamefont {K.}~\bibnamefont {Belczynski}}, \bibinfo {author}
  {\bibfnamefont {C.}~\bibnamefont {Fryer}}, \bibinfo {author} {\bibfnamefont
  {D.~E.}\ \bibnamefont {Holz}}, \bibinfo {author} {\bibfnamefont
  {T.}~\bibnamefont {Bulik}},\ and\ \bibinfo {author} {\bibfnamefont
  {F.}~\bibnamefont {Pannarale}},\ }\href@noop {} {\bibfield  {journal}
  {\bibinfo  {journal} {The Astrophysical Journal}\ }\textbf {\bibinfo {volume}
  {806}},\ \bibinfo {pages} {263} (\bibinfo {year} {2015})}\BibitemShut
  {NoStop}%
\bibitem [{\citenamefont {Postnov}\ and\ \citenamefont
  {Kuranov}(2019)}]{postnov2019black}%
  \BibitemOpen
  \bibfield  {author} {\bibinfo {author} {\bibfnamefont {K.}~\bibnamefont
  {Postnov}}\ and\ \bibinfo {author} {\bibfnamefont {A.}~\bibnamefont
  {Kuranov}},\ }\href@noop {} {\bibfield  {journal} {\bibinfo  {journal}
  {Monthly Notices of the Royal Astronomical Society}\ }\textbf {\bibinfo
  {volume} {483}},\ \bibinfo {pages} {3288} (\bibinfo {year}
  {2019})}\BibitemShut {NoStop}%
\bibitem [{\citenamefont {Belczynski}\ \emph {et~al.}(2020)\citenamefont
  {Belczynski}, \citenamefont {Klencki}, \citenamefont {Fields}, \citenamefont
  {Olejak}, \citenamefont {Berti}, \citenamefont {Meynet}, \citenamefont
  {Fryer}, \citenamefont {Holz}, \citenamefont {O'Shaughnessy}, \citenamefont
  {Brown} \emph {et~al.}}]{belczynski2020evolutionary}%
  \BibitemOpen
  \bibfield  {author} {\bibinfo {author} {\bibfnamefont {K.}~\bibnamefont
  {Belczynski}}, \bibinfo {author} {\bibfnamefont {J.}~\bibnamefont {Klencki}},
  \bibinfo {author} {\bibfnamefont {C.}~\bibnamefont {Fields}}, \bibinfo
  {author} {\bibfnamefont {A.}~\bibnamefont {Olejak}}, \bibinfo {author}
  {\bibfnamefont {E.}~\bibnamefont {Berti}}, \bibinfo {author} {\bibfnamefont
  {G.}~\bibnamefont {Meynet}}, \bibinfo {author} {\bibfnamefont
  {C.}~\bibnamefont {Fryer}}, \bibinfo {author} {\bibfnamefont
  {D.}~\bibnamefont {Holz}}, \bibinfo {author} {\bibfnamefont {R.}~\bibnamefont
  {O'Shaughnessy}}, \bibinfo {author} {\bibfnamefont {D.}~\bibnamefont
  {Brown}}, \emph {et~al.},\ }\href@noop {} {\bibfield  {journal} {\bibinfo
  {journal} {Astronomy \& Astrophysics}\ }\textbf {\bibinfo {volume} {636}},\
  \bibinfo {pages} {A104} (\bibinfo {year} {2020})}\BibitemShut {NoStop}%
\bibitem [{\citenamefont {Zwart}\ and\ \citenamefont
  {McMillan}(1999)}]{zwart1999black}%
  \BibitemOpen
  \bibfield  {author} {\bibinfo {author} {\bibfnamefont {S.~F.~P.}\
  \bibnamefont {Zwart}}\ and\ \bibinfo {author} {\bibfnamefont {S.~L.}\
  \bibnamefont {McMillan}},\ }\href@noop {} {\bibfield  {journal} {\bibinfo
  {journal} {The Astrophysical Journal Letters}\ }\textbf {\bibinfo {volume}
  {528}},\ \bibinfo {pages} {L17} (\bibinfo {year} {1999})}\BibitemShut
  {NoStop}%
\bibitem [{\citenamefont {O'leary}\ \emph {et~al.}(2006)\citenamefont
  {O'leary}, \citenamefont {Rasio}, \citenamefont {Fregeau}, \citenamefont
  {Ivanova},\ and\ \citenamefont {O'Shaughnessy}}]{o2006binary}%
  \BibitemOpen
  \bibfield  {author} {\bibinfo {author} {\bibfnamefont {R.~M.}\ \bibnamefont
  {O'leary}}, \bibinfo {author} {\bibfnamefont {F.~A.}\ \bibnamefont {Rasio}},
  \bibinfo {author} {\bibfnamefont {J.~M.}\ \bibnamefont {Fregeau}}, \bibinfo
  {author} {\bibfnamefont {N.}~\bibnamefont {Ivanova}},\ and\ \bibinfo {author}
  {\bibfnamefont {R.}~\bibnamefont {O'Shaughnessy}},\ }\href@noop {} {\bibfield
   {journal} {\bibinfo  {journal} {The Astrophysical Journal}\ }\textbf
  {\bibinfo {volume} {637}},\ \bibinfo {pages} {937} (\bibinfo {year}
  {2006})}\BibitemShut {NoStop}%
\bibitem [{\citenamefont {Miller}\ and\ \citenamefont
  {Lauburg}(2009)}]{miller2009mergers}%
  \BibitemOpen
  \bibfield  {author} {\bibinfo {author} {\bibfnamefont {M.~C.}\ \bibnamefont
  {Miller}}\ and\ \bibinfo {author} {\bibfnamefont {V.~M.}\ \bibnamefont
  {Lauburg}},\ }\href@noop {} {\bibfield  {journal} {\bibinfo  {journal} {The
  Astrophysical Journal}\ }\textbf {\bibinfo {volume} {692}},\ \bibinfo {pages}
  {917} (\bibinfo {year} {2009})}\BibitemShut {NoStop}%
\bibitem [{\citenamefont {Banerjee}\ \emph {et~al.}(2010)\citenamefont
  {Banerjee}, \citenamefont {Baumgardt},\ and\ \citenamefont
  {Kroupa}}]{banerjee2010stellar}%
  \BibitemOpen
  \bibfield  {author} {\bibinfo {author} {\bibfnamefont {S.}~\bibnamefont
  {Banerjee}}, \bibinfo {author} {\bibfnamefont {H.}~\bibnamefont
  {Baumgardt}},\ and\ \bibinfo {author} {\bibfnamefont {P.}~\bibnamefont
  {Kroupa}},\ }\href@noop {} {\bibfield  {journal} {\bibinfo  {journal}
  {Monthly Notices of the Royal Astronomical Society}\ }\textbf {\bibinfo
  {volume} {402}},\ \bibinfo {pages} {371} (\bibinfo {year}
  {2010})}\BibitemShut {NoStop}%
\bibitem [{\citenamefont {Downing}\ \emph {et~al.}(2010)\citenamefont
  {Downing}, \citenamefont {Benacquista}, \citenamefont {Giersz},\ and\
  \citenamefont {Spurzem}}]{downing2010compact}%
  \BibitemOpen
  \bibfield  {author} {\bibinfo {author} {\bibfnamefont {J.}~\bibnamefont
  {Downing}}, \bibinfo {author} {\bibfnamefont {M.}~\bibnamefont
  {Benacquista}}, \bibinfo {author} {\bibfnamefont {M.}~\bibnamefont
  {Giersz}},\ and\ \bibinfo {author} {\bibfnamefont {R.}~\bibnamefont
  {Spurzem}},\ }\href@noop {} {\bibfield  {journal} {\bibinfo  {journal}
  {Monthly Notices of the Royal Astronomical Society}\ }\textbf {\bibinfo
  {volume} {407}},\ \bibinfo {pages} {1946} (\bibinfo {year}
  {2010})}\BibitemShut {NoStop}%
\bibitem [{\citenamefont {Ziosi}\ \emph {et~al.}(2014)\citenamefont {Ziosi},
  \citenamefont {Mapelli}, \citenamefont {Branchesi},\ and\ \citenamefont
  {Tormen}}]{ziosi2014dynamics}%
  \BibitemOpen
  \bibfield  {author} {\bibinfo {author} {\bibfnamefont {B.~M.}\ \bibnamefont
  {Ziosi}}, \bibinfo {author} {\bibfnamefont {M.}~\bibnamefont {Mapelli}},
  \bibinfo {author} {\bibfnamefont {M.}~\bibnamefont {Branchesi}},\ and\
  \bibinfo {author} {\bibfnamefont {G.}~\bibnamefont {Tormen}},\ }\href@noop {}
  {\bibfield  {journal} {\bibinfo  {journal} {Monthly Notices of the Royal
  Astronomical Society}\ }\textbf {\bibinfo {volume} {441}},\ \bibinfo {pages}
  {3703} (\bibinfo {year} {2014})}\BibitemShut {NoStop}%
\bibitem [{\citenamefont {Rodriguez}\ \emph {et~al.}(2015)\citenamefont
  {Rodriguez}, \citenamefont {Morscher}, \citenamefont {Pattabiraman},
  \citenamefont {Chatterjee}, \citenamefont {Haster},\ and\ \citenamefont
  {Rasio}}]{rodriguez2015binary}%
  \BibitemOpen
  \bibfield  {author} {\bibinfo {author} {\bibfnamefont {C.~L.}\ \bibnamefont
  {Rodriguez}}, \bibinfo {author} {\bibfnamefont {M.}~\bibnamefont {Morscher}},
  \bibinfo {author} {\bibfnamefont {B.}~\bibnamefont {Pattabiraman}}, \bibinfo
  {author} {\bibfnamefont {S.}~\bibnamefont {Chatterjee}}, \bibinfo {author}
  {\bibfnamefont {C.-J.}\ \bibnamefont {Haster}},\ and\ \bibinfo {author}
  {\bibfnamefont {F.~A.}\ \bibnamefont {Rasio}},\ }\href@noop {} {\bibfield
  {journal} {\bibinfo  {journal} {Physical Review Letters}\ }\textbf {\bibinfo
  {volume} {115}},\ \bibinfo {pages} {051101} (\bibinfo {year}
  {2015})}\BibitemShut {NoStop}%
\bibitem [{\citenamefont {Samsing}\ and\ \citenamefont
  {Ramirez-Ruiz}(2017)}]{samsing2017assembly}%
  \BibitemOpen
  \bibfield  {author} {\bibinfo {author} {\bibfnamefont {J.}~\bibnamefont
  {Samsing}}\ and\ \bibinfo {author} {\bibfnamefont {E.}~\bibnamefont
  {Ramirez-Ruiz}},\ }\href@noop {} {\bibfield  {journal} {\bibinfo  {journal}
  {The Astrophysical Journal Letters}\ }\textbf {\bibinfo {volume} {840}},\
  \bibinfo {pages} {L14} (\bibinfo {year} {2017})}\BibitemShut {NoStop}%
\bibitem [{\citenamefont {Samsing}\ and\ \citenamefont
  {D’Orazio}(2018)}]{samsing2018black}%
  \BibitemOpen
  \bibfield  {author} {\bibinfo {author} {\bibfnamefont {J.}~\bibnamefont
  {Samsing}}\ and\ \bibinfo {author} {\bibfnamefont {D.~J.}\ \bibnamefont
  {D’Orazio}},\ }\href@noop {} {\bibfield  {journal} {\bibinfo  {journal}
  {Monthly Notices of the Royal Astronomical Society}\ }\textbf {\bibinfo
  {volume} {481}},\ \bibinfo {pages} {5445} (\bibinfo {year}
  {2018})}\BibitemShut {NoStop}%
\bibitem [{\citenamefont {Rodriguez}\ \emph {et~al.}(2018)\citenamefont
  {Rodriguez}, \citenamefont {Amaro-Seoane}, \citenamefont {Chatterjee},\ and\
  \citenamefont {Rasio}}]{rodriguez2018post}%
  \BibitemOpen
  \bibfield  {author} {\bibinfo {author} {\bibfnamefont {C.~L.}\ \bibnamefont
  {Rodriguez}}, \bibinfo {author} {\bibfnamefont {P.}~\bibnamefont
  {Amaro-Seoane}}, \bibinfo {author} {\bibfnamefont {S.}~\bibnamefont
  {Chatterjee}},\ and\ \bibinfo {author} {\bibfnamefont {F.~A.}\ \bibnamefont
  {Rasio}},\ }\href@noop {} {\bibfield  {journal} {\bibinfo  {journal}
  {Physical Review Letters}\ }\textbf {\bibinfo {volume} {120}},\ \bibinfo
  {pages} {151101} (\bibinfo {year} {2018})}\BibitemShut {NoStop}%
\bibitem [{\citenamefont {Gond{\'a}n}\ \emph {et~al.}(2018)\citenamefont
  {Gond{\'a}n}, \citenamefont {Kocsis}, \citenamefont {Raffai},\ and\
  \citenamefont {Frei}}]{gondan2018eccentric}%
  \BibitemOpen
  \bibfield  {author} {\bibinfo {author} {\bibfnamefont {L.}~\bibnamefont
  {Gond{\'a}n}}, \bibinfo {author} {\bibfnamefont {B.}~\bibnamefont {Kocsis}},
  \bibinfo {author} {\bibfnamefont {P.}~\bibnamefont {Raffai}},\ and\ \bibinfo
  {author} {\bibfnamefont {Z.}~\bibnamefont {Frei}},\ }\href@noop {} {\bibfield
   {journal} {\bibinfo  {journal} {The Astrophysical Journal}\ }\textbf
  {\bibinfo {volume} {860}},\ \bibinfo {pages} {5} (\bibinfo {year}
  {2018})}\BibitemShut {NoStop}%
\bibitem [{\citenamefont {{Liu}}\ and\ \citenamefont
  {{Lai}}(2020{\natexlab{a}})}]{bin_misc1}%
  \BibitemOpen
  \bibfield  {author} {\bibinfo {author} {\bibfnamefont {B.}~\bibnamefont
  {{Liu}}}\ and\ \bibinfo {author} {\bibfnamefont {D.}~\bibnamefont {{Lai}}},\
  }\href@noop {} {\bibfield  {journal} {\bibinfo  {journal} {arXiv}\ ,\
  \bibinfo {eid} {arXiv:2009.10068}} (\bibinfo {year} {2020}{\natexlab{a}})},\
  \Eprint {https://arxiv.org/abs/2009.10068} {arXiv:2009.10068 [astro-ph.HE]}
  \BibitemShut {NoStop}%
\bibitem [{\citenamefont {{Liu}}\ and\ \citenamefont
  {{Lai}}(2020{\natexlab{b}})}]{bin_misc2}%
  \BibitemOpen
  \bibfield  {author} {\bibinfo {author} {\bibfnamefont {B.}~\bibnamefont
  {{Liu}}}\ and\ \bibinfo {author} {\bibfnamefont {D.}~\bibnamefont {{Lai}}},\
  }\href {https://doi.org/10.1103/PhysRevD.102.023020} {\bibfield  {journal}
  {\bibinfo  {journal} {Physical Review D}\ }\textbf {\bibinfo {volume}
  {102}},\ \bibinfo {eid} {023020} (\bibinfo {year} {2020}{\natexlab{b}})},\
  \Eprint {https://arxiv.org/abs/2004.10205} {arXiv:2004.10205 [astro-ph.HE]}
  \BibitemShut {NoStop}%
\bibitem [{\citenamefont {{Liu}}\ \emph
  {et~al.}(2019{\natexlab{a}})\citenamefont {{Liu}}, \citenamefont {{Lai}},\
  and\ \citenamefont {{Wang}}}]{bin_misc3}%
  \BibitemOpen
  \bibfield  {author} {\bibinfo {author} {\bibfnamefont {B.}~\bibnamefont
  {{Liu}}}, \bibinfo {author} {\bibfnamefont {D.}~\bibnamefont {{Lai}}},\ and\
  \bibinfo {author} {\bibfnamefont {Y.-H.}\ \bibnamefont {{Wang}}},\ }\href
  {https://doi.org/10.3847/2041-8213/ab40c0} {\bibfield  {journal} {\bibinfo
  {journal} {The Astrophysical Journal Letters}\ }\textbf {\bibinfo {volume}
  {883}},\ \bibinfo {eid} {L7} (\bibinfo {year} {2019}{\natexlab{a}})},\
  \Eprint {https://arxiv.org/abs/1906.07726} {arXiv:1906.07726 [astro-ph.HE]}
  \BibitemShut {NoStop}%
\bibitem [{\citenamefont {{Liu}}\ \emph
  {et~al.}(2019{\natexlab{b}})\citenamefont {{Liu}}, \citenamefont {{Lai}},\
  and\ \citenamefont {{Wang}}}]{bin_misc4}%
  \BibitemOpen
  \bibfield  {author} {\bibinfo {author} {\bibfnamefont {B.}~\bibnamefont
  {{Liu}}}, \bibinfo {author} {\bibfnamefont {D.}~\bibnamefont {{Lai}}},\ and\
  \bibinfo {author} {\bibfnamefont {Y.-H.}\ \bibnamefont {{Wang}}},\ }\href
  {https://doi.org/10.3847/1538-4357/ab2dfb} {\bibfield  {journal} {\bibinfo
  {journal} {The Astrophysical Journal Letters}\ }\textbf {\bibinfo {volume}
  {881}},\ \bibinfo {eid} {41} (\bibinfo {year} {2019}{\natexlab{b}})},\
  \Eprint {https://arxiv.org/abs/1905.00427} {arXiv:1905.00427 [astro-ph.HE]}
  \BibitemShut {NoStop}%
\bibitem [{\citenamefont {{Liu}}\ and\ \citenamefont
  {{Lai}}(2019)}]{bin_misc5}%
  \BibitemOpen
  \bibfield  {author} {\bibinfo {author} {\bibfnamefont {B.}~\bibnamefont
  {{Liu}}}\ and\ \bibinfo {author} {\bibfnamefont {D.}~\bibnamefont {{Lai}}},\
  }\href {https://doi.org/10.1093/mnras/sty3432} {\bibfield  {journal}
  {\bibinfo  {journal} {Monthly Notices of the Royal Astronomical Society}\
  }\textbf {\bibinfo {volume} {483}},\ \bibinfo {pages} {4060} (\bibinfo {year}
  {2019})},\ \Eprint {https://arxiv.org/abs/1809.07767} {arXiv:1809.07767
  [astro-ph.HE]} \BibitemShut {NoStop}%
\bibitem [{\citenamefont {Blaes}\ \emph {et~al.}(2002)\citenamefont {Blaes},
  \citenamefont {Lee},\ and\ \citenamefont {Socrates}}]{blaes2002kozai}%
  \BibitemOpen
  \bibfield  {author} {\bibinfo {author} {\bibfnamefont {O.}~\bibnamefont
  {Blaes}}, \bibinfo {author} {\bibfnamefont {M.~H.}\ \bibnamefont {Lee}},\
  and\ \bibinfo {author} {\bibfnamefont {A.}~\bibnamefont {Socrates}},\
  }\href@noop {} {\bibfield  {journal} {\bibinfo  {journal} {The Astrophysical
  Journal}\ }\textbf {\bibinfo {volume} {578}},\ \bibinfo {pages} {775}
  (\bibinfo {year} {2002})}\BibitemShut {NoStop}%
\bibitem [{\citenamefont {Miller}\ and\ \citenamefont
  {Hamilton}(2002)}]{miller2002four}%
  \BibitemOpen
  \bibfield  {author} {\bibinfo {author} {\bibfnamefont {M.~C.}\ \bibnamefont
  {Miller}}\ and\ \bibinfo {author} {\bibfnamefont {D.~P.}\ \bibnamefont
  {Hamilton}},\ }\href@noop {} {\bibfield  {journal} {\bibinfo  {journal} {The
  Astrophysical Journal}\ }\textbf {\bibinfo {volume} {576}},\ \bibinfo {pages}
  {894} (\bibinfo {year} {2002})}\BibitemShut {NoStop}%
\bibitem [{\citenamefont {Wen}(2003)}]{wen2003eccentricity}%
  \BibitemOpen
  \bibfield  {author} {\bibinfo {author} {\bibfnamefont {L.}~\bibnamefont
  {Wen}},\ }\href@noop {} {\bibfield  {journal} {\bibinfo  {journal} {The
  Astrophysical Journal}\ }\textbf {\bibinfo {volume} {598}},\ \bibinfo {pages}
  {419} (\bibinfo {year} {2003})}\BibitemShut {NoStop}%
\bibitem [{\citenamefont {Antonini}\ and\ \citenamefont
  {Perets}(2012)}]{antonini2012secular}%
  \BibitemOpen
  \bibfield  {author} {\bibinfo {author} {\bibfnamefont {F.}~\bibnamefont
  {Antonini}}\ and\ \bibinfo {author} {\bibfnamefont {H.~B.}\ \bibnamefont
  {Perets}},\ }\href@noop {} {\bibfield  {journal} {\bibinfo  {journal} {The
  Astrophysical Journal}\ }\textbf {\bibinfo {volume} {757}},\ \bibinfo {pages}
  {27} (\bibinfo {year} {2012})}\BibitemShut {NoStop}%
\bibitem [{\citenamefont {Antonini}\ \emph {et~al.}(2017)\citenamefont
  {Antonini}, \citenamefont {Toonen},\ and\ \citenamefont
  {Hamers}}]{antonini2017binary}%
  \BibitemOpen
  \bibfield  {author} {\bibinfo {author} {\bibfnamefont {F.}~\bibnamefont
  {Antonini}}, \bibinfo {author} {\bibfnamefont {S.}~\bibnamefont {Toonen}},\
  and\ \bibinfo {author} {\bibfnamefont {A.~S.}\ \bibnamefont {Hamers}},\
  }\href@noop {} {\bibfield  {journal} {\bibinfo  {journal} {The Astrophysical
  Journal}\ }\textbf {\bibinfo {volume} {841}},\ \bibinfo {pages} {77}
  (\bibinfo {year} {2017})}\BibitemShut {NoStop}%
\bibitem [{\citenamefont {Silsbee}\ and\ \citenamefont
  {Tremaine}(2017)}]{silsbee2016lidov}%
  \BibitemOpen
  \bibfield  {author} {\bibinfo {author} {\bibfnamefont {K.}~\bibnamefont
  {Silsbee}}\ and\ \bibinfo {author} {\bibfnamefont {S.}~\bibnamefont
  {Tremaine}},\ }\href@noop {} {\bibfield  {journal} {\bibinfo  {journal} {The
  Astrophysical Journal}\ }\textbf {\bibinfo {volume} {836}},\ \bibinfo {pages}
  {39} (\bibinfo {year} {2017})}\BibitemShut {NoStop}%
\bibitem [{\citenamefont {Liu}\ and\ \citenamefont {Lai}(2017)}]{bin1}%
  \BibitemOpen
  \bibfield  {author} {\bibinfo {author} {\bibfnamefont {B.}~\bibnamefont
  {Liu}}\ and\ \bibinfo {author} {\bibfnamefont {D.}~\bibnamefont {Lai}},\
  }\href@noop {} {\bibfield  {journal} {\bibinfo  {journal} {The Astrophysical
  Journal Letters}\ }\textbf {\bibinfo {volume} {846}},\ \bibinfo {pages} {L11}
  (\bibinfo {year} {2017})}\BibitemShut {NoStop}%
\bibitem [{\citenamefont {Liu}\ and\ \citenamefont {Lai}(2018)}]{bin2}%
  \BibitemOpen
  \bibfield  {author} {\bibinfo {author} {\bibfnamefont {B.}~\bibnamefont
  {Liu}}\ and\ \bibinfo {author} {\bibfnamefont {D.}~\bibnamefont {Lai}},\
  }\href@noop {} {\bibfield  {journal} {\bibinfo  {journal} {The Astrophysical
  Journal}\ }\textbf {\bibinfo {volume} {863}},\ \bibinfo {pages} {68}
  (\bibinfo {year} {2018})}\BibitemShut {NoStop}%
\bibitem [{\citenamefont {Randall}\ and\ \citenamefont
  {Xianyu}(2018)}]{randall2018induced}%
  \BibitemOpen
  \bibfield  {author} {\bibinfo {author} {\bibfnamefont {L.}~\bibnamefont
  {Randall}}\ and\ \bibinfo {author} {\bibfnamefont {Z.-Z.}\ \bibnamefont
  {Xianyu}},\ }\href@noop {} {\bibfield  {journal} {\bibinfo  {journal} {The
  Astrophysical Journal}\ }\textbf {\bibinfo {volume} {853}},\ \bibinfo {pages}
  {93} (\bibinfo {year} {2018})}\BibitemShut {NoStop}%
\bibitem [{\citenamefont {Hoang}\ \emph {et~al.}(2018)\citenamefont {Hoang},
  \citenamefont {Naoz}, \citenamefont {Kocsis}, \citenamefont {Rasio},\ and\
  \citenamefont {Dosopoulou}}]{hoang2018black}%
  \BibitemOpen
  \bibfield  {author} {\bibinfo {author} {\bibfnamefont {B.-M.}\ \bibnamefont
  {Hoang}}, \bibinfo {author} {\bibfnamefont {S.}~\bibnamefont {Naoz}},
  \bibinfo {author} {\bibfnamefont {B.}~\bibnamefont {Kocsis}}, \bibinfo
  {author} {\bibfnamefont {F.~A.}\ \bibnamefont {Rasio}},\ and\ \bibinfo
  {author} {\bibfnamefont {F.}~\bibnamefont {Dosopoulou}},\ }\href@noop {}
  {\bibfield  {journal} {\bibinfo  {journal} {The Astrophysical Journal}\
  }\textbf {\bibinfo {volume} {856}},\ \bibinfo {pages} {140} (\bibinfo {year}
  {2018})}\BibitemShut {NoStop}%
\bibitem [{\citenamefont {Schmidt}\ \emph {et~al.}(2015)\citenamefont
  {Schmidt}, \citenamefont {Ohme},\ and\ \citenamefont
  {Hannam}}]{schmidt2015towards}%
  \BibitemOpen
  \bibfield  {author} {\bibinfo {author} {\bibfnamefont {P.}~\bibnamefont
  {Schmidt}}, \bibinfo {author} {\bibfnamefont {F.}~\bibnamefont {Ohme}},\ and\
  \bibinfo {author} {\bibfnamefont {M.}~\bibnamefont {Hannam}},\ }\href@noop {}
  {\bibfield  {journal} {\bibinfo  {journal} {Physical Review D}\ }\textbf
  {\bibinfo {volume} {91}},\ \bibinfo {pages} {024043} (\bibinfo {year}
  {2015})}\BibitemShut {NoStop}%
\bibitem [{\citenamefont {Zackay}\ \emph {et~al.}(2019)\citenamefont {Zackay},
  \citenamefont {Venumadhav}, \citenamefont {Dai}, \citenamefont {Roulet},\
  and\ \citenamefont {Zaldarriaga}}]{zackay2019highly}%
  \BibitemOpen
  \bibfield  {author} {\bibinfo {author} {\bibfnamefont {B.}~\bibnamefont
  {Zackay}}, \bibinfo {author} {\bibfnamefont {T.}~\bibnamefont {Venumadhav}},
  \bibinfo {author} {\bibfnamefont {L.}~\bibnamefont {Dai}}, \bibinfo {author}
  {\bibfnamefont {J.}~\bibnamefont {Roulet}},\ and\ \bibinfo {author}
  {\bibfnamefont {M.}~\bibnamefont {Zaldarriaga}},\ }\href@noop {} {\bibfield
  {journal} {\bibinfo  {journal} {Physical Review D}\ }\textbf {\bibinfo
  {volume} {100}},\ \bibinfo {pages} {023007} (\bibinfo {year}
  {2019})}\BibitemShut {NoStop}%
\bibitem [{\citenamefont {Venumadhav}\ \emph {et~al.}(2020)\citenamefont
  {Venumadhav}, \citenamefont {Zackay}, \citenamefont {Roulet}, \citenamefont
  {Dai},\ and\ \citenamefont {Zaldarriaga}}]{venumadhav2020new}%
  \BibitemOpen
  \bibfield  {author} {\bibinfo {author} {\bibfnamefont {T.}~\bibnamefont
  {Venumadhav}}, \bibinfo {author} {\bibfnamefont {B.}~\bibnamefont {Zackay}},
  \bibinfo {author} {\bibfnamefont {J.}~\bibnamefont {Roulet}}, \bibinfo
  {author} {\bibfnamefont {L.}~\bibnamefont {Dai}},\ and\ \bibinfo {author}
  {\bibfnamefont {M.}~\bibnamefont {Zaldarriaga}},\ }\href@noop {} {\bibfield
  {journal} {\bibinfo  {journal} {Physical Review D}\ }\textbf {\bibinfo
  {volume} {101}},\ \bibinfo {pages} {083030} (\bibinfo {year}
  {2020})}\BibitemShut {NoStop}%
\bibitem [{\citenamefont {Abbott}\ \emph
  {et~al.}(2020{\natexlab{a}})\citenamefont {Abbott}, \citenamefont {Abbott},
  \citenamefont {Abbott}, \citenamefont {Abernathy}, \citenamefont {Acernese},
  \citenamefont {Ackley}, \citenamefont {Adams}, \citenamefont {Adams},
  \citenamefont {Addesso}, \citenamefont {Adhikari} \emph {et~al.}}]{GW190412}%
  \BibitemOpen
  \bibfield  {author} {\bibinfo {author} {\bibfnamefont {B.~P.}\ \bibnamefont
  {Abbott}}, \bibinfo {author} {\bibfnamefont {R.}~\bibnamefont {Abbott}},
  \bibinfo {author} {\bibfnamefont {T.}~\bibnamefont {Abbott}}, \bibinfo
  {author} {\bibfnamefont {M.}~\bibnamefont {Abernathy}}, \bibinfo {author}
  {\bibfnamefont {F.}~\bibnamefont {Acernese}}, \bibinfo {author}
  {\bibfnamefont {K.}~\bibnamefont {Ackley}}, \bibinfo {author} {\bibfnamefont
  {C.}~\bibnamefont {Adams}}, \bibinfo {author} {\bibfnamefont
  {T.}~\bibnamefont {Adams}}, \bibinfo {author} {\bibfnamefont
  {P.}~\bibnamefont {Addesso}}, \bibinfo {author} {\bibfnamefont
  {R.}~\bibnamefont {Adhikari}}, \emph {et~al.},\ }\href@noop {} {\bibfield
  {journal} {\bibinfo  {journal} {Physical Review D}\ }\textbf {\bibinfo
  {volume} {102}},\ \bibinfo {pages} {043015} (\bibinfo {year}
  {2020}{\natexlab{a}})}\BibitemShut {NoStop}%
\bibitem [{\citenamefont {Abbott}\ \emph
  {et~al.}(2020{\natexlab{b}})\citenamefont {Abbott}, \citenamefont {Abbott},
  \citenamefont {Abraham}, \citenamefont {Acernese}, \citenamefont {Ackley},
  \citenamefont {Adams}, \citenamefont {Adhikari}, \citenamefont {Adya},
  \citenamefont {Affeldt}, \citenamefont {Agathos} \emph {et~al.}}]{190521}%
  \BibitemOpen
  \bibfield  {author} {\bibinfo {author} {\bibfnamefont {R.}~\bibnamefont
  {Abbott}}, \bibinfo {author} {\bibfnamefont {T.}~\bibnamefont {Abbott}},
  \bibinfo {author} {\bibfnamefont {S.}~\bibnamefont {Abraham}}, \bibinfo
  {author} {\bibfnamefont {F.}~\bibnamefont {Acernese}}, \bibinfo {author}
  {\bibfnamefont {K.}~\bibnamefont {Ackley}}, \bibinfo {author} {\bibfnamefont
  {C.}~\bibnamefont {Adams}}, \bibinfo {author} {\bibfnamefont
  {R.}~\bibnamefont {Adhikari}}, \bibinfo {author} {\bibfnamefont
  {V.}~\bibnamefont {Adya}}, \bibinfo {author} {\bibfnamefont {C.}~\bibnamefont
  {Affeldt}}, \bibinfo {author} {\bibfnamefont {M.}~\bibnamefont {Agathos}},
  \emph {et~al.},\ }\href@noop {} {\bibfield  {journal} {\bibinfo  {journal}
  {The Astrophysical Journal Letters}\ }\textbf {\bibinfo {volume} {900}},\
  \bibinfo {pages} {L13} (\bibinfo {year} {2020}{\natexlab{b}})}\BibitemShut
  {NoStop}%
\bibitem [{\citenamefont {Liu}\ \emph {et~al.}(2019)\citenamefont {Liu},
  \citenamefont {Lai},\ and\ \citenamefont {Wang}}]{bin3}%
  \BibitemOpen
  \bibfield  {author} {\bibinfo {author} {\bibfnamefont {B.}~\bibnamefont
  {Liu}}, \bibinfo {author} {\bibfnamefont {D.}~\bibnamefont {Lai}},\ and\
  \bibinfo {author} {\bibfnamefont {Y.-H.}\ \bibnamefont {Wang}},\ }\href@noop
  {} {\bibfield  {journal} {\bibinfo  {journal} {The Astrophysical Journal}\
  }\textbf {\bibinfo {volume} {881}},\ \bibinfo {pages} {41} (\bibinfo {year}
  {2019})}\BibitemShut {NoStop}%
\bibitem [{\citenamefont {Antonini}\ \emph {et~al.}(2018)\citenamefont
  {Antonini}, \citenamefont {Rodriguez}, \citenamefont {Petrovich},\ and\
  \citenamefont {Fischer}}]{antonini2018precessional}%
  \BibitemOpen
  \bibfield  {author} {\bibinfo {author} {\bibfnamefont {F.}~\bibnamefont
  {Antonini}}, \bibinfo {author} {\bibfnamefont {C.~L.}\ \bibnamefont
  {Rodriguez}}, \bibinfo {author} {\bibfnamefont {C.}~\bibnamefont
  {Petrovich}},\ and\ \bibinfo {author} {\bibfnamefont {C.~L.}\ \bibnamefont
  {Fischer}},\ }\href@noop {} {\bibfield  {journal} {\bibinfo  {journal}
  {Monthly Notices of the Royal Astronomical Society: Letters}\ }\textbf
  {\bibinfo {volume} {480}},\ \bibinfo {pages} {L58} (\bibinfo {year}
  {2018})}\BibitemShut {NoStop}%
\bibitem [{\citenamefont {Yu}\ \emph {et~al.}(2020)\citenamefont {Yu},
  \citenamefont {Ma}, \citenamefont {Giesler},\ and\ \citenamefont
  {Chen}}]{yu2020spin}%
  \BibitemOpen
  \bibfield  {author} {\bibinfo {author} {\bibfnamefont {H.}~\bibnamefont
  {Yu}}, \bibinfo {author} {\bibfnamefont {S.}~\bibnamefont {Ma}}, \bibinfo
  {author} {\bibfnamefont {M.}~\bibnamefont {Giesler}},\ and\ \bibinfo {author}
  {\bibfnamefont {Y.}~\bibnamefont {Chen}},\ }\href@noop {} {\bibfield
  {journal} {\bibinfo  {journal} {arXiv}\ } (\bibinfo {year}
  {2020})}\BibitemShut {NoStop}%
\bibitem [{\citenamefont {Liu}\ \emph {et~al.}(2015)\citenamefont {Liu},
  \citenamefont {Mu{\~n}oz},\ and\ \citenamefont {Lai}}]{bin_diego}%
  \BibitemOpen
  \bibfield  {author} {\bibinfo {author} {\bibfnamefont {B.}~\bibnamefont
  {Liu}}, \bibinfo {author} {\bibfnamefont {D.~J.}\ \bibnamefont {Mu{\~n}oz}},\
  and\ \bibinfo {author} {\bibfnamefont {D.}~\bibnamefont {Lai}},\ }\href@noop
  {} {\bibfield  {journal} {\bibinfo  {journal} {Monthly Notices of the Royal
  Astronomical Society}\ }\textbf {\bibinfo {volume} {447}},\ \bibinfo {pages}
  {747} (\bibinfo {year} {2015})}\BibitemShut {NoStop}%
\bibitem [{\citenamefont {Anderson}\ \emph {et~al.}(2016)\citenamefont
  {Anderson}, \citenamefont {Storch},\ and\ \citenamefont
  {Lai}}]{anderson2016formation}%
  \BibitemOpen
  \bibfield  {author} {\bibinfo {author} {\bibfnamefont {K.~R.}\ \bibnamefont
  {Anderson}}, \bibinfo {author} {\bibfnamefont {N.~I.}\ \bibnamefont
  {Storch}},\ and\ \bibinfo {author} {\bibfnamefont {D.}~\bibnamefont {Lai}},\
  }\href@noop {} {\bibfield  {journal} {\bibinfo  {journal} {Monthly Notices of
  the Royal Astronomical Society}\ }\textbf {\bibinfo {volume} {456}},\
  \bibinfo {pages} {3671} (\bibinfo {year} {2016})}\BibitemShut {NoStop}%
\bibitem [{\citenamefont {Kinoshita}(1993)}]{kinoshita}%
  \BibitemOpen
  \bibfield  {author} {\bibinfo {author} {\bibfnamefont {H.}~\bibnamefont
  {Kinoshita}},\ }\href@noop {} {\bibfield  {journal} {\bibinfo  {journal}
  {Celestial Mechanics and Dynamical Astronomy}\ }\textbf {\bibinfo {volume}
  {57}},\ \bibinfo {pages} {359} (\bibinfo {year} {1993})}\BibitemShut
  {NoStop}%
\bibitem [{\citenamefont {Storch}\ and\ \citenamefont {Lai}(2015)}]{storch}%
  \BibitemOpen
  \bibfield  {author} {\bibinfo {author} {\bibfnamefont {N.~I.}\ \bibnamefont
  {Storch}}\ and\ \bibinfo {author} {\bibfnamefont {D.}~\bibnamefont {Lai}},\
  }\href@noop {} {\bibfield  {journal} {\bibinfo  {journal} {Monthly Notices of
  the Royal Astronomical Society}\ }\textbf {\bibinfo {volume} {448}},\
  \bibinfo {pages} {1821} (\bibinfo {year} {2015})}\BibitemShut {NoStop}%
\bibitem [{\citenamefont {Storch}\ \emph {et~al.}(2014)\citenamefont {Storch},
  \citenamefont {Anderson},\ and\ \citenamefont {Lai}}]{storch2014chaotic}%
  \BibitemOpen
  \bibfield  {author} {\bibinfo {author} {\bibfnamefont {N.~I.}\ \bibnamefont
  {Storch}}, \bibinfo {author} {\bibfnamefont {K.~R.}\ \bibnamefont
  {Anderson}},\ and\ \bibinfo {author} {\bibfnamefont {D.}~\bibnamefont
  {Lai}},\ }\href@noop {} {\bibfield  {journal} {\bibinfo  {journal} {Science}\
  }\textbf {\bibinfo {volume} {345}},\ \bibinfo {pages} {1317} (\bibinfo {year}
  {2014})}\BibitemShut {NoStop}%
\bibitem [{\citenamefont {Storch}\ \emph {et~al.}(2017)\citenamefont {Storch},
  \citenamefont {Lai},\ and\ \citenamefont {Anderson}}]{storch2017dynamics}%
  \BibitemOpen
  \bibfield  {author} {\bibinfo {author} {\bibfnamefont {N.~I.}\ \bibnamefont
  {Storch}}, \bibinfo {author} {\bibfnamefont {D.}~\bibnamefont {Lai}},\ and\
  \bibinfo {author} {\bibfnamefont {K.~R.}\ \bibnamefont {Anderson}},\
  }\href@noop {} {\bibfield  {journal} {\bibinfo  {journal} {Monthly Notices of
  the Royal Astronomical Society}\ }\textbf {\bibinfo {volume} {465}},\
  \bibinfo {pages} {3927} (\bibinfo {year} {2017})}\BibitemShut {NoStop}%
\bibitem [{\citenamefont {Floquet}(1883)}]{floquet1883equations}%
  \BibitemOpen
  \bibfield  {author} {\bibinfo {author} {\bibfnamefont {G.}~\bibnamefont
  {Floquet}},\ }in\ \href@noop {} {\emph {\bibinfo {booktitle} {Annales
  scientifiques de l'{\'E}cole normale sup{\'e}rieure}}},\ Vol.~\bibinfo
  {volume} {12}\ (\bibinfo {year} {1883})\ pp.\ \bibinfo {pages}
  {47--88}\BibitemShut {NoStop}%
\bibitem [{\citenamefont {Chicone}(2006)}]{chicone2006ordinary}%
  \BibitemOpen
  \bibfield  {author} {\bibinfo {author} {\bibfnamefont {C.}~\bibnamefont
  {Chicone}},\ }\href@noop {} {\emph {\bibinfo {title} {Ordinary differential
  equations with applications}}},\ Vol.~\bibinfo {volume} {34}\ (\bibinfo
  {publisher} {Springer Science \& Business Media},\ \bibinfo {year}
  {2006})\BibitemShut {NoStop}%
\bibitem [{Note1()}]{Note1}%
  \BibitemOpen
  \bibinfo {note} {More formally, $\protect \boldsymbol {\protect \mathbf
  {\protect \tilde {M}}} = \protect \boldsymbol {\protect \mathbf {\protect
  \tilde {\Phi }}}(P_{\protect \rm LK})$ where $\protect \boldsymbol {\protect
  \mathbf {\protect \tilde {\Phi }}}(t)$ is the \protect \emph {principal
  fundamental matrix solution}: the columns of $\protect \boldsymbol {\protect
  \mathbf {\protect \tilde {\Phi }}}$ are solutions to Eq.~\protect \textup
  {\hbox {\mathsurround \z@ \protect \normalfont (\ignorespaces \ref
  {eq:dsdt_weff}\unskip \@@italiccorr )}} and $\protect \boldsymbol {\protect
  \mathbf {\protect \tilde {\Phi }}}(0)$ is the identity. By linearity, the
  columns of $\protect \boldsymbol {\protect \mathbf {\protect \tilde {\Phi
  }}}(t)$ remain orthonormal, while its determinant does not change, so
  $\protect \boldsymbol {\protect \mathbf {\protect \tilde {M}}}$ is a proper
  orthogonal matrix, or a rotation matrix.}\BibitemShut {Stop}%
\bibitem [{\citenamefont {Zwillinger}(2002)}]{zwillinger2002crc}%
  \BibitemOpen
  \bibfield  {author} {\bibinfo {author} {\bibfnamefont {D.}~\bibnamefont
  {Zwillinger}},\ }\href@noop {} {\emph {\bibinfo {title} {CRC standard
  mathematical tables and formulae}}}\ (\bibinfo  {publisher} {CRC press},\
  \bibinfo {year} {2002})\BibitemShut {NoStop}%
\bibitem [{Note2()}]{Note2}%
  \BibitemOpen
  \bibinfo {note} {This approximation is suitable for the problem studied in
  the main text because the only characteristic frequency scale is $j_{\protect
  \qopname \relax m{min}}^{-1} \gg 1$, so all Fourier harmonics $\protect
  \boldsymbol {\protect \mathbf {\Omega }}_{\protect \rm eN}$ for $N \lesssim
  j_{\protect \qopname \relax m{min}}^{-1}$ are similar.}\BibitemShut {Stop}%
\bibitem [{\citenamefont {Fuller}\ and\ \citenamefont
  {Ma}(2019)}]{fuller2019most}%
  \BibitemOpen
  \bibfield  {author} {\bibinfo {author} {\bibfnamefont {J.}~\bibnamefont
  {Fuller}}\ and\ \bibinfo {author} {\bibfnamefont {L.}~\bibnamefont {Ma}},\
  }\href@noop {} {\bibfield  {journal} {\bibinfo  {journal} {The Astrophysical
  Journal Letters}\ }\textbf {\bibinfo {volume} {881}},\ \bibinfo {pages} {L1}
  (\bibinfo {year} {2019})}\BibitemShut {NoStop}%
\bibitem [{\citenamefont {Magnus}(1954)}]{magnus1954exponential}%
  \BibitemOpen
  \bibfield  {author} {\bibinfo {author} {\bibfnamefont {W.}~\bibnamefont
  {Magnus}},\ }\href@noop {} {\bibfield  {journal} {\bibinfo  {journal}
  {Communications on pure and applied mathematics}\ }\textbf {\bibinfo {volume}
  {7}},\ \bibinfo {pages} {649} (\bibinfo {year} {1954})}\BibitemShut {NoStop}%
\end{thebibliography}%

\clearpage
\appendix

\section{Floquet Theory Analysis}\label{app:floquet}

In this appendix, we provide an alternative approach to analyzing the BH spin
dynamics based on Floquet theory \citep{chicone2006ordinary} that provides
results complementary to those in the main text. Although the resulting equations
cannot be analytically solved, they place strong constraints on the allowed
behavior of the system. Additionally, Eq.~\eqref{eq:harmonic_dqeff} has a
natural interpretation in this formulation, and its accuracy is numerically
tested in Appendix~\ref{app:numeric}. We again work in the corotating frame and
neglect GW dissipation, so the equation of motion is given by
Eq.~\eqref{eq:dsdt_weff}. If we define the matrix operator $\tilde{\bv{A}}$
satisfying $\tilde{\bv{A}}\bv{S} = \bv{\Omega}_{\rm e} \times \bv{S}$, then the
equation of motion is
\begin{align}
    \p{\rd{\bv{S}}{t}}_{\rm rot}
        &= \tilde{\bv{A}} \bv{S}\label{eq:app_dsdt_weff}.
\end{align}
Here, $\tilde{\bv{A}}$ is periodic with period $P_{\rm LK}$ [see
Eq.~\eqref{eq:PLK_def}].

\subsection{Without Nutation}\label{app:a1}

First, for simplicity, let us assume that $\bv{\Omega}_{\rm e}$ does not nutate,
so its orientation is fixed. In this case, Eq.~\eqref{eq:app_dsdt_weff} admits
an exact conserved quantity:
\begin{equation}
    \rd{}{t}\s{e^{-\Phi}\bv{S}} = 0,\label{eq:app_cons}
\end{equation}
where
\begin{equation}
    \Phi(t) \equiv \int\limits^t \tilde{\bv{A}}\;\mathrm{d}t.
\end{equation}
Separately, since Eq.~\eqref{eq:app_dsdt_weff} is linear and has periodic
coefficients, Floquet theory tells us that $\bv{S}\p{t + P_{\rm LK}}$ is related
to $\bv{S}(t)$ by the monodromy matrix $\tilde{\bv{M}}$:
\begin{equation}
    \bv{S}\p{t + P_{\rm LK}} = \tilde{\bv{M}}\bv{S}(t).\label{eq:app_M}
\end{equation}
Comparing Eqs.~\eqref{eq:app_cons} and~\eqref{eq:app_M}, we immediately find
that
\begin{align}
    \tilde{\bv{M}} &= \exp\s{
            \int\limits_0^{P_{\rm LK}}\tilde{\bv{A}}\;\mathrm{d}t}
            ,\nonumber\\
        &= \exp\s{P_{\rm LK} \overline{\tilde{\bv{A}}}},\label{eq:app_Mbase}
\end{align}
where again the overbar denotes time averaging. Note that $\tilde{\bv{M}}$ is a
rotation matrix, so it must have exactly one eigenvector with eigenvalue $1$;
call this eigenvector $\bv{R}$. But $\bv{R} = \hat{\bv{\Omega}}_{\rm e}$, so
$\bv{S} \cdot \hat{\bv{\Omega}}_{\rm e}$ must be constant for every $t = NP_{\rm
LK}$.

This example is somewhat trivial: since $\bv{\Omega}_{\rm e}$ does not nutate,
$\bv{S}$ just precesses around fixed $\bv{R} = \hat{\bv{\Omega}}_{\rm e}$
at a variable rate, and $\bv{S} \cdot \hat{\bv{\Omega}}_{\rm e}$ is conserved.
However, the equation of motion studied in Section~\ref{s:fast_merger}
[Eq.~\eqref{eq:dsdt_0only}] neglects nutation yet provides a good description of
the evolution of $\bar{\theta}_{\rm e}$. For the fiducial LK-induced merger,
Fig.~\ref{fig:4sim_90_350_zoom} shows that $\bv{\Omega}_{\rm e}$ nutates
substantially within a LK period when $\mathcal{A} \simeq 1$ (nutation is
equivalent to $\Omega_{\rm e1} \neq 0$ and $\Delta I_{\rm e1} \neq 0^\circ$).
We infer that, even when $\bv{\Omega}_{\rm e}$ does nutate appreciably, the
nutation can sometimes be neglected to good approximation.

In Section~\ref{s:harmonic}, we showed that being close to a resonance
$\overline{\Omega}_{\rm e} \approx M\Omega_{\rm LK}$ results in non-conservation
of $\bar{\theta}_{\rm e}$. However, the converse is not obviously true: our
approximate analysis does not prove that being far from these resonances
guarantees good conservation of $\bar{\theta}_{\rm e}$. In the next section, we
argue that, for the dynamics studied in this paper, this converse is likely true
as well.

\subsection{With Nutation}\label{app:a2}

When $\bv{\Omega}_{\rm e}$ is allowed to nutate within $P_{\rm LK}$, the
quantity given by Eq.~\eqref{eq:app_cons} is no longer conserved, as
\begin{align}
    \rd{}{t}\s{e^{-\Phi}\bv{S}} = e^{-\Phi}\rd{\bv{S}}{t}
        - \tilde{A}e^{-\Phi}\bv{S} \neq e^{-\Phi}\s{\rd{\bv{S}}{t} -
        \tilde{\bv{A}}\bv{S}} = 0.
\end{align}
Instead, we define two new quantities $\Phi'$ and $\tilde{\bv{B}}$ via
\begin{align}
    \Phi'(t) &\equiv \int\limits^t
        \p{\tilde{\bv{A}} + \tilde{\bv{B}}}\;\mathrm{d}t,\\
    \rd{}{t}\s{e^{-\Phi'}\bv{S}} &= 0.
\end{align}
This requires
\begin{equation}
    \tilde{\bv{B}} = \s{e^{-\Phi'}, \tilde{\bv{A}}}
        e^{\Phi'},\label{eq:app_bexact}
\end{equation}
where the square brackets denote the commutator. The monodromy matrix is then
\begin{equation}
    \tilde{\bv{M}} = \exp\s{
        \int\limits_0^{P_{\rm LK}}\p{\tilde{\bv{A}} + \tilde{\bv{B}}}
            \;\mathrm{d}t}.\label{eq:app_Mgen}
\end{equation}

We next want to understand when Eq.~\eqref{eq:app_Mgen} can be well approximated
by Eq.~\eqref{eq:app_Mbase}. We first expand the matrix
exponential using the Zassenhaus formula \citep[the inverse of the well-known
Baker-Campbell-Hausdorff formula, see e.g.][]{magnus1954exponential}
\begin{equation}
    \tilde{\bv{M}} = e^{P_{\rm LK}\overline{\tilde{\bv{A}}}}
        e^{P_{\rm LK}\overline{\tilde{\bv{B}}}}
        \exp \Bigg[-\frac{P_{\rm LK}^2\s{\overline{\tilde{\bv{A}}},
            \overline{\tilde{\bv{B}}}}}{2} +\dots\Bigg].\label{eq:m_gen}
\end{equation}
Note that since $\tilde{\bv{M}}$ is a rotation matrix (see
Section~\ref{ss:monodromy}), and $\exp\s{P_{\rm LK}\overline{\tilde{\bv{A}}}}$
is also a rotation matrix ($\tilde{\bv{A}}$ is skew-symmetric), the remainder
of the right-hand side above must also be a rotation matrix (as the rotation
matrices are closed under matrix multiplication). For convenience, define
\begin{align}
    \tilde{\bv{R}}_{\rm A} &\equiv
        e^{P_{\rm LK}\overline{\tilde{\bv{A}}}}&
    \tilde{\bv{R}}_{\rm B} &\equiv
        e^{P_{\rm LK}\overline{\tilde{\bv{B}}}}
        \exp\Bigg[-\frac{P_{\rm LK}^2\s{\overline{\tilde{\bv{A}}},
            \overline{\tilde{\bv{B}}}}}{2} +\dots\Bigg],
\end{align}
where $\tilde{\bv{R}}_{\rm A}$ and $\tilde{\bv{R}}_{\rm B}$ are rotation
matrices that effect rotations by angles $\theta_{\rm A}$ and $\theta_{\rm B}$
about their respective axes.

When can $\tilde{\bv{R}}_{\rm B}$ be neglected? From Eq.~\eqref{eq:app_bexact},
we see that $\tilde{\bv{B}} = 0$ vanishes ($\theta_{\rm B} = 0$) when
$\s{\tilde{\bv{A}}, \Phi} = 0$, which occurs when $\bv{\Omega}_{\rm e}$ does not
nutate, and we recover Eq.~\eqref{eq:app_Mbase}. In fact, we will argue later
that $\theta_{\rm B}$ is generally small for the spin dynamics studied in the
main text. However, a small $\theta_{\rm B}$ alone is not sufficient to
guarantee $\tilde{\bv{M}} \approx \tilde{\bv{R}}_{\rm A}$. To see this, note if
$\theta_{\rm B}$ is small, then $\tilde{\bv{R}}_{\rm B} \approx \bv{1}$ where
$\bv{1}$ is the $3\times3$ identity matrix. On the other hand, $\theta_{\rm A} =
\overline{\Omega}_{\rm e}T_{\rm LK}$. Then, if $\theta_{\rm A}$ is not too near
an integer multiple of $2\pi$, $\tilde{\bv{M}} \approx \tilde{\bv{R}}_{\rm A}$
and $\tilde{\bv{M}} \approx \tilde{\bv{R}}_{\rm A}$ as before. However, if
$\overline{\Omega}_{\rm e}T_{\rm LK} \approx 2\pi M$ for integer $M$, then
$\tilde{\bv{R}}_{\rm A}$ itself is near the identity as well, and
$\tilde{\bv{R}}_{\rm B}$ \emph{cannot} be neglected when calculating
$\tilde{\bv{M}}$. The criterion for neglecting $\tilde{\bv{R}}_{\rm B}$ is then
clear: $\theta_{\rm B}$ must be much closer to an integer multiple of $2\pi M$
than $\theta_{\rm A}$.

To complete this picture, we argue that, for the spin dynamics studied in the
main text, $\theta_{\rm B}$ is small, so generally $\tilde{\bv{R}}_{\rm B}
\approx \bv{1}$, and $\overline{\Omega}_{\rm e} \approx M\Omega_{\rm LK}$ is a
\emph{necessary} condition for $\bv{R}$ to differ significantly from
$\hat{\overline{\bv{\Omega}}}_{\rm e}$. To do this, we recall that
$\overline{\Omega}_{\rm e} P_{\rm LK} \lesssim 2\pi$ (Figs.~\ref{fig:dWs}
and~\ref{fig:dWs_inner}), and we seek to show that $\theta_{\rm B}$ must
generally be small compared to $\theta_{\rm A}$, which would imply $\theta_{\rm
B} \ll 2\pi$. We first approximate that $e^{\Phi'} \approx e^{\Phi}$ in
Eq.~\eqref{eq:app_bexact} (requiring $\theta_{\rm B} \ll \theta_{\rm A}$, which
we will verify retroactively, and being far from resonance, $\theta_{\rm A} \neq
2\pi$), which gives
\begin{equation}
    \tilde{\bv{B}} \approx
        \s{e^{-\Phi},\tilde{\bv{A}}}e^{\Phi}.\label{eq:bapprox}
\end{equation}
Next, recall that $\Phi$ is the integral of $-\tilde{\bv{A}}$, and so the
magnitude of $\overline{\tilde{\bv{B}}}$ [$\tilde{\bv{B}}$ primarily affects
$\tilde{\bv{M}}$ via its average, see Eq.~\eqref{eq:app_Mgen}] depends on the
average misalignment between the vectors $\bv{\Omega}_{\rm e}$ and $\int^t
\bv{\Omega}_{\rm e}\;\mathrm{d}t \simeq \overline{\bv{\Omega}}_{\rm e}$. There
are a few possible regimes to consider: (i) if $\mathcal{A} \gg 1$, then LK
oscillations are frozen, and $\bv{\Omega}_{\rm e}$ does not nutate; (ii) if
$\mathcal{A} \ll 1$, then $\uv{\Omega}_{\rm e} \approx \uv{L}_{\rm tot} \approx
\hat{\overline{\bv{\Omega}}}_{\rm e}$ for almost all of $T_{\rm Lk}$; and (iii)
if $\mathcal{A} \simeq 1$, then $e_{\max}$ cannot be too large
[Eq.~\eqref{eq:jemaxepsgr}], and so $\Omega_{\rm L}$ and $\Omega_{\rm SL}$
cannot vary too much within an LK cycle and the nutation of $\bv{\Omega}_{\rm
e}$ is limited. This analysis suggests that, at least far from resonance, the
nutation of $\bv{\Omega}_{\rm e}$ is limited, and the commutator in
Eq.~\eqref{eq:bapprox} is small in the sense that $\theta_{\rm B} \ll
\theta_{\rm A}$, justifying our earlier claim. While this analysis is not
rigorous, it suggests that the only resonances present in the system are near
$\overline{\Omega}_{\rm e} = M\Omega_{\rm LK}$ and that otherwise $\bv{R}
\parallel \overline{\bv{\Omega}}_{\rm e}$, in agreement with the results of the
main text.

\subsection{Quantitative Effect}\label{app:numeric}

Above, we have given a qualitative analysis of the exact solution for the
monodromy matrix $\tilde{\bv{M}}$. In this section, we aim to reconcile this
with the quantitative, approximate results in the text and suggest that the
results in the text constitute a complete characterization of the spin dynamics.

In Section~\ref{s:harmonic}, we found that one effect of the Fourier harmonics
in Eq.~\eqref{eq:dsdt_fullft} are fluctuations in $\bar{\theta}_{\rm e}$ when
$\overline{\Omega}_{\rm e} \approx M\Omega_{\rm LK}$ for some integer $M$ with
amplitude given by Eq.~\eqref{eq:harmonic_dqeff}. On the other hand, in
Appendix~\ref{app:a2}, we found that $\bv{R}$ is aligned with
$\overline{\bv{\Omega}}_{\rm e}$ except when $\overline{\Omega}_{\rm e}$ is
sufficiently close to $M\Omega_{\rm LK}$ that the nutation of $\bv{\Omega}_{\rm
e}$ becomes important, but we were not able to determine a closed-form
expression for the misalignment. In this section, we show numerically that the
formulas given in the main text give good predictions for the orientation of
$\bv{R}$.
\begin{figure}
    \centering
    \includegraphics[width=0.5\colummwidth]{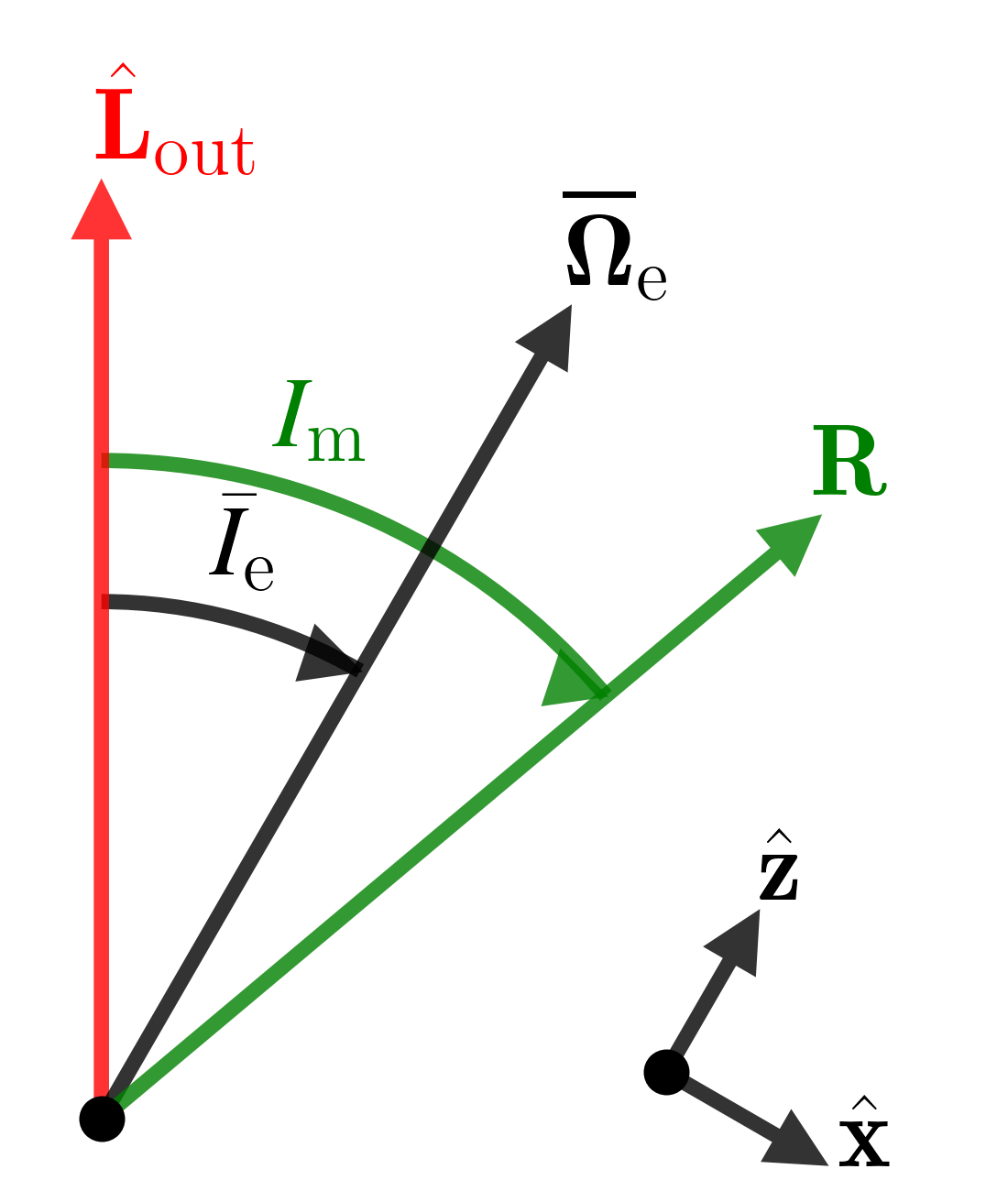}
    \caption{Definition of angles for numerical study of the monodromy matrix
    rotation axis. $\bv{R}$ is the eigenvector of the monodromy matrix
    $\tilde{\bv{M}}$ with eigenvalue $1$.}\label{fig:app_vecs}
\end{figure}

To validate the analytic prediction given by Eq.~\eqref{eq:harmonic_dqeff}, we
numerically compute $\tilde{\bv{M}}$. We study the $\eta \neq 0$, LK-enhanced
regime ($m_1 = m_2 = m_3 = 30M_{\odot}$, $a_0 = 0.1\;\mathrm{AU}$, $e_{\rm
0} = 10^{-3}$,$\tilde{a}_{\rm out} = 3\;\mathrm{AU}$). We still neglect GW
dissipation in order for Floquet analysis to be applicable. For $2000$ different
$I_0$ of the inner binary, we construct $\tilde{\bv{M}}$ by evolving the spin
equation of motion Eq.~\eqref{eq:dsdt_weff} starting with the three initial
conditions $\bv{S}_0 = \uv{x}$, $\bv{S}_0 = \uv{y}$, and $\bv{S}_0 = \bv{z}$
(see Fig.~\ref{fig:app_vecs}) over a single LK period, then using
\begin{equation}
    \tilde{\bv{M}} = \bv{\Phi}\p{P_{\rm LK}} \bv{\Phi}^{-1}(0)
        = \bv{\Phi}\p{P_{\rm LK}},
\end{equation}
where $\bv{\Phi}(t)$ is the \emph{principal fundamental matrix solution} whose
columns are solutions to Eq.~\eqref{eq:dsdt_weff} and $\bv{\Phi}(0)$ is the
identity. $\bv{R}$ is then the eigenvector that has eigenvalue $1$. Note that if
$\bv{v}$ is an eigenvector, so too is $-\bv{v}$; we choose convention that
$\bv{R}$ points in the same direction as $\overline{\bv{\Omega}}_{\rm e}$, i.e.\
$\Delta I_{\rm m} \equiv \abs{I_{\rm m} - \bar{I}_{\rm e}} < 90^\circ$.

\begin{figure}
    \centering
    \includegraphics[width=\colummwidth]{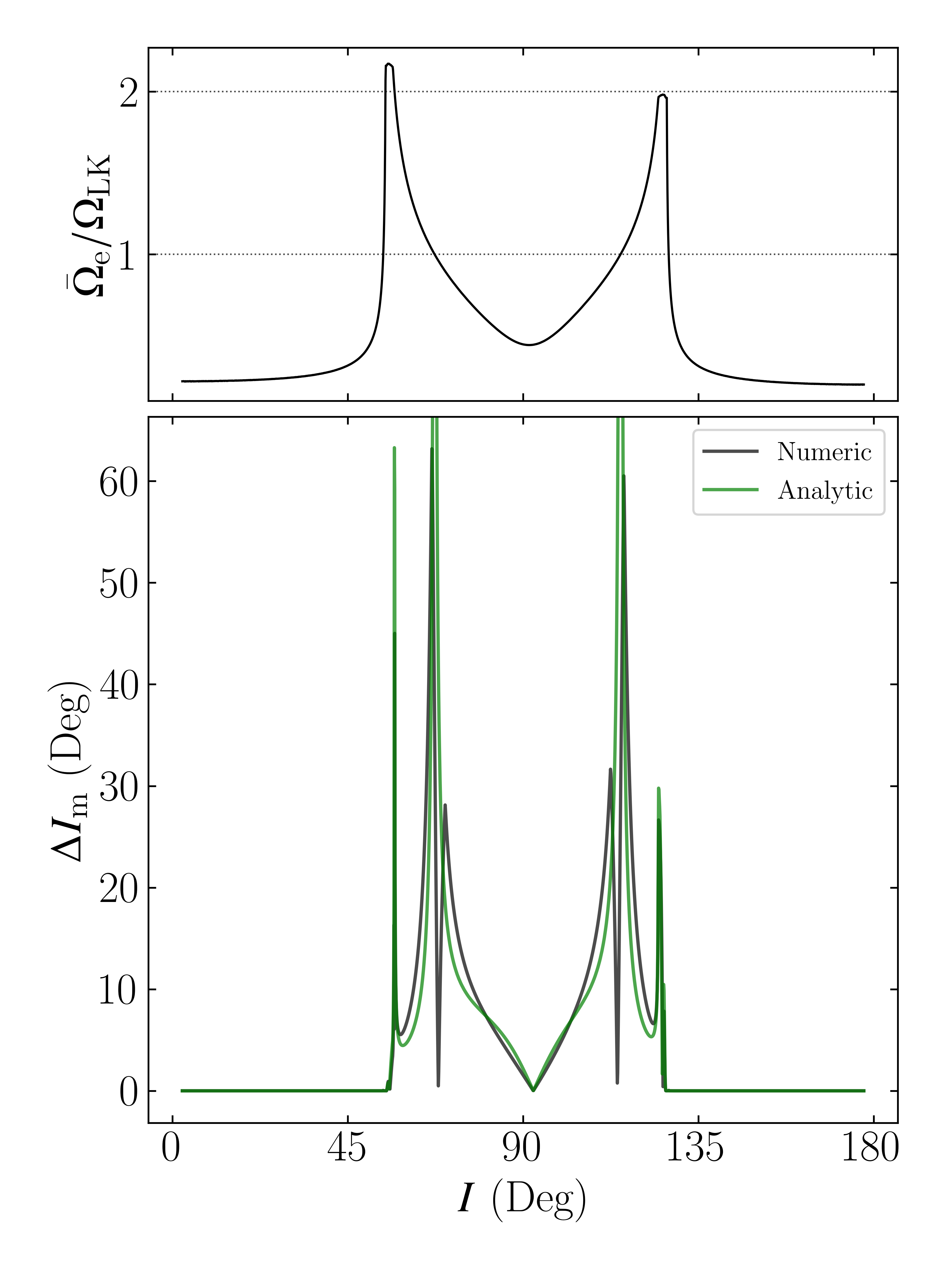}
    \caption{Comparison of the orientation $\bv{R}$ obtained from numerical
    simulations of Eq.~\eqref{eq:dsdt_weff} in the $\eta \neq 0$, LK-enhanced
    parameter regime ($m_1 = m_2 = m_3 = 30M_{\odot}$, $a_0 =
    0.1\;\mathrm{AU}$, $e_0 = 10^{-3}$,$\tilde{a}_{\rm out} =
    3\;\mathrm{AU}$) with the analytic resonance formula given by
    Eq.~\eqref{eq:app_im_pred} as a function of initial inclination, in the
    absence of GW dissipation. The top panel shows the ratio
    $\overline{\Omega}_{\rm e} / \Omega_{\rm LK}$ is shown as the solid black
    line, while the horizontal dashed lines denote $\overline{\Omega}_{\rm e} =
    \Omega_{\rm LK}$ and $\overline{\Omega}_{\rm e} = 2\Omega_{\rm LK}$. The
    bottom panel shows the misalignment angle between the numerically-computed
    $\bv{R}$ and $\overline{\bv{\Omega}}_{\rm e}$ as the black line. Separately,
    the predicted misalignment $\Delta I_{\rm m}$ due to interaction with the
    Fourier harmonics is given by Eq.~\eqref{eq:app_im_pred}. We see that the
    scaling of the misalignment angle near resonances is well captured
    by our analytic formula, but the numerical misalignment angle crosses $0$
    within the $M = 1$ resonance, which is not predicted by our simple
    theory.}\label{fig:finite_bin_comp_m}
\end{figure}

The orientation of $\bv{R}$ is related to the $\abs{\Delta \bar{\theta}_{\rm
e}}$ predicted by Eq.~\eqref{eq:harmonic_dqeff}: if $\bv{R}$ and
$\overline{\bv{\Omega}}_{\rm e}$ are misaligned by angle $\Delta I_{\rm m}$,
then $\bar{\theta}_{\rm e}$ oscillates with semi-amplitude $\Delta I_{\rm m}$.
Thus, we can infer $\Delta I_{\rm m}$ in the vicinity of each
$\overline{\Omega}_{\rm e} = M\Omega_{\rm LK}$ resonance from
Eq.~\eqref{eq:harmonic_dqeff}, and we obtain by linearity that
\begin{equation}
    \Delta I_{\rm m} \sim
        \sum\limits_{M = 1}^{\infty} \frac{d(M)}{2}
        \abs{\frac{\Omega_{\rm eM}\sin \Delta I_{\rm eM}
        }{\overline{\Omega}_{\rm e} - M\Omega_{\rm LK}}}.
        \label{eq:app_im_pred}
\end{equation}
Figure~\ref{fig:finite_bin_comp_m} shows that this calculation predicts the
numeric $\Delta I_{\rm m}$ well. In particular, the amplitude of both resonances
are well predicted, suggesting that the approximate factor $d(M)$ introduced in
Eq.~\eqref{eq:harmonic_dqeff} is accurate. This supports our assertion that
$\bv{R}$ deviates from $\hat{\overline{\bv{\Omega}}}_{\rm e}$ at resonances
$\overline{\Omega}_{\rm e} \approx M \Omega_{\rm LK}$, and the misalignment is
captured by Eq.~\eqref{eq:harmonic_dqeff}.

\end{document}